\documentclass[english,pra,preprint,nofootinbib,superscriptaddress]{revtex4-1}
\usepackage[T1]{fontenc}
\usepackage[latin9]{inputenc}
\usepackage{geometry}
\usepackage{color}
\usepackage{babel}
\usepackage{amsmath}
\usepackage{amssymb}
\usepackage{graphicx}
\usepackage{esint}
\usepackage[unicode=true,pdfusetitle,
 bookmarks=true,bookmarksnumbered=false,bookmarksopen=false,
 breaklinks=false,pdfborder={0 0 0},backref=false,colorlinks=true]
 {hyperref}
 
\begin{document}
\global\long\def\x{\times}
\global\long\def\t{\cdot}
\global\long\def\d{\mathrm{d}}
\global\long\def\ket#1{\left|#1\right\rangle }
\global\long\def\bra#1{\left\langle #1\right|}
\global\long\def\braket#1#2{\langle#1|#2\rangle}
\global\long\def\braoket#1#2#3{\left\langle #1\middle\vert#2\middle\vert#3\right\rangle }
\global\long\def\i{\mathrm{i}}
\global\long\def\e{\mathrm{e}}

\title{Propensity rules in photoelectron circular dichroism 
in chiral molecules 
II: General picture}

\author{Andres F. Ordonez}
\email{ordonez@mbi-berlin.de}
\affiliation{Max-Born-Institut, Berlin, Germany}
\affiliation{Technische Universit\"at Berlin, Berlin, Germany}
\author{Olga Smirnova}
\email{smirnova@mbi-berlin.de}
\affiliation{Max-Born-Institut, Berlin, Germany}
\affiliation{Technische Universit\"at Berlin, Berlin, Germany}

\begin{abstract}
Photoelectron circular dichroism results from one-photon ionization
of chiral molecules by circularly polarized light and manifests itself
in forward-backward asymmetry of electron emission in the direction
orthogonal to the light polarization plane. %What is the physical mechanism
%underlying asymmetric electron ejection? How ``which
%way'' information builds up in a chiral molecule and
%maps into forward-backward asymmetry? 
%We identify and intersect the geometrical and dynamical (in companion paper) origins of chiral response in photoionization.
To expose the physical mechanism
responsible for asymmetric electron ejection, we first establish a rigorous relation between 
the responses of unaligned and partially or perfectly aligned molecules. Next, we identify a propensity field, which is responsible for the chiral response in the electric-dipole approximation, i.e. a chiral response without magnetic interactions. %We show that the forward-backward asymmetry as a function photoelectron momentum $k$ is generated by the flux of the chiral field through the surface of the sphere of radius $k$ in momentum space.
We find that this propensity field, up to notations, is equivalent to the Berry curvature in a two-band solid. The propensity field directly encodes optical propensity rules, extending our conclusions regarding the role of propensity rules in defining the sign of forward-backward asymmetry from the specific case of chiral hydrogen \cite{ordonez_2018_hydrogen} to generic chiral systems.
Optical propensity rules underlie the 
%dynamical origin of the 
chiral response in photoelectron circular dichroism.
The enantiosensitive flux of the propensity field through the sphere in momentum space determines the forward-backward asymmetry in unaligned molecules and suggests a %complementary, 
geometrical origin of the chiral response. % By construction the flux  is similar to the Chern number in band insulators. By its function, it 
This flux has opposite sign for opposite enantiomers and vanishes for achiral molecules.
%Similarly, the Chern number discriminates topological insulators (non-zero Chern number) from trivial ones, where the Chern number is zero. 
%To address its dynamical origin we introduce
%instances of bound chiral wave functions resulting from stationary
%superpositions of states in a hydrogen atom. We show that the chiral
%response in one-photon ionization originates from two propensity rules:
%(i) sensitivity of ionization to the sense of electron rotation in
%polarization plane (ii) sensitivity of ionization to the direction
%of charge displacement or stationary current orthogonal to polarization
%plane. %The interference of two ionization pathways involving these
%selection rules is necessary to observe photoelectron circular dichroism.
%We also show why our conclusions derived for exotic chiral states
%remain valid for arbitrary chiral molecules.%, and derive general expressions
%for photoelectron circular dichroism in the presence of molecular alignment. 
\end{abstract}
\maketitle

\section{Introduction}
%We describe geometrical aspects of 
%photoelectron circular dichroism (PECD), i.e. circular dichroism in photoionization of chiral molecules \cite{ritchie_theory_1976, powis00, bowering_asymmetry_2001}.
Photoelectron circular dichroism (PECD) \cite{ritchie_theory_1976, powis00, bowering_asymmetry_2001} is an extremely efficient method of chiral discrimination, due to the very high value of circular dichroism, several orders of magnitude higher than in conventional optical methods, such as absorption circular dichroism or optical rotation (see e.g. \cite{condon_theories_1937}). 
PECD is intimately related \cite{ordonez_generalized_2018} to other phenomena where a chiral response arises already in the electric-dipole approximation, such as methods based on exciting rotational \cite{patterson_enantiomer-specific_2013,patterson_sensitive_2013,yurchenko_2016,patterson_2017},
electronic, and vibronic \cite{fischer2001isotropic,beaulieu_PXCD}
chiral dynamics without relying on relatively weak interactions with magnetic fields.
PECD is not only a promising technique of chiral discrimination but  also a powerful tool for studying ultrafast chiral dynamics in molecules as  documented in several experimental \cite{bowering_asymmetry_2001, garcia_2003, turchini_2004, Hergenhahn_2004, Lischke_2004, Stranges_2005, Giardini_2005, Harding_2005, nahon_determination_2006, Contini_2007, Garcia_2008, ulrich_giant_2008, Powis_2008_CPC_9, Powis_2008_PRA_78, Turchini_2009, Garcia_2010, Nahon_2010, Daly_2011, Catone_2012, Catone_2013, garcia13, Turchini_2013, Tia_2013, Powis_2014, Tia_2014, Nahon2016_det, Garcia_2016, Catone_2017} and theoretical
\cite{ritchie_theory_1976, cherepkov_circular_1982, powis00, Powis_lAlanine_2000, stener_density_2004, Harding_2006, Tommaso_2006, stener2006theoretical,  dreissigacker_photoelectron_2014, Artemyev_2015, Koch2017, Ilchen_2017, Tia_2017, Daly_2017, Miles_2017, ordonez_generalized_2018}
studies. PECD was recently extended to the multiphoton \cite{lux_circular_2012, lehmann_imaging_2013, Rafiee_2014, Rafiee_2015, lux_photoelectron_2015,  Lux_2016, rafiee2016wavelength, Kastner_2016, beaulieu_science_2017, Kastner_2017},
pump-probe \cite{comby_relaxation_2016} and strong-field ionization
regimes \cite{beaulieu_universality_2016, Beaulieu_2016_Faraday}.

%SOME WORDS ABOUT ALIGNMENT 

In this and in the companion paper \cite{ordonez_2018_hydrogen}  we focus on physical mechanisms underlying the chiral response
in one-photon ionization at the level of electrons. 
%The physical mechanism can be scrutinized at has two origins: the dynamical, pertinent to 
While the physical mechanism itself is the same for perfectly aligned, partially aligned, and randomly oriented ensembles of chiral molecules, the chiral response will have a different magnitude and may have a different sign in each case (see e.g. \cite{Tia_2017}). 
In our companion paper \cite{ordonez_2018_hydrogen} we have considered an example of chiral electronic states in hydrogen to identify %the physical mechanism of PECD relevant for aligned molecules. 
physical mechanisms relevant for PECD in aligned molecules.
%Now we will connect the chiral response of randomly oriented and aligned ensembles of enantiomers. 
Here we will expose the connection between the chiral response of aligned and unaligned molecular ensembles, 
%A similar connection is highly non-trivial in case of photoionization of non-chiral molecules. However, 
and show that since handedness is a rotationally invariant property,
the basic  structure of the molecular pseudoscalar remains the same in aligned and unaligned ensembles, providing a robust link between photoionization chiral observables in the two cases.

The rotationally invariant molecular pseudoscalar underlying the chiral response of randomly oriented ensembles \cite{ordonez_generalized_2018} is a scalar triple product of three vectors: the  photoionization dipole, its complex conjugate, and the photoelectron momentum. 
We find that the vector product of the photoionization dipole and its complex conjugate counterpart describes a propensity field in momentum space which underlies the chiral response in photoionization, and up to notations coincides with the Berry curvature in a two-band solid \cite{yao2008}. Similarly to the latter, this field explicitly reflects absorption circular dichroism resolved on photoelectron momentum and implicitly encodes optical propensity rules. 
%While the chiral response in unaligned molecules is governed by its flux through a sphere in momentum space, the chiral response in aligned molecules is only sensitive to specific components of this field.
Its flux through a sphere in momentum space determines the chiral response in PECD, and the effect of each of its components on the chiral response can be either enhanced or suppressed via molecular alignment. 
%The strength and direction of this field 
%are determined by 
%encode optical propensity rules. 
This way, we extend the
%conclusions obtained 
ideas presented
in our companion paper for chiral states in hydrogen \cite{ordonez_2018_hydrogen} to the general case of arbitrary chiral molecules. 
%While optical propensity rules 
%constitute the dynamical origin of 
%underlie the emergence of
%PECD, 
%The chiral response of unaligned molecules is proportional to the flux of molecular field
%the emergence of the flux of the propensity field points to %a geometrical origin of the chiral response.
The remarkable appearance of a flux of a Berry-curvature-like field in the description of  PECD points to the role of geometry in the emergence of the chiral response.

This paper is organized as follows: In Sec. \ref{sec:chiral_field} we introduce the propensity field and the chiral flux and discuss the interplay between dynamical and geometrical aspects of the chiral response. In Sec. \ref{sec:sign_FBA_unaligned} we establish the connection between the chiral response in unaligned and aligned molecules. In Sec. \ref{sec:PECD_aligned} we analyze the chiral response in aligned molecules in terms of the propensity field and the chiral flux density. Sec. \ref{sec:conclusions} concludes the paper.
%In aligned ensembles, the triple product is supplemented by similar contracts containing 

%Thus, to get to the core of the physical mechanism, pertinent for a single enatiomer, we first establish the relation between the chiral response of randomly oriented and aligned ensembles of enatiomers.
%This relation reveals the geometrical origin of chiral response in photoionization and is completely general.

\section{The physical meaning of the triple product in PECD and the propensity field. \label{sec:chiral_field}}

%What is the connection between the chiral response of aligned and
%unaligned molecules? 
Recently, we derived a simple and general expression
for PECD in unaligned (i.e. randomly oriented) molecular ensembles
\cite{ordonez_generalized_2018}. In this section we will begin by
inspecting this expression further in order to gain more insight into
its meaning.% and at the same time reveal its connection to the perfectly-oriented and perfectly-aligned cases. 

The expression for the orientation-averaged net photoelectron current density
in the lab frame resulting from photoionization of a randomly-oriented
molecular ensemble via an electric field circularly polarized in the
$x^{\mathrm{L}}y^{\mathrm{L}}$ plane is {[}see Eq. (13) in
Ref. \cite{ordonez_generalized_2018}{]}
%\begin{eqnarray}
%\vec{j}^{\mathrm{L}}\left(k\right) & = & \left\{ \frac{1}{6}\int\mathrm{d}\Omega_{k}^{\mathrm{M}}\left[\i\left(\vec{D}^{\mathrm{M}*}\x\vec{D}^{\mathrm{M}}\right)\t\vec{k}^{\mathrm{M}}\right]\right\} \left\{ \sigma\left|\tilde{\mathcal{E}}\right|^{2}\hat{z}^{\mathrm{L}}\right\} ,\label{eq:j}
%\end{eqnarray}
\begin{eqnarray}
\vec{j}^{\mathrm{L}}\left(k\right) & = & \left\{ \frac{1}{6}\int\mathrm{d}\Omega_{k}^{\mathrm{M}}\left[\i\left(\vec{D}^{\mathrm{M}*}\x\vec{D}^{\mathrm{M}}\right)\t\vec{k}^{\mathrm{M}}\right]\right\} \left\{\tilde{\vec{\mathcal{E}}}_x^{\mathrm{L}}\x\tilde{\vec{\mathcal{E}}}_y^{\mathrm{L}}\right\} ,\label{eq:jnew}
\end{eqnarray}
where the L and M superscripts indicate vectors expressed in the lab
and molecular frames, respectively. $\vec{D}\equiv\langle\vec{k}^{\mathrm{M}}\vert\hat{\vec{d}}\vert g\rangle$
is the transition dipole between the ground state and the scattering
state with photoelectron momentum $\vec{k}^{\mathrm{M}}$. 
$\tilde{\vec{\mathcal{E}}} = \tilde{\mathcal{E}}(\hat{x}^{\mathrm{L}}+\mathrm{i}\sigma\hat{y}^{\mathrm{L}})/\sqrt{2} \equiv(\tilde{\mathcal{E}}_{x}^{\mathrm{L}} + \mathrm{i}\tilde{\mathcal{E}}_{y}^{\mathrm{L}})/\sqrt{2}$,
 is the Fourier transform of the field at the transition frequency and $\sigma=\pm1$
defines the rotation direction of the field.

Equation \eqref{eq:jnew} shows that $\vec{j}^{\mathrm{L}}\left(k\right)$
can be factored into a molecule-specific rotationally-invariant pseudoscalar
and a field-specific pseudovector.
%\footnote{In Ref. \cite{ordonez_generalized_2018} we showed that for an arbitrary field
%we must replace $\mathrm{i}\{\tilde{\vec{\mathcal{E}}}_x^{\mathrm{L}}\x\tilde{\vec{\mathcal{E}}}_y^{\mathrm{L}}\}$
%by $\{\tilde{\vec{\mathcal{E}}}^{\mathrm{L}*}\x\tilde{\vec{\mathcal{E}}}^{\mathrm{L}}\}$.
%Here we will limit the discussion to the circularly polarized case.}. 
As shown in \cite{ordonez_generalized_2018}, the term $\i\vec{D}^{\mathrm{M}*}\x\vec{D}^{\mathrm{M}}$
has its origins in the interference between the transitions caused
by the $\hat{x}^{\mathrm{L}}$ and $\hat{y}^{\mathrm{L}}$ components
of the field, and it is the only ``part'' of $\vec{D}^{\mathrm{M}}$
that remains after averaging over all possible molecular orientations.
It is instructive to rewrite this term as
% To clarify the meaning of this term further, consider the absolute
% value squared of the spherical components $\pm$ of an arbitrary complex
% vector $\vec{v}$:

% \begin{align}
% \left|v_{\pm}\right|^{2} & =\left|\vec{v}\t\frac{\hat{x}\pm\i\hat{y}}{\sqrt{2}}\right|^{2},\nonumber \\
%  & =\frac{1}{2}\left[v_{x}^{2}+v_{y}^{2}\pm\left(\i v_{x}^{*}v_{y}-\i v_{x}v_{y}^{*}\right)\right],\nonumber \\
%  & =\frac{1}{2}\left[v_{x}^{2}+v_{y}^{2}\pm\left(\i\vec{v}^{*}\x\vec{v}\right)_{z}\right].
% \end{align}

% This shows how the interference term can be written as the $z$ component
% of a cross product and suggests to consider also the spherical components
% with respect to the $x$ and $y$ axes to obtain 

% \begin{equation}
% \i\vec{v}^{*}\x\vec{v}=\frac{1}{2}\left(\begin{array}{c}
% \left|\vec{v}\t\left(\hat{y}+\i\hat{z}\right)\right|^{2}-\left|\vec{v}\t\left(\hat{y}-\i\hat{z}\right)\right|^{2}\\
% \left|\vec{v}\t\left(\hat{z}+\i\hat{x}\right)\right|^{2}-\left|\vec{v}\t\left(\hat{z}-\i\hat{x}\right)\right|^{2}\\
% \left|\vec{v}\t\left(\hat{x}+\i\hat{y}\right)\right|^{2}-\left|\vec{v}\t\left(\hat{x}-\i\hat{y}\right)\right|^{2}
% \end{array}\right).\label{eq:iv*xv}
% \end{equation}

% %UPDATE TO GENERAL CASE BETWEEN TWO DIFFERENT VECTORS?

% When we apply this  formula, valid for any vector, to our case
% of interest we obtain

\begin{equation}
\i\left(\vec{D}^{\mathrm{M}*}\x\vec{D}^{\mathrm{M}}\right)=\frac{1}{2}\left(\begin{array}{c}
\left|\vec{D}^{\mathrm{M}}\t\left(\hat{y}^{\mathrm{M}}+\i\hat{z}^{\mathrm{M}}\right)\right|^{2}-\left|\vec{D}^{\mathrm{M}}\t\left(\hat{y}^{\mathrm{M}}-\i\hat{z}^{\mathrm{M}}\right)\right|^{2}\\
\left|\vec{D}^{\mathrm{M}}\t\left(\hat{z}^{\mathrm{M}}+\i\hat{x}^{\mathrm{M}}\right)\right|^{2}-\left|\vec{D}^{\mathrm{M}}\t\left(\hat{z}^{\mathrm{M}}-\i\hat{x}^{\mathrm{M}}\right)\right|^{2}\\
\left|\vec{D}^{\mathrm{M}}\t\left(\hat{x}^{\mathrm{M}}+\i\hat{y}^{\mathrm{M}}\right)\right|^{2}-\left|\vec{D}^{\mathrm{M}}\t\left(\hat{x}^{\mathrm{M}}-\i\hat{y}^{\mathrm{M}}\right)\right|^{2}
\end{array}\right),
\label{eq:CD}
\end{equation}

which shows that each component of $\i(\vec{D}^{\mathrm{M}*}\x\vec{D}^{\mathrm{M}})$
corresponds to the interference term that would arise if the molecule
(with fixed orientation) interacts with light circularly polarized
in the plane perpendicular to each molecular axis. %This ``geometrical'' factor will be important for establishing the connection between the chiral responses of aligned and unaligned molecules (Sec. \ref{sec:sign_FBA_unaligned}).

Equation \eqref{eq:CD} leads to several important conclusions: First,
the $i$-th component of $\i(\vec{D}^{\mathrm{M}*}\x\vec{D}^{\mathrm{M}})$
is simply the ``local'' (i.e. $\vec{k}^\mathrm{M}$-resolved) absorption circular dichroism for light circularly polarized
with respect to the $i$-th molecular axis (for a fixed molecular orientation). 
%Local circular dichroism, i.e.  difference in absorption of left and right circularly polarized light for a given photoelectron momentum $\vec{k}^{\mathrm{M}}$, manifests itself via photoionization propensity rules. 
Second, the $i$-th component  of $\i(\vec{D}^{\mathrm{M}*}\x\vec{D}^{\mathrm{M}})$ is non-zero only in the absence of rotational symmetry around the $i$-th axis. %If every component of $\i(\vec{D}^{\mathrm{M}*}\x\vec{D}^{\mathrm{M}})$ is non-zero, the system is chiral. 
Third, the $\vec{k}^{\mathrm{M}}$-dependent field $\i(\vec{D}^{\mathrm{M}*}\x\vec{D}^{\mathrm{M}})$ 
%emerges as a consequence of photoionization propensity rules 
encodes photoionization propensity rules and 
%represents an effective ``propensity'' field 
is analogous to the Berry curvature in two-band solids as we will demonstrate below. For comparison purposes, until the end of this section we will write $\hbar$, and the mass $m$ and charge of the electron $-e$ explicitly.

%Our result has been derived for dipoles 
Equation \eqref{eq:jnew} was derived in the length gauge. Since  for any two stationary states of the Hamiltonian we have that $\vec{p}_{fi} \equiv \mathrm{i}m\omega_{fi}\vec{r}_{fi}$, 
% \begin{equation}
% \bra{f}\vec{p}\ket{i} \equiv \frac{\mathrm{i}m}{\hbar}(E_f-E_i)\bra{f}\vec{r}\ket{i},
% \end{equation}
then the photoionization dipole defined above can be rewritten as
\begin{equation}\vec{D}^{\mathrm{M}}(\vec{k}^{\mathrm{M}}) \equiv \frac{\mathrm{i}\hbar e }{m(E(k)-E_g)}\vec{P}^{\mathrm{M}} (\vec{k}^{\mathrm{M}}),
\label{P}
\end{equation}
where $E(k)-E_g$ is the energy ``gap'' between the ground state of the molecule and the energy of photoelectron and $\vec{P}\equiv\langle\vec{k}^{\mathrm{M}}\vert\hat{\vec{p}}\vert g\rangle$
is the transition dipole between the ground state and the scattering
state with photoelectron momentum $\hbar \vec{k}^{\mathrm{M}}$, now defined in the velocity gauge. % as it is customary, for example, in solid state problems.
This simple relationship will allow us to uncover another interesting property of the vector product discussed above.
%is, up to a constant $-e^2$, where $e$ is an electron charge, equivalent to Berry curvature. 

Let us formally introduce a propensity  field $\vec{B}^{\mathrm{M}}(\vec{k}^{\mathrm{M}})$:
%\begin{equation}
%\vec{B}_{ch}(\vec{k}^{\mathrm{M}})={\i(\vec{D}^{\mathrm{M}*}\x\vec{D}^{\mathrm{M}})}\equiv -e^2 \vec{\Omega_{ch}},
%\end{equation}
\begin{align}
\vec{B}^{\mathrm{M}}(\vec{k}^{\mathrm{M}}) & \equiv -\frac{1}{e^2}\i\left[\vec{D}^{\mathrm{M}*}(\vec{k}^{\mathrm{M}})\x\vec{D}^{\mathrm{M}}(\vec{k}^{\mathrm{M}})\right] \label{eq:B}\\
& \equiv \mathrm{i}\frac{\hbar^2}{m^2}\frac{\left[\vec{P}^{\mathrm{M}}(\vec{k}^{\mathrm{M}})\x\vec{P}^{\mathrm{M}*}(\vec{k}^{\mathrm{M}})\right]}{\left(E(k)-E_g\right)^2}.
\end{align}
Note that, up to notation,  $\vec{B}^{\mathrm{M}}(\vec{k}^{\mathrm{M}})$  is equivalent to the Berry curvature $\Omega(\vec{k})$ of the upper band in a two-band solid (see e.g. \cite{yao2008})
% Indeed, the Berry curvature that reflects intercellular current circulation in the upper Bloch band, for the simple case of two bands (see e.g. \cite{yao2008}) is given by
\begin{equation}
\vec{\Omega}(\vec{k}) = \mathrm{i}\frac{\hbar^2}{m^2}\frac{\left[\vec{P}^{\mathrm{ci}}(\vec{k})\x\vec{P}^{\mathrm{ic}}(\vec{k})\right]}{\left(E_c(\vec{k})-E_i(\vec{k})\right)^2},
\end{equation}
where  $\vec{P}^{\mathrm{ci}}(\vec{k})=\vec{P}^{\mathrm{ic}*}(\vec{k})$ is the transition dipole matrix element between the two bands, and $E_i(\vec{k})$ and  $E_c(\vec{k})$ are the lower and upper band dispersions, respectively.
%Note that the enatiosensitive current $\vec{j}_z(k)$ is proportional to the flux of the chiral field through the surface of the sphere with radius $k$ in momentum space:
%\begin{eqnarray}
%\vec{j}^{\mathrm{L}}\left(k\right) & = &  \frac{1}{6}\left\{ \sigma\left|\tilde{\mathcal{E}}\right|^{2}\hat{z}^{\mathrm{L}}\right\}\frac{-e^2}{k}\int\mathrm{d}\vec{S} \vec{B}_{ch}(\vec{k}^{\mathrm{M}})  ,\label{eq:flux}
%\end{eqnarray}

The enantiosensitive net current $\vec{j}^{\mathrm{L}}(k)=j_z^{\mathrm{L}}\hat{z}^{\mathrm{L}}$ can be understood as arising due to an anisotropic enantiosensitive conductivity $\sigma_{z,xy}^{\chi}(k)$:
\begin{equation}
\vec{j}^{\mathrm{L}}(k) =  \sigma_{z,xy}^{\chi}(k) \left\{\tilde{\vec{\mathcal{E}}}_x^{\mathrm{L}}\x\tilde{\vec{\mathcal{E}}}_y^{\mathrm{L}}\right\}.\label{eq:cond}
\end{equation}
The conductivity $\sigma_{z,xy}^{\chi}(k)$ is proportional to to the flux of the propensity field through the surface of the sphere of radius $k$ in momentum space [cf. Eqs. \eqref{eq:jnew} and \eqref{eq:B}]:
\begin{eqnarray}
\sigma^{\chi}_{z,xy}(k) & \equiv &  \frac{e^3}{6\hbar k m } \int\mathrm{d}\vec{S}^{\mathrm{M}} \cdot \vec{B}^{\mathrm{M}}(\vec{k}^{\mathrm{M}}) ,\label{eq:flux}
\end{eqnarray}
where $\mathrm{d}\vec{S}^{\mathrm{M}}=k^2\mathrm{d}\Omega_{k}^{\mathrm{M}} (\vec{k}^{\mathrm{M}}/k)$ is the surface element, and the continuum wave functions used to calculate the transition dipoles are $k$-normalized\footnote{When $\hbar$, $m$, and $-e$ are written explicitly one must include a factor of $-e/(\hbar m)$ in Eq. \eqref{eq:jnew}}. The enantiosensitive flux 
\begin{eqnarray}
\Phi^{\chi}(k)\equiv \int\mathrm{d}\vec{S}^{\mathrm{M}} \cdot \vec{B}^{\mathrm{M}}(\vec{k}^{\mathrm{M}})
\label{eq:ch_flux}
\end{eqnarray}
is a molecular pseudoscalar, which defines the handedness of the enantiomer, i.e. the flux has opposite sign for opposite enantiomers. The relation between the propensity field and the enantiosensitive conductivity in Eq. (\ref{eq:flux}) is reminiscent of the one between the Berry curvature and the Hall conductivity (see e.g. \cite{yao2008}). 
%albeit $\vec{B}^{\mathrm{M}}$ is defined in three-dimensional $k$-space. 
Similarly, the relation between the enantiosensitive flux and the propensity field in Eq. \eqref{eq:ch_flux} is reminiscent of the relation between the Chern number and the Berry curvature of a given band in a two-dimensional solid. 

The propensity  field $\vec{B}^{\mathrm{M}}$ is related to the angular momentum of the photoelectron as follows:
\begin{align}
 \vec{L}^{\mathrm{M}}(\vec{k}^{\mathrm{M}}) & \equiv
\braoket{\vec{k}^{\mathrm{M}}}{\big[\vec{r}\x\vec{p}\big]}{\vec{k}^{\mathrm{M}}} \nonumber \\
 & = \sum_n 
 \braoket{\vec{k}^{\mathrm{M}}}{\vec{r}}{n} \x \braoket{n}{\vec{p}}{\vec{k}^{\mathrm{M}}} \nonumber \\
 & = \frac{m}{\hbar e^2}\sum_n (E_k-E_n)\i(\vec{D}_n^{\mathrm{M}*}\x\vec{D}_n^{\mathrm{M}}) \nonumber \\
 & = -\frac{m}{\hbar}\sum_n (E_k-E_n)\vec{B}_n^{\mathrm{M}}(\vec{k}^{\mathrm{M}}),
 \label{L}
\end{align}
where the sum is over all bound and continuum eigenstates $\vert n\rangle$ of the Hamiltonian, $\vec{D}_n^{\mathrm{M}}\equiv 
\langle\vec{k}^{\mathrm{M}}\vert\vec{r}\vert n\rangle$, and $\vec{B}_n^{\mathrm{M}}(\vec{k}^{\mathrm{M}})\equiv-\i(\vec{D}_n^{\mathrm{M}*}\x\vec{D}_n^{\mathrm{M}})/e^2$ in analogy with Eq. \eqref{eq:B}. Introducing the angular momentum  $\vec{L}_n^{\mathrm{M}}(\vec{k}^{\mathrm{M}})$ associated with the transition from a specific state $n$: 
\begin{equation}
  \vec{L}_n^{\mathrm{M}}(\vec{k}^{\mathrm{M}})\equiv  -\frac{m}{\hbar}
  (E_k-E_n)\vec{B}_n^{\mathrm{M}}(\vec{k}^{\mathrm{M}}),
\end{equation}
we find that the propensity field $\vec{B}^{\mathrm{M}}(\vec{k}^{\mathrm{M}})$ reflects the angular momentum $\vec{L}_g^{\mathrm{M}}(\vec{k}^{\mathrm{M}})$ associated with photoionization from the ground state. Since such angular momentum arises due to selection rules, its connection to the propensity field is natural. Thus, Eqs. \eqref{eq:CD}, \eqref{eq:cond} and \eqref{eq:flux} show that the enantiosensitve net current emerges as a result of  propensity rules. A specific example, explicitly demonstrating the interplay of two propensity rules has been described in the companion paper \cite{ordonez_2018_hydrogen}.

The helicity of a (spinless) photoelectron is given by the projection of its angular momentum on the direction of electron momentum: $\eta(\vec{k}^{\mathrm{M}})=\vec{L}^{\mathrm{M}}_g\cdot\frac{\vec{k}^{\mathrm{M}}}{\hbar k}$. 
% Evidently, the molecular pseudoscalar in Eq. \eqref{eq:jnew} is then proportional to the photoelectron helicity  integrated over all directions of photelectron momentum $\vec{k}^{\mathrm{M}}$:
Evidently, the molecular pseudoscalar in Eq. \eqref{eq:jnew}, the enantiosensitive conductivity \eqref{eq:flux} and flux \eqref{eq:ch_flux}, and the angle integrated photoelectron helicity, are all proportional to each other:
% \begin{equation}
% \phi(k)=\int\mathrm{d}\Omega_{k}^{\mathrm{M}}\phi^{\mathrm{M}}(\vec{k}^{\mathrm{M}}).
% \label{helicity}
% \end{equation}
% The helicity $\phi(k)$, up to constant, is equal to chiral flux:
% \begin{equation}
% \Phi^{\chi}(k)=e\phi(k).
% \label{flux_helic}
% \end{equation}
\begin{equation}
    \sigma_{z,xy}^{\chi}(k) = \frac{e^3}{6\hbar km} \Phi^{\chi}(k) = -\frac{e^3\hbar k}{6 m^2 (E_k - E_g)} \int \mathrm{d} \Omega_k^{\mathrm{M}} \eta(\vec{k}^{\mathrm{M}}).
    \label{eq:molecular_pseudoscalar}
\end{equation}
%Note that the factor $k^{-1}$ appears on the right hand side of Eq. (\ref{eq:flux}) due to the fact that $\vec{j}^{\mathrm{L}}\left(k\right)$ represents current density in momentum space, i.e. the total current $j^{tot}$ obtains once we integrate $\vec{j}^{\mathrm{L}}\left(k\right)$ over all photoelectron momenta $k$:  $\int \vec{j}^{\mathrm{L}}\left(k\right)k^2 dk=j^{tot}$. 

The propensity field $\vec{B}^{\mathrm{M}}(\vec{k}^{\mathrm{M}})$ and the chiral flux $\Phi^{\chi}(k)$ emphasize  different molecular properties.
The pseudovector field $\vec{B}^{\mathrm{M}}(\vec{k}^{\mathrm{M}})$ determines the local absorption circular dichroism, is proportional to the  angular momentum of the photoelectron $\vec{L}_g^{\mathrm{M}}(\vec{k}^{\mathrm{M}})$ associated with the ionization from the ground state, and can be non-zero even in achiral systems. 
On the other hand, the pseudoscalar flux $\Phi^{\chi}(k)$
determines the enantiosensitivity of PECD,
is proportional to the average helicity of the photoelectrons with energy $E_k$, and 
is non-zero only in chiral systems.  %In parallel to the Chern number in solids, which is zero for trivial and non-zero for topological insulators, the flux $\\Phi^{\chi}(k)$  is zero  for achiral molecules  and is non-zero for chiral molecules. 
%The emergence of an enantiosensitive flux points to 
%Its emergence suggests a geometrical origin for the chiral response in PECD. 
Its emergence emphasizes the importance of geometry in the chiral response in PECD. 

Further aspects underlying the connection between the enantiosensitive net current and the propensity field will be addressed in our forthcoming publication. Now we will show how the propensity field $ \vec{B}^{\mathrm{M}}$ underlying the response of unaligned molecules manifests itself in the chiral response of aligned molecules.

\section{The connection
between PECD in aligned and unaligned molecules. 
\label{sec:sign_FBA_unaligned}}
In the following we use atomic units everywhere, and we take $-e=1$.
We first rewrite Eq. \eqref{eq:jnew} in an equivalent form using Eqs. (\ref{eq:B}), (\ref{eq:cond}), (\ref{eq:flux}), (\ref{eq:ch_flux}), and explicitly evaluating the vector product of field components:
\begin{eqnarray}
\vec{j}^{\mathrm{L}}\left(k\right) & \equiv & \left\{ \frac{1}{6k}\int\mathrm{d}\vec{S}^{\mathrm{M}}\cdot\vec{B}^{\mathrm{M}}(\vec{k}^{\mathrm{M}})\right\} \left\{ \sigma\left|\tilde{\mathcal{E}}\right|^{2}\hat{z}^{\mathrm{L}}\right\}\equiv \frac{1}{6k} \Phi^{\chi}\left\{ \sigma\left|\tilde{\mathcal{E}}\right|^{2}\hat{z}^{\mathrm{L}}\right\}. \label{eq:j}
\end{eqnarray}
%\begin{eqnarray}
%\vec{j}^{\mathrm{L}}\left(k\right) & = & \left\{ \frac{1}{6}\int\mathrm{d}\Omega_{k}^{\mathrm{M}}\left[\i\left(\vec{D}^{\mathrm{M}*}\x\vec{D}^{\mathrm{M}}\right)\t\vec{k}^{\mathrm{M}}\right]\right\} \left\{ \sigma\left|\tilde{\mathcal{E}}\right|^{2}\hat{z}^{\mathrm{L}}\right\} ,\label{eq:j}
%\end{eqnarray}
We will focus on the analysis of  the  chiral flux 
%$\Phi^{\chi}=\int\mathrm{d}\vec{S} \vec{B}^{\mathrm{M}}(\vec{k}^{\mathrm{M}})$ 
and specifically on the  flux  of each cartesian component of  the propensity field, i.e. on 
\begin{equation}
\Phi^{\chi}_i\equiv\int \mathrm{d}{S}_i^\mathrm{M} {B}^{\mathrm{M}}_i(\vec{k}^{\mathrm{M}}), \quad i=x,y,z. 
\label{eq:flux_i}
\end{equation}
%Now that we have recognized the physical meaning of $\i(\vec{D}^{\mathrm{M}*}\x\vec{D}^{\mathrm{M}})$,
%it is a simple matter to obtain a physical interpretation of the molecular
%term in Eq. \eqref{eq:j}. If we multiply the $i$-th component of
%$\i(\vec{D}^{\mathrm{M}*}\x\vec{D}^{\mathrm{M}})$ by the $i$-th
%component of $\vec{k}^{\mathrm{M}}$ and integrate over the directions
%of $\vec{k}^{\mathrm{M}}$, then
If we pick a specific direction, given by the $i$-th component of propensity field in the molecular frame,  we obtain the difference between
the net photoelectron currents along the $i$-th molecular axis
resulting from left and right circularly polarized light (defined
with respect to the same axis), for a fixed molecular orientation.
For example, for the flux of the $x$ component of the propensity field we obtain [see Eq. \eqref{eq:CD}:
\begin{eqnarray}
\frac{1}{k}\int\mathrm{d}{S}_x\vec{B}^{\mathrm{M}}_x(\vec{k}^{\mathrm{M}}) & \equiv & \int\mathrm{d}\Omega_{k}^{\mathrm{M}}\i\left(\vec{D}^{\mathrm{M}*}\x\vec{D}^{\mathrm{M}}\right)_{x}k_{x}^{\mathrm{M}} \nonumber \\
& = &\int\mathrm{d}\Omega_{k}^{\mathrm{M}}\left|\vec{D}^{\mathrm{M}}\t\frac{\hat{y}^{\mathrm{M}}+\i\hat{z}^{\mathrm{M}}}{\sqrt{2}}\right|^{2}k_{x}^{\mathrm{M}} %\nonumber \\
% & & 
 -\int\mathrm{d}\Omega_{k}^{\mathrm{M}}\left|\vec{D}^{\mathrm{M}}\t\frac{\hat{y}^{\mathrm{M}}-\i\hat{z}^{\mathrm{M}}}{\sqrt{2}}\right|^{2}k_{x}^{\mathrm{M}} \nonumber \\
& = &\frac{1}{\left|\tilde{\mathcal{E}}\right|^{2}}\left[j_{x}^{\mathrm{M}}\left(+_{x}\right)-j_{x}^{\mathrm{M}}\left(-_{x}\right)\right],\label{eq:j(+)-j(-)}
\end{eqnarray}
where the subscript of the plus and of the minus indicates the axis
with respect to which the light is left or right circularly polarized.
An analogous result is obtained for the flux of the $y$ and $z$ components of propensity field.
Then, the chiral flux in Eq. \eqref{eq:j} is simply the sum of
the differences \eqref{eq:j(+)-j(-)} along each molecular axis, normalized
by the intensity of the Fourier component of the field at the transition
frequency, and we can write the net photoelectron current in the
lab frame in terms of the photoelectron currents in the molecular
frame as\footnote{We drop the argument $k$ of the currents in the lab and molecular
frames for simplicity.}

\begin{eqnarray}
\vec{j}^{\mathrm{L}} & = & \left\{ \frac{1}{6}\sum_{i=x,y,z}\left[j_{i}^{\mathrm{M}}\left(+_{i}\right)-j_{i}^{\mathrm{M}}\left(-_{i}\right)\right]\right\} \left\{ \sigma\hat{z}^{\mathrm{L}}\right\} .\label{eq:j_1_orientation_6_fields}
\end{eqnarray}

\begin{figure}
\noindent \begin{centering}
\includegraphics[scale=0.7]{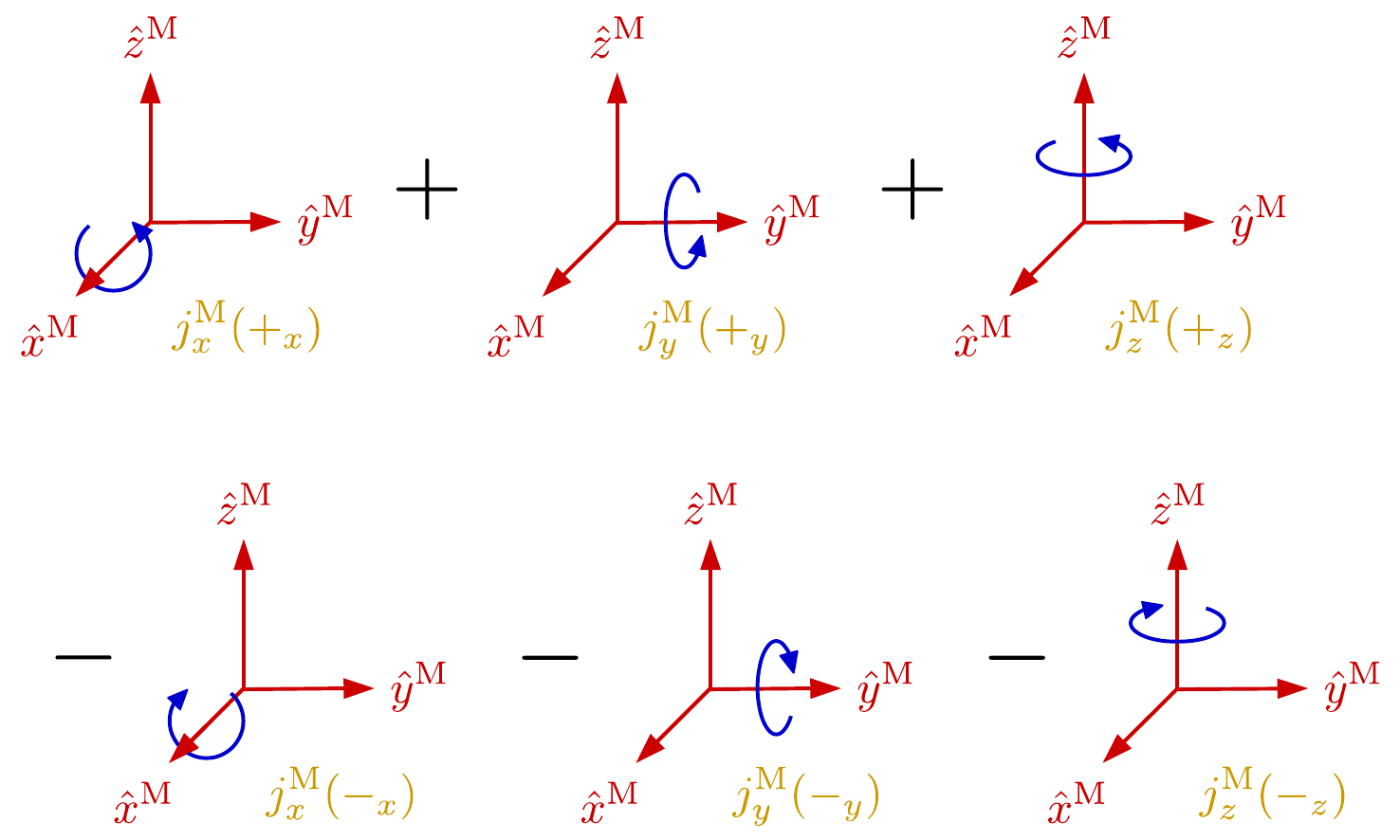}
\par\end{centering}

\caption{Scheme of the right hand side of Eq. \eqref{eq:j_1_orientation_6_fields}
depicting the 6 different field geometries (circular blue arrows)
in the molecular frame contributing to the net photoelectron current
in the lab frame. For each field geometry only the component of the
photoelectron current perpendicular to the polarization plane is taken
into account. \label{fig:j_1_orientation_6_fields}}
\end{figure}

\begin{figure}
\noindent \begin{centering}
\includegraphics[scale=0.7]{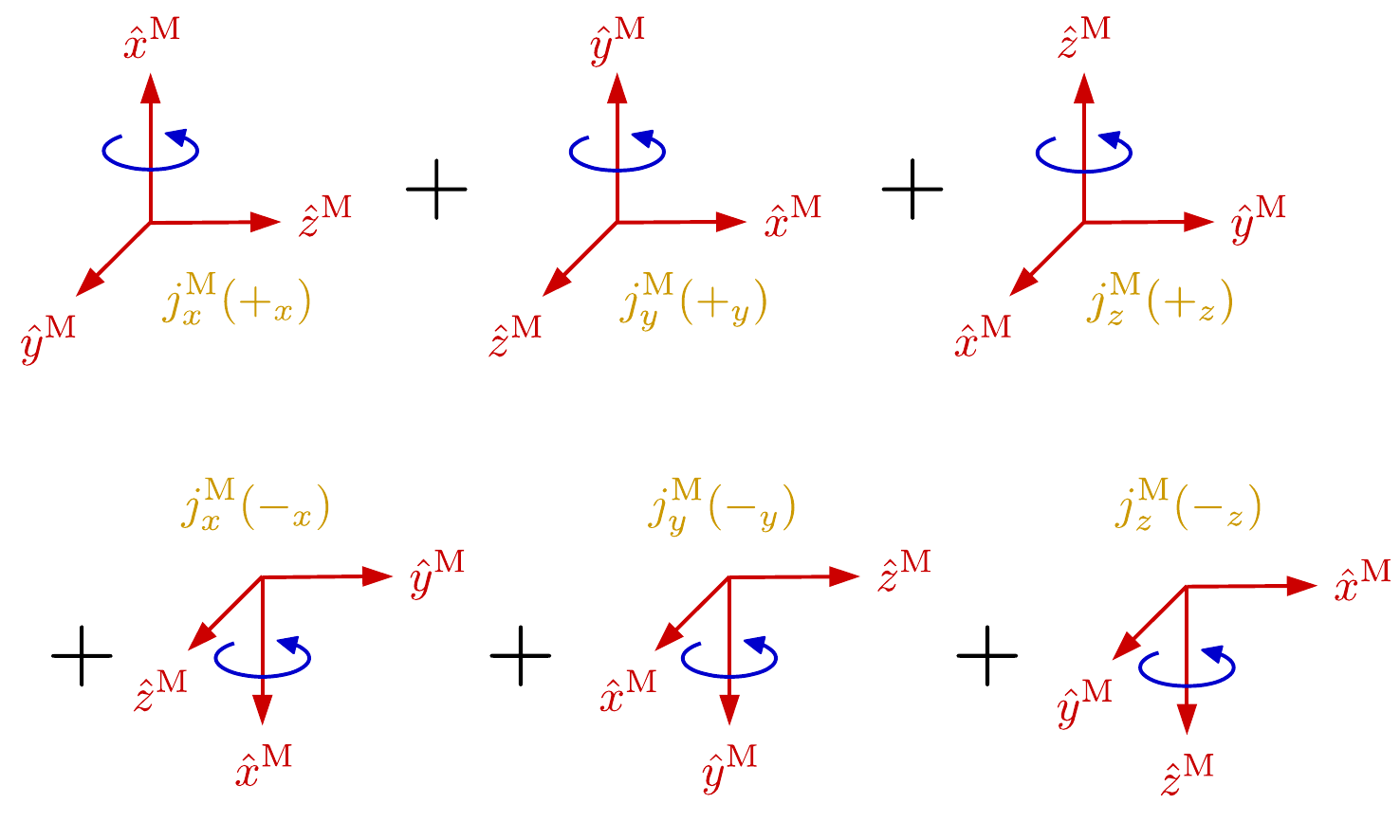}
\par\end{centering}

\caption{Scheme of the right hand side of Eq. \eqref{eq:j_1_field_6_orientations}
depicting the 6 different orientations of the molecular frame contributing
to the net photoelectron current in the lab frame. The curved blue
arrows indicate the field in the lab frame. For each orientation only
the component of the photoelectron current perpendicular to the polarization
plane is taken into account. These orientations are unique only up
to a rotation around the axis perpendicular to the polarization plane.
\label{fig:j_6_orientations_1_field}}
\end{figure}

The right hand side of Eq. \eqref{eq:j_1_orientation_6_fields} is
depicted in Fig. \ref{fig:j_1_orientation_6_fields}, which shows
the different field geometries and the corresponding components of
the current in the molecular frame that account for the net current
in the lab frame. This figure immediately suggests the equivalent
but somewhat more natural picture shown in Fig. \ref{fig:j_6_orientations_1_field},
where the field geometry is kept fixed and the molecule assumes the
six different orientations in which $\hat{x}^{\mathrm{M}}$, $-\hat{x}^{\mathrm{M}}$,
$\hat{y}^{\mathrm{M}}$, $-\hat{y}^{\mathrm{M}}$, $\hat{z}^{\mathrm{M}}$,
and $-\hat{z}^{\mathrm{M}}$, coincide with $\hat{z}^{\mathrm{L}}$.
To reflect this picture Eq. \eqref{eq:j_1_orientation_6_fields} can be rewritten as follows:

\begin{equation}
\vec{j}^{\mathrm{L}}=\left\{ \frac{1}{6}\sum_{i=x,y,z}\left[j_{z}^{\mathrm{L}}\left(\sigma,\lambda_{i}\right)+j_{z}^{\mathrm{L}}\left(\sigma,\lambda_{-i}\right)\right]\right\} \hat{z}^{\mathrm{L}},\label{eq:j_1_field_6_orientations}
\end{equation}

where $\lambda_{i}$ and $\lambda_{-i}$ are the Euler angles specifying
the orientation for which the $i$-th molecular axis is parallel to
$\hat{z}^{\mathrm{L}}$ and $-\hat{z}^{\mathrm{L}}$, respectively.
This change of picture corresponds to the substitutions: $j_{z}^{\mathrm{L}}\left(\sigma,\lambda_{\pm i}\right)=\pm j_{i}^{\mathrm{M}}\left((\pm\sigma)_{i}\right)$
which directly follow from comparing Figs. \ref{fig:j_1_orientation_6_fields}
and \ref{fig:j_6_orientations_1_field}. The Euler angles $\lambda_{\pm i}$
are not unique because the $z$ component of the current $\vec{j}^{\mathrm{L}}\left(\sigma,\lambda_{\pm i}\right)$
is of course invariant with respect to rotations of the molecular
frame about $\hat{z}^{\mathrm{L}}$, and therefore the specific orientation
of the molecular axes that lie on the polarization plane is irrelevant.
Furthermore, the definition of the orientation of the molecular frame
with respect to the nuclei that form the molecule is also arbitrary.
Thus, what Eq. \eqref{eq:j_1_field_6_orientations} really says is
that the orientation-averaged photoelectron current for a randomly-oriented
ensemble is equivalent to the average over six molecular orientations,
where each orientation corresponds to having one of the six spatial
directions in the molecular frame pointing along $\hat{z}^{\mathrm{L}}$. 

%In comparison with Eq. \eqref{eq:j} where there is only one molecular
%orientation, Eq. \eqref{fig:j_6_orientations_1_field} is not an improvement
%in terms of the number of orientations that have to be taken into
%account, however, its derivation and the result obtained provide significant
%insight into the physical meaning and content of Eq. \eqref{eq:j}.

We can work a bit more on Eq. (\ref
{eq:j_1_field_6_orientations})
to avoid the ambiguity of $\lambda_{\pm i}$ mentioned above. If for
a given orientation $\lambda_{i}$ the current in the lab frame is
$\vec{j}^{\mathrm{L}}\left(\sigma,\lambda_{i}\right)$, then the average
of $\vec{j}^{\mathrm{L}}\left(\sigma,\lambda_{i}\right)$ over all
the orientations $\lambda_{i}\left(\phi\right)$, that yield the same
orientation as $\lambda_{i}$ up to a rotation by $\phi$ of the molecular
frame around $\hat{z}^{\mathrm{L}}$, yields the $z$ component of
$\vec{j}^{\mathrm{L}}\left(\sigma,\lambda_{i}\right)$, i.e.

\begin{equation}
\frac{1}{2\pi}\int_{0}^{2\pi}\mathrm{d}\phi\vec{j}^{\mathrm{L}}\left(\sigma,\lambda_{i}\left(\phi_{i}\right)\right)=j_{z}^{\mathrm{L}}\left(\sigma,\lambda_{i}\right)\hat{z}^{\mathrm{L}}.
\end{equation}
This means that we can rewrite the net orientation-averaged photoelectron
current {[}Eq. \eqref{eq:j_1_field_6_orientations}{]} in the more
symmetric form

\begin{equation}
\vec{j}^{\mathrm{L}}\left(k\right)=\frac{1}{3}\sum_{i=x,y,z}\frac{1}{2}\left\{ \frac{1}{2\pi}\int_{0}^{2\pi}\mathrm{d}\phi\vec{j}^{\mathrm{L}}\left(\lambda_{i}\left(\phi\right)\right)+\frac{1}{2\pi}\int_{0}^{2\pi}\mathrm{d}\phi\vec{j}^{\mathrm{L}}\left(\lambda_{-i}\left(\phi\right)\right)\right\} .\label{eq:j_isotropic_aligned}
\end{equation}

This equation provides the relationship between the isotropically-oriented-ensemble
PECD and the aligned-ensemble PECD that we were looking for. The $i$-th term
in the summation corresponds to the average photoelectron current that
a molecular ensemble yields when its $i$-th molecular axis is perfectly
aligned (parallel and anti-parallel) along the normal to the polarization
plane, and the other two molecular axes take all possible orientations
in the polarization plane. That is, Eq. \eqref{eq:j_isotropic_aligned}
shows that the net photoelectron current for an isotropically-oriented
ensemble is simply the average of the three different aligned-ensemble
cases.

\section{PECD in aligned molecular ensembles \label{sec:PECD_aligned}}

From Eq. \eqref{eq:j_isotropic_aligned} we can infer that the introduction
of partial alignment along an axis perpendicular to the polarization
plane in an otherwise isotropic ensemble will simply change the weight
factors of the aligned-ensemble contributions in favor of the molecular
axis which is being aligned. In this section we will confirm that
this is indeed the case by deriving an exact formula for the net
photoelectron current in the lab frame resulting from photoionization
via circularly polarized light of a molecular ensemble exhibiting
an arbitrary degree of alignment with respect to the normal to the
polarization plane. We will also derive an analogous formula for the
case in which the alignment axis is in the plane of polarization, 
%which although less symmetric, is of experimental relevance because
which corresponds to the standard experimental set-up when the laser field used
to align the sample co-propagates with the ionizing field. %(CHECK
%LAST SENTENCE). 
But first we will briefly discuss some general symmetry properties that explain
enantiosensitivity and dichroism in these ensembles from a purely
geometrical point of view.

\subsection{Symmetry considerations for aligned and oriented ensembles\label{sub:Symmetry-considerations}}

\begin{figure}
\noindent \begin{centering}
\includegraphics[scale=0.8]{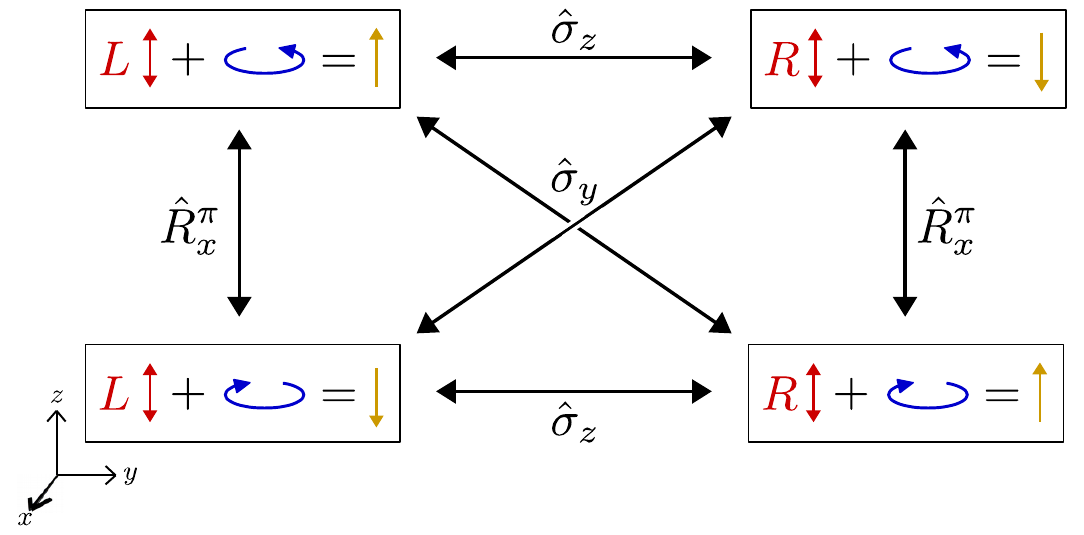} 
\par\end{centering}

\caption{Symmetry properties of an ensemble of chiral molecules interacting
with circularly polarized light in the electric-dipole approximation. The ensemble
is partially (or totally) aligned along the ($z$) axis perpendicular to
the ($xy$) polarization plane of the light. The box represents the ``enantiomer+field''
system. Inside the box: the red letters $L$ and $R$ specify the
enantiomer, the red double-headed vertical arrow specifies the direction
along which the molecules are aligned, the blue curved arrow specifies
the direction of rotation of a field circularly polarized in the $xy$
plane, and the golden vertical arrow stands for a polar vector observable
$\vec{v}=v_{z}\hat{z}$ displaying asymmetry with respect to the polarization
plane $xy$. A reflection $\hat{\sigma}_{z}$ with respect to the
$xy$ plane, leaves the field invariant, but swaps the enantiomer
and flips $\vec{v}$ (enantiosensitivity). A rotation $\hat{R}_{a}^{\pi}$ by $\pi$ radians
around any vector $\vec{a}$ contained in the $xy$ plane leaves the
enantiomer invariant, but swaps
the polarization and flips $\vec{v}$ (dichroism). Note that a rotation $\hat{R}_{x}^{\pi}$
($\hat{R}_{y}^{\pi}$) followed by a reflection $\hat{\sigma}_{z}$
is equivalent to a reflection $\hat{\sigma}_{y}$ ($\hat{\sigma}_{x}$)
and leaves $\vec{v}$ invariant but swaps both the enantiomer and
the polarization. Thus, except for very specific cases (see Fig. \ref{fig:alignment_induced_chirality}), FBA in aligned ensembles is a signature of molecular chirality (see also Fig. \ref{fig:symmetry_aligned_in_plane}). \label{fig:symmetry_aligned}}
\end{figure}

\begin{figure}
\noindent \begin{centering}
\includegraphics[scale=0.3]{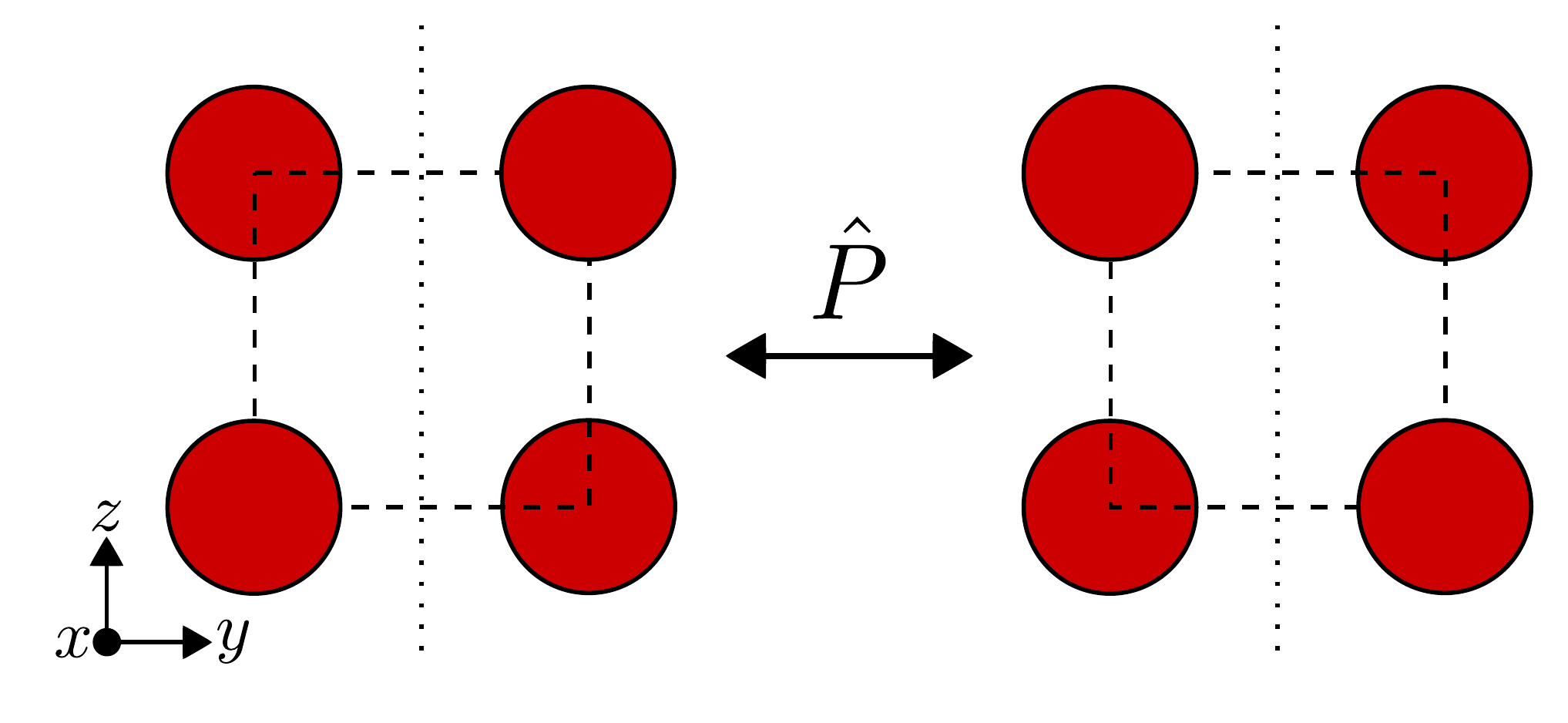}
\par\end{centering}

\caption{Left: an achiral molecule consisting of four identical atoms with Cartesian coordinates $(-a,-b,b)$, $(a, b, b)$, $(a, -b, -b)$, and $(-a,b,-b)$. Right: spatial inversion of the molecule on the left. A rotation by $\pi/2$ around the $x$ axis yields the molecule on the left. However, molecular alignment restricts available rotations. Thus if we consider a sample aligned along the vertical dotted line, the rotation by $\pi/2$ is not allowed and the aligned sample becomes effectively chiral. \label{fig:alignment_induced_chirality}}
\end{figure}

\begin{figure}
\noindent \begin{centering}
\includegraphics[scale=0.8]{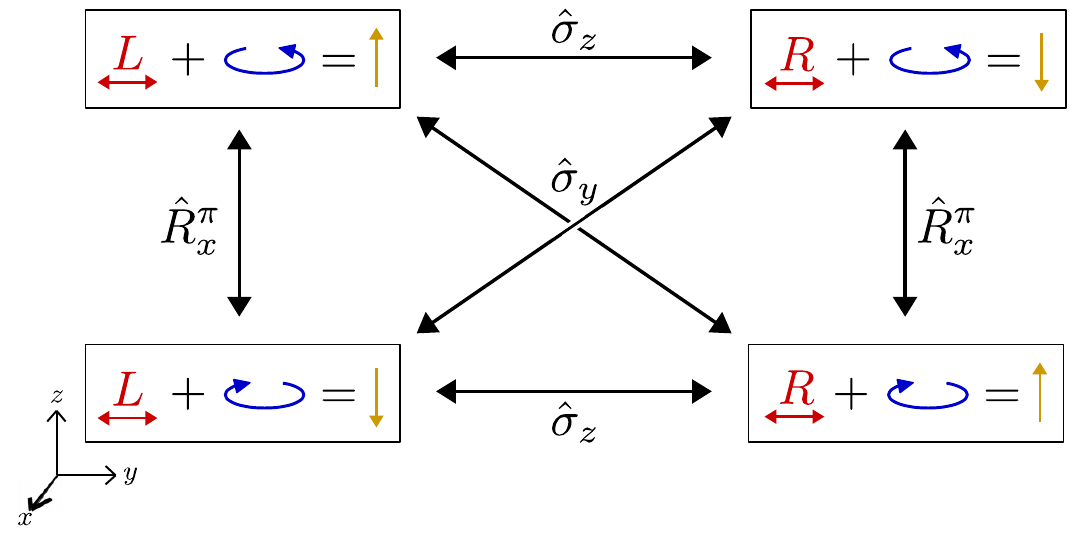} 
\par\end{centering}

\caption{%Same as Fig. \ref{fig:symmetry_aligned} but for an alignment axis
%contained in the polarization plane.
Symmetry properties of an ensemble of chiral molecules interacting with circularly
polarized light in the electric-dipole approximation. The ensemble is partially (or totally) aligned
along an axis ($y$) contained in the ($xy$) polarization plane of the light. Notations are described in Fig. \ref{fig:symmetry_aligned}. This shows that, except for very specific cases (see Fig. \ref{fig:alignment_induced_chirality}), FBA in aligned ensembles is a signature of molecular chirality (see also Fig. \ref{fig:symmetry_aligned}). 
\label{fig:symmetry_aligned_in_plane}}
\end{figure}

\begin{figure}
\noindent \begin{centering}
\includegraphics[scale=0.8]{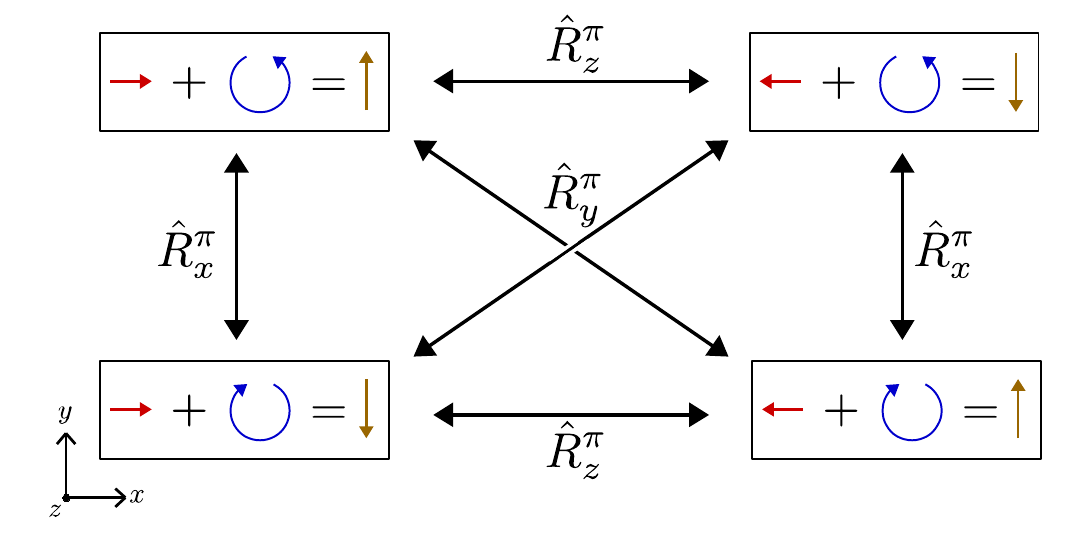} 
\par\end{centering}

\caption{%Same as Fig. \ref{fig:symmetry_aligned} but 
Symmetry considerations for an ensemble of oriented
(chiral or achiral) molecules. Notations are described in Fig. \ref{fig:symmetry_aligned}. 
The orientation axis ($x$) is in the polarization
plane of the light ($xy$). A rotation $\hat{R}_{z}^{\pi}$ by $\pi$ radians
around the $z$ axis leaves the field invariant, but flips both the
molecular orientation and $v_y\hat{y}$ (orientation sensitivity). A rotation $\hat{R}_{x}^{\pi}$
by $\pi$ radians around the $x$ axis leaves the orientation invariant,
but swaps the polarization and flips $v_y\hat{y}$ (dichroism). Note that a rotation
$\hat{R}_{z}^{\pi}$ ($\hat{R}_{x}^{\pi}$) followed by a rotation
$\hat{R}_{x}^{\pi}$ ($\hat{R}_{z}^{\pi}$) is equivalent to a rotation
$\hat{R}_{y}^{\pi}$ and leaves $v_y\hat{y}$ invariant but flips the
orientation and the polarization. For achiral molecules reflection symmetry forbids signals perpendicular to the polarization plane, i.e. $v_z=0$. For chiral molecules such FBA signal is not symmetry forbidden and it is enantiosensitive and dichroic. \label{fig:symmetry_oriented}}
\end{figure}

The relevant symmetry properties of an \emph{aligned} ensemble of
chiral molecules interacting with circularly polarized light in the 
electric-dipole approximation are 
summarized in Fig. \ref{fig:symmetry_aligned} (cf. Fig. 2 in 
\cite{ordonez_generalized_2018}) for the case of alignment perpendicular
to the polarization plane. In this case, the cylindrical symmetry
of the ``aligned-enantiomer + field'' system enforces cylindrical symmetry
on the observables and therefore limits the asymmetry of the photoelectron
angular distribution to be along the axis perpendicular to the 
polarization plane, i.e. forward-backward asymmetry\footnote{We note that the term FBA can be misleading
because it seems to imply that the direction of propagation of the
light, i.e. the sign of the wave vector plays a role. This is clearly
not the case as the effect is within the electric-dipole approximation, and
therefore whether the light propagates in the $+\hat{z}$ direction
or the $-\hat{z}$ direction is completely irrelevant. The only thing
that matters is the rotation direction of the light, i.e. the spin of the 
photon and not its helicity (see Appendix 1 of Ref. \cite{ordonez_generalized_2018}).} (FBA). 
Furthermore, if the aligned sample is achiral then 
the ``enantiomer + field'' system becomes invariant with respect to 
reflections through the polarization plane, and the forward-backward
asymmetry disappears. That is, like in the isotropic case, in 
the aligned case the FBA is also a signature of the chirality of the 
sample. 

However, unlike in the isotropic case, in the aligned case one must be careful of 
distinguishing between the chirality of the aligned sample and the chirality of the molecules that make up the sample. Although an aligned sample of chiral molecules is always chiral, a chiral aligned sample is not necessarily made out of chiral molecules. The reason is that restricting the degrees of freedom of the molecular orientation may lead to suppression of the orientation corresponding to the reflection of an allowed orientation and therefore induce chirality. An example of how this may occur is shown in Fig. \ref{fig:alignment_induced_chirality}, where we can see that such alignment-induced chirality seems to require a very particular interplay between the molecular symmetry and the alignment axis. In the absence of such particular conditions, alignment does not induces chirality in the sample and FBA can be traced back to the chirality of the molecules. 

Figure \ref{fig:symmetry_aligned_in_plane} shows a symmetry
diagram analogous to that in Fig. \ref{fig:symmetry_aligned} for the setup in which the molecular alignment axis is in
the plane of polarization of the ionizing light. In this case the
molecular alignment breaks the cylindrical symmetry. Nevertheless
the system remains invariant with respect to rotations by $\pi$ around
the $z$ axis and the vector observable is again constrained to the
$\hat{z}$ direction. Like for the previous case, the dichroic and
enantiosensitive FBA is a signature of the chirality of the molecular sample.

%, namely that the achiral molecule has $S_{4n}$ (but no $C_{4n}$) symmetry about an axis perpendicular to the alignment axis. In the absence of such symmetry, alignment does not induces chirality in the sample and FBA can be traced back to the chirality of the molecules. 

% A molecular sample is chiral if either its molecules are chiral, or 
% if the molecular orientations of the achiral molecules that make up the sample are restricted in such a way that the reduced dimensionality induces chirality. An extreme example of such restriction are uniaxially oriented samples of achiral molecules, where one may think of one orientation as corresponding to one enantiomer, and the opposite orientation to the other enantiomer. Dichroism!...

It is important to distinguish the FBA discussed here from the dichroic asymmetry 
observed in Refs.  \cite{dubs_circular_1985,dubs_circular_1985-1,holmegaard_photoelectron_2010} in oriented achiral samples. While the former is with respect to the polarization plane and is a hallmark of the chirality of the sample, the latter is with respect to the plane containing the spin of the photon and the orientation axis (a plane perpendicular to the polarization plane), and takes place even for achiral samples. Figure \ref{fig:symmetry_oriented} shows how such dichroic asymmetry can emerge in uniaxially oriented chiral and achiral systems. The $x$ component of $\vec{v}$ is not shown because it is not dichroic and the $z$ component reflecting FBA is zero in achiral systems because of reflection symmetry with respect to the $xy$ plane.

\subsection{Connection between chiral current and molecular field for aligned ensembles\label{sub:Formulas-aligned}}

Now that we have established a  general starting point based
on the symmetry properties of the ``aligned-enantiomer + field'' system,
we will proceed to the derivation of the lab-frame net photoelectron current
for such an ensemble for the case of one-photon absorption. The molecular alignment can
be introduced in the orientation-averaging procedure via a weight
function $w\left(\lambda\right)$ that depends on the Euler angles
$\lambda\equiv\alpha\beta\gamma$, which are the angles that determine
the relative orientation between the lab frame and the molecular frame.
In the ZYZ convention, $\beta$ determines the angle between the $z$
axes of the two frames, so that to describe molecular alignment we
can use a distribution $w\left(\beta\right)$ that only depends on
this angle and that is symmetric with respect to $\beta=\pi/2$. With the
molecular alignment defined along the $\hat{z}^{\mathrm{L}}$ axis
(or viceversa), we can consider that the circularly polarized field
is in the $x^{\mathrm{L}}y^{\mathrm{L}}$ plane or in the $y^{\mathrm{L}}z^{\mathrm{L}}$
plane, depending on whether we are interested in the setup where the
molecular alignment is perpendicular or parallel to the light polarization
plane, respectively.

\subsubsection{Alignment perpendicular to the plane of polarization}

For light circularly polarized in the $x^{\mathrm{L}}y^{\mathrm{L}}$
plane the photoelectron current density in the lab frame corresponding
to a given photoelectron momentum $\vec{k}^{\mathrm{M}}$ in the molecular
frame and a given molecular orientation $\lambda\equiv(\alpha,\beta,\gamma)$
is \cite{ordonez_generalized_2018}

\begin{align}
\vec{j}^{\mathrm{L}}\left(\vec{k}^{\mathrm{M}},\lambda\right) & =\frac{\left|\tilde{\mathcal{E}}\right|^{2}}{2}\left|\vec{D}^{\mathrm{L}}\t\left(\hat{x}^{\mathrm{L}}+\sigma\i\hat{y}^{\mathrm{L}}\right)\right|^{2}\vec{k}^{\mathrm{L}}\nonumber \\
 & =\frac{\left|\tilde{\mathcal{E}}\right|^{2}}{2}\left[\left|S\vec{D}^{\mathrm{M}}\t\hat{x}^{\mathrm{L}}\right|^{2}+\left|S\vec{D}^{\mathrm{M}}\t\hat{y}^{\mathrm{L}}\right|^{2}+\sigma\i S\left(\vec{D}^{\mathrm{M}*}\x\vec{D}^{\mathrm{M}}\right)\t\hat{z}^{\mathrm{L}}\right]S\vec{k}^{\mathrm{M}},\label{eq:j_veck_lambda_perp}
\end{align}

where $S(\lambda)$ is the rotation matrix that takes vectors from
the molecular frame to the lab frame, i.e. $\vec{v}^{\mathrm{L}}=S\left(\lambda\right)\vec{v}^{\mathrm{M}}$,
$\tilde{\mathcal{E}}$ is the Fourier transform of the electric field
evaluated at the transition frequency, and $\sigma=\pm1$ stands for
left($+$)/right($-$) circularly polarized light. Before moving on to the
case at hand, Eq. \eqref{eq:j_veck_lambda_perp} gives us the opportunity
to briefly point out another reason why only the coherent term survives
the orientation averaging in both isotropically-oriented and aligned
ensembles. For each orientation $\lambda_{i}$ of the molecular frame
there will be another orientation $\lambda_{-i}$ related to it by
a rotation by $\pi$ around (for example) $\hat{y}^{\mathrm{L}}$
that will change the sign of the $x^{\mathrm{L}}$ and $z^{\mathrm{L}}$
components of all molecular vectors. Therefore, if we consider the
average of $j_{z}^{\mathrm{L}}(\vec{k}^{\mathrm{M}},\lambda)$ over
those two orientations, $[j_{z}^{\mathrm{L}}(\vec{k}^{\mathrm{M}},\lambda_{i})+j_{z}^{\mathrm{L}}(\vec{k}^{\mathrm{M}},\lambda_{-i})]/2$,
we can see from \eqref{eq:j_veck_lambda_perp} that the incoherent
terms $\vert S\vec{D}^{\mathrm{M}}\t\hat{x}^{\mathrm{L}}\vert^{2}(S\vec{k}^{\mathrm{M}}\t\hat{z}^\mathrm{L})$
and $\vert S\vec{D}^{\mathrm{M}}\t\hat{y}^{\mathrm{L}}\vert^{2}(S\vec{k}^{\mathrm{M}}\t\hat{z}^\mathrm{L})$
will cancel because they have opposite signs for opposite orientations,
while the coherent term $\sigma[S(\i\vec{D}^{\mathrm{M}*}\x\vec{D}^{\mathrm{M}})\t\hat{z}^{\mathrm{L}}](S\vec{k}^{\mathrm{M}}\t\hat{z}^\mathrm{L})$
will not because it is the same for both orientations. That
is, while obviously each term of $j_{z}^{\mathrm{L}}(\vec{k}^{\mathrm{M}},\lambda)$
is invariant with respect to rotations of the molecular frame by $\pi$
around $\hat{z}^{\mathrm{L}}$, only the coherent term is invariant
with respect to rotations by $\pi$ with respect to any axis. Thus,
either for isotropically-oriented samples or aligned samples (with
molecular alignment perpendicular to the polarization plane or not),
the incoherent terms will always cancel by pairs in the orientation
averaging while the coherent term term will not.

For a distribution of orientations $w\left(\beta\right)$, the net
photoelectron current in the lab frame takes the form:

\begin{eqnarray}
\vec{j}^{\mathrm{L}}\left(k\right) & = & \int\mathrm{d}\Omega_{k}^{\mathrm{M}}\int\mathrm{d}\lambda w\left(\beta\right)\vec{j}^{\mathrm{L}}\left(\vec{k}^{\mathrm{M}},\lambda\right),\label{eq:j_lab_w_unsolved-1}
\end{eqnarray}

where $\int\mathrm{d}\lambda\equiv\int_{0}^{2\pi}\mathrm{d}\alpha\int_{0}^{\pi}\mathrm{d}\beta\int_{0}^{2\pi}\mathrm{d}\gamma\sin\beta/8\pi^{2}$
is the integral over molecular orientations, and $\int\mathrm{d}\Omega_{k}^{\mathrm{M}}\equiv\int_{0}^{2\pi}\mathrm{d}\phi_{k}^{\mathrm{M}}\int_{0}^{\pi}\mathrm{d}\theta_{k}^{\mathrm{M}}\sin\theta_{k}^{\mathrm{M}}$
is the integral over directions of the photoelectron momentum $\vec{k}^{\mathrm{M}}$.
For an alignment distribution $w\left(\beta\right)\propto\cos^{2}\beta$,
Eq. (\ref{eq:j_lab_w_unsolved-1}) becomes equivalent to the photoelectron
current found in the case where a pump linearly polarized along $\hat{z}^{\mathrm{L}}$
resonantly excites the molecule via a transition dipole parallel to
$\hat{z}^{\mathrm{M}}$ into a bound excited electronic state and
is then photoionized from the latter by a circularly polarized probe
pulse. Therefore, for such a distribution we could simply make use
of Eq. (31) derived in \cite{ordonez_generalized_2018} in the context of
the generalized PXECD (see Appendix \ref{sub:Orientation-averaging-in-aligned-ensembles}).
This equivalence reveals the close relation between aligned ensembles
where the molecular orientation is anisotropic and isotropically-oriented
ensembles that have been electronically excited. This happens because the field imprints its anisotropy on the originally isotropic molecular ensemble via the excitation.

In the following we will make no assumption about $w\left(\beta\right)$
except that it is symmetric with respect to $\beta=\pi/2$, which simply imposes the condition of alignment. The first
two terms in Eq. \eqref{eq:j_veck_lambda_perp} describe interaction
with a linearly polarized field and therefore, from symmetry considerations\footnote{For example, for polarization along $\hat{x}^{\mathrm{L}}$, the system
is invariant with respect to a rotation by $\pi$ around $\hat{x}^{\mathrm{L}}$
which means that $j_{y}^{\mathrm{L}}=j_{z}^{\mathrm{L}}=0$, and also
with respect to rotation by $\pi$ around $\hat{z}^{\mathrm{L}}$
which means that $j_{x}^{\mathrm{L}}=j_{y}^{\mathrm{L}}=0$.}, they lead to $\vec{j}^{\mathrm{L}}\left(k\right)=0$. The integral
over orientations of the third term in Eq.  \eqref{eq:j_veck_lambda_perp}
can be carried out with the help of Eq. \eqref{eq:(a.z)b} derived
in Appendix \ref{sub:Orientation-averaging-in-aligned-ensembles}
and yields

\begin{multline}
\vec{j}^{\mathrm{L}}\left(k\right)=\frac{\sigma\left|\tilde{\mathcal{E}}\right|^{2}}{2}\bigg[\frac{1}{3}w_{i}\int\mathrm{d}\Omega_{k}^{\mathrm{M}}\left(\i\vec{D}^{\mathrm{M}*}\x\vec{D}^{\mathrm{M}}\right)\t\vec{k}^{\mathrm{M}}\\
+\left(1-w_{i}\right)\int\mathrm{d}\Omega_{k}^{\mathrm{M}}\left(\i\vec{D}^{\mathrm{M}*}\x\vec{D}^{\mathrm{M}}\right)_{z}k_{z}^{\mathrm{M}}\bigg]\hat{z}^{\mathrm{L}},\label{eq:j_lab_w_solved_perpendicular}
\end{multline}
where we assumed that $w\left(\beta\right)$ is properly normalized
{[}see Eq. \eqref{eq:w(beta)_norm}{]} and we defined $w_{i}$ as
\begin{equation}
w_{i}\equiv\frac{3}{4}\int_{0}^{\pi}\mathrm{d}\beta w\left(\beta\right)\sin^{3}\beta.
\end{equation}
$w_{i}$ %corresponds to the weight of the isotropic response and is
is a parameter 
determined exclusively by $w(\beta)$. 
Equation (\ref{eq:j_lab_w_solved_perpendicular}) can be written in an equivalent form [cf. Eqs. \eqref{eq:B} \eqref{eq:ch_flux}, and \eqref{eq:flux_i}]:
\begin{equation}
\vec{j}^{\mathrm{L}}\left(k\right)=\frac{\sigma\left|\tilde{\mathcal{E}}\right|^{2}}{2k}\bigg[\frac{1}{3}w_{i}\Phi^{\chi}(k)
+\left(1-w_{i}\right)
%\int\mathrm{d}S_z B^{\mathrm{M}}_z(\vec{k}^{\mathrm{M}})
\Phi^{\chi}_{z}(k)
\bigg]\hat{z}^{\mathrm{L}}.\label{eq:j_lab_w_solved_perpendicular_new}
\end{equation}
% For an isotropically-oriented
% distribution  $w(\beta)=1$, $w_{i}=1$, and Eq. %\eqref{eq:j_lab_w_solved_perpendicular} and
% \eqref{eq:j_lab_w_solved_perpendicular_new}
% reduces to Eq. \eqref{eq:j}, while for a perfectly aligned sample
% we have\footnote{The factor of 2 comes from having the Dirac-deltas centered at the
% extremes of the integration interval.} $w(\beta)=2[\delta(\beta)+\delta(\beta-\pi)]/\sin\beta$, $w_{i}=0$,
% and we are left only with
% %the $z$ component of the scalar product in Eq. \eqref{eq:j},
% the contribution of the $z$ component of the propensity field $\vec{B}^\mathrm{M}$ to the flux in Eq. \eqref{eq:j},
%  in full agreement\footnote{The factor of 3 comes from the normalization of $w(\beta)$.} with our discussion in Sec. \ref{sec:sign_FBA_unaligned},
% where we identified the $x$, $y$, and $z$ components of the scalar product
% in Eq. \eqref{eq:j} with the photoelectron current resulting from
% an ensemble with its $\hat{x}^{\mathrm{M}}$, $\hat{y}^{\mathrm{M}}$,
% and $\hat{z}^{\mathrm{M}}$ axis, respectively, aligned along $\hat{z}^{\mathrm{L}}$
% {[}see Eq. \eqref{eq:j_isotropic_aligned}{]}.
Table \ref{tab:alignment} shows $w(\beta)$, $w_i$, and $j_z^\mathrm{L}$ for perfectly aligned, isotropic, and perfectly antialigned samples. While the isotropic case reduces to Eq. \eqref{eq:j} and gives the average of $\Phi_x^\chi$, $\Phi_y^\chi$, and $\Phi_z^\chi$ [i.e. the total flux, see Eqs. \eqref{eq:ch_flux} and \eqref{eq:flux_i}], the perfectly aligned case singles out the $\Phi_z^\chi$, in full agreement
%\footnote{The factor of 3 comes from the normalization of $w(\beta)$.} 
with our discussion in Sec. \ref{sec:sign_FBA_unaligned}. On the other hand, the perfectly antialigned case, where the molecular $z$ axis is constrained to be perpendicular to the laboratory $z$ axis, prevents $\Phi_z^\chi$ from contributing to the photoelectron current. 

% One can always define
% the orientation of the molecular frame so that its $\hat{z}^{\mathrm{M}}$
% axis coincides with the axis being aligned and therefore there is
% no need to consider separately the case where the $\hat{x}^{\mathrm{M}}$
% or the $\hat{y}^{\mathrm{M}}$ axis is the one being aligned along
% $\hat{z}^{\mathrm{L}}$. 
%Thus, when $\hat{z}^{\mathrm{M}}$ is aligned along $\hat{z}^{\mathrm{L}}$ ($w_{i}=0$), the enatiosensitive current  only stems from $z$-component of molecular field  $\vec{B}^{\mathrm{M}}(\vec{k}^{\mathrm{M}})$, since the first term vanishes for $w_{i}=0$. 

As shown in Appendix \ref{sub:Orientation-averaging-in-aligned-ensembles},
for the case  $w\left(\beta\right)\propto\cos^{2}\beta$, Eq. \eqref{eq:j_lab_w_solved_perpendicular}
coincides with the predictions of the generalized PXECD formula derived
in \cite{ordonez_generalized_2018} and discussed above.

\begin{table}
\setlength{\tabcolsep}{10pt}
\renewcommand{\arraystretch}{1.5}
\noindent \begin{center}
\begin{tabular}{c|cccc}
\multicolumn{1}{c}{} &  &  & $\tilde{\vec{\mathcal{E}}}=\frac{\tilde{\mathcal{E}}}{\sqrt{2}}\left(\hat{x}+\sigma i\hat{y}\right)$ & $\tilde{\vec{\mathcal{E}}}=\frac{\tilde{\mathcal{E}}}{\sqrt{2}}\left(\hat{y}+\sigma i\hat{z}\right)$\tabularnewline
 & $w\left(\beta\right)$ & $w_{i}$ & $j_z/\frac{\sigma\left|\tilde{\mathcal{E}}\right|^{2}}{2k}$ & $j_x/\frac{\sigma\left|\tilde{\mathcal{E}}\right|^{2}}{2k}$\tabularnewline
\hline 
aligned & $\frac{2\left[\delta\left(\beta\right)+\delta\left(\beta-\pi\right)\right]}{\sin\beta}$ & $0$ & $\Phi_{z}$ & $\frac{1}{2}\left(\Phi_{x}+\Phi_{y}\right)$\tabularnewline
isotropic & $1$ & $1$ & $\frac{1}{3}\left(\Phi_{x}+\Phi_{y}+\Phi_{z}\right)$ & $\frac{1}{3}\left(\Phi_{x}+\Phi_{y}+\Phi_{z}\right)$\tabularnewline
antialigned & $\frac{2\delta\left(\beta-\frac{\pi}{2}\right)}{\sin\beta}$ & $\frac{3}{2}$ & $\frac{1}{2}\left(\Phi_{x}+\Phi_{y}\right)$ & $\frac{1}{4}\left(\Phi_{x}+\Phi_{y}+2\Phi_{z}\right)$\tabularnewline
\end{tabular}
\par\end{center}
\caption{Photoelectron current density in aligned, isotropic, and antialigned samples for circular polarization perpendicular [fourth column, Eq. \eqref{eq:j_lab_w_solved_perpendicular_new}] and parallel [fifth column, Eq. \eqref{eq:j_lab_w_solved_parallel_new}] to the alignment axis $z$. We have dropped the L and the $\chi$ superscripts for simplicity. \label{tab:alignment}}
\end{table}

\subsubsection{Alignment parallel to the plane of polarization}

The derivation for the setup in which the molecular alignment axis
is contained in the polarization plane follows analogously with only
subtle differences. This time we define the orientation of the lab
frame such that the molecular alignment remains along the $\hat{z}^\mathrm{L}$ axis
but now the light is polarized in the $y^\mathrm{L}z^\mathrm{L}$ plane, and therefore we
have that the photoelectron current in the lab frame corresponding
to a given photoelectron momentum $\vec{k}^{\mathrm{M}}$ in the molecular
frame and a given molecular orientation $\lambda\equiv(\alpha,\beta,\gamma)$
reads as

\begin{equation}
\vec{j}^{\mathrm{L}}\left(\vec{k}^{\mathrm{M}},\lambda\right)=\frac{\left|\tilde{\mathcal{E}}\right|^{2}}{2}\left[\left|S\vec{D}^{\mathrm{M}}\t\hat{y}^{\mathrm{L}}\right|^{2}+\left|S\vec{D}^{\mathrm{M}}\t\hat{z}^{\mathrm{L}}\right|^{2}+\sigma\i S\left(\vec{D}^{\mathrm{M}*}\x\vec{D}^{\mathrm{M}}\right)\t\hat{x}^{\mathrm{L}}\right]S\vec{k}^{\mathrm{M}}.\label{eq:j_veck_lambda_parallel}
\end{equation}

With the help of Eq. \eqref{eq:(a.x)b} derived in Appendix \ref{sub:Orientation-averaging-in-aligned-ensembles}
we obtain 

\begin{multline}
\vec{j}^{\mathrm{L}}\left(k\right)=\frac{\sigma\left|\tilde{\mathcal{E}}\right|^{2}}{2}\bigg[\frac{1}{3}\frac{\left(3-w_{i}\right)}{2}\int\mathrm{d}\Omega_{k}^{\mathrm{M}}\left(\i\vec{D}^{\mathrm{M}*}\x\vec{D}^{\mathrm{M}}\right)\t\vec{k}^{\mathrm{M}}\\
+\frac{1}{2}\left(w_{i}-1\right)\int\mathrm{d}\Omega_{k}^{\mathrm{M}}\left(\i\vec{D}^{\mathrm{M}*}\x\vec{D}^{\mathrm{M}}\right)_{z}k_{z}^{\mathrm{M}}\bigg]\hat{x}^{\mathrm{L}}.\label{eq:j_lab_w_solved_parallel}
\end{multline}
Like in the previous case, and as follows from the symmetry considerations
of Sec. \ref{sub:Symmetry-considerations}, the current is directed
along the direction perpendicular to the polarization plane of the
incident field. Comparing with Eq. \eqref{eq:j_lab_w_solved_perpendicular}
we can see that the factors in front of the isotropic and anisotropic
contributions are slightly different from what we obtained in the
previous case.  We can rewrite this equation in an equivalent form [cf. Eqs. \eqref{eq:B}, \eqref{eq:ch_flux}, and \eqref{eq:flux_i}]:
\begin{equation}
\vec{j}^{\mathrm{L}}\left(k\right)=\frac{\sigma\left|\tilde{\mathcal{E}}\right|^{2}}{2k}\bigg[\frac{1}{3}\frac{\left(3-w_{i}\right)}{2}\Phi^{\chi}(k)
+\frac{1}{2}\left(w_{i}-1\right)
%\int\mathrm{d}S_z B^{\mathrm{M}}_z(\vec{k}^{\mathrm{M}})
\Phi^{\chi}_{z}(k)
\bigg]\hat{x}^{\mathrm{L}}.\label{eq:j_lab_w_solved_parallel_new}
\end{equation}
Some limiting cases of this equation are shown in the last column of Table \ref{tab:alignment}, where we can see that, as expected, this formula reduces to Eq. \eqref{eq:j} in the isotropic case. We can also see that the aligned case with the alignment parallel to the polarization plane yields the same result as the antialigned case with the alignment perpendicular to the polarization plane, as expected from symmetry, and that antialignment doubles the weight of $\Phi_z^\chi$ with respect to that of $\Phi_x^\chi$ and $\Phi_y^\chi$.
Appendix \ref{sub:Orientation-averaging-in-aligned-ensembles}
shows how Eq. \eqref{eq:j_lab_w_solved_parallel} can also be derived
from the generalized PXECD formulas derived in \cite{ordonez_generalized_2018}
when $w\left(\beta\right)\propto\cos^{2}\beta$.

Equations \eqref{eq:j_lab_w_solved_perpendicular_new} and \eqref{eq:j_lab_w_solved_parallel_new}
suggest that choosing the alignment properly could lead to an increase
of the PECD signal. Such increase has been recently discovered both theoretically and experimentally in Ref. \cite{Tia_2017}.
The increase can be rationalized in terms of the propensity  field $\vec{B}^{\mathrm{M}}(\vec{k}^{\mathrm{M}})$ and its strength along different $\vec{k}^{\mathrm{M}}$ directions. For example, if a molecule is such that 
$|\Phi^{\chi}_{z}| > |\Phi^{\chi}_{x}|$ and 
$|\Phi^{\chi}_{z}| > |\Phi^{\chi}_{y}|$,
% $\Phi^{\chi}_z=\int\mathrm{d}S_z B^{\mathrm{M}}_z(\vec{k}^{\mathrm{M}})$ %$\int\mathrm{d}\Omega_{k}^{\mathrm{M}}(\i\vec{D}^{\mathrm{M}*}\x\vec{D}^{\mathrm{M}})_{z}k_{z}^{\mathrm{M}}\equiv \int\mathrm{d}\Omega_{k}^{\mathrm{M}}B^{ch}_{z}k_{z}^{\mathrm{M}}$
% is bigger than $\Phi^{\chi}_x=\int\mathrm{d}S_x B^{\mathrm{M}}_x(\vec{k}^{\mathrm{M}})$%$\int\mathrm{d}\Omega_{k}^{\mathrm{M}}(\i\vec{D}^{\mathrm{M}*}\x\vec{D}^{\mathrm{M}})_{x}k_{x}^{\mathrm{M}}\equiv \int\mathrm{d}\Omega_{k}^{\mathrm{M}}B^{ch}_{x}k_{x}^{\mathrm{M}}$
% and $\Phi^{\chi}_y=\int\mathrm{d}S_y B^{\mathrm{M}}_y(\vec{k}^{\mathrm{M}})$  %$\int\mathrm{d}\Omega_{k}^{\mathrm{M}}(\i\vec{D}^{\mathrm{M}*}\x\vec{D}^{\mathrm{M}})_{y}k_{y}^{\mathrm{M}}\equiv \int\mathrm{d}\Omega_{k}^{\mathrm{M}}B^{ch}_{y}k_{y}^{\mathrm{M}}$,
and the $z$ molecular axis can be aligned, then Eq. \eqref{eq:j_lab_w_solved_perpendicular_new}
shows that the PECD signal will increase with the alignment. Similarly,
if for example $\Phi^{\chi}_z$
has an opposite sign to that of $\Phi^{\chi}_x$
and $\Phi^{\chi}_y$,
then Eq. \eqref{eq:j_lab_w_solved_perpendicular_new} shows that the PECD
signal will also benefit from the alignment.

\section{Conclusions \label{sec:conclusions}}

The enantiosensitive photoelectron current, or in other words, the forward-backward asymmetry in photoelectron circular dichroism (PECD), 
is determined by the the propensity field, which is analogous to the Berry curvature in a two-band solid.
This field is independent of light properties, is defined in the molecular frame, and is unique to each molecule. 
The enantiosensitive photoelectron current stemming from aligned ensembles of chiral molecules is only sensitive to specific components of the propensity field and therefore the increase or decrease of the chiral response vs. molecular alignment depends on the %spatial
structure of this field.
%All three components of the field are simultaneously non-zero only for chiral systems, because these components probe the absence of rotational symmetry in respective orthogonal planes. 
Each component of the propensity field reflects photoelectron-momentum-resolved absorption circular dichroism and is only non-zero 
%due to optical propensity rules for photoionization.
in the absence of rotational symmetry about the corresponding axis. 
The propensity field underlies the emergence of PECD.
%How exactly the interplay of propensity rules maps into forward-backward assymtry is desctibed in our companion paper \cite{ordonez_2018_hydrogen}.
 %Much like Berry curvature in solids records the handedness of Bloch electrons and may define optical selection rules \cite{yao2008}, the chiral field encodes optical selection rules.
Thus, in this paper we have generalized the ideas presented in our companion paper \cite{ordonez_2018_hydrogen}, which illustrates the role of optical propensity rules in PECD in aligned molecular ensembles for specific examples of chiral states. 

In the case of unaligned molecular ensembles, the enantiosensitive photoelectron current for a given absolute value $k$ of the photoelectron momentum is proportional to the flux of the propensity field  through the sphere  of radius $k$. The flux is a pseudoscalar and has opposite sign for opposite enantiomers. Molecular alignment allows one to probe the flux generated by specific components of the propensity field. 

%It is similar to the Chern number in band insulators and highlights 
%another, geometrical, 
%the role of geometry in the emergence of the chiral response in PECD.

\section{Acknowledgements}
The authors thank Alvaro Jim\'enez-Gal\'an,  Rui Silva, David Ayuso and Misha Ivanov for illuminating discussions and essential collaboration on this topic. The authors gratefully acknowledge the MEDEA project, which has received funding from the European Union's Horizon 2020 research and innovation programme under the Marie Sk\l{}odowska-Curie grant agreement 641789. The authors gratefully  acknowledge support from the DFG SPP 1840 ``Quantum Dynamics in Tailored Intense Fields'' and DFG grant SM 292/5-2.

\section{Appendix}

\subsection{Orientation averaging in aligned ensembles \label{sub:Orientation-averaging-in-aligned-ensembles}}

In this appendix we will derive the orientation averaged net photoelectron
current in the lab frame for the aligned ensembles considered in Sec.
\eqref{sub:Formulas-aligned}. Before deriving the expression for
an arbitrary distribution $w(\beta)$, we will consider the particular
distribution $w\left(\beta\right)=3\cos^{2}\beta$ in order to draw
some connections between the results obtained in a randomly oriented
sample and an aligned sample. In this case the net photoelectron
current can be written as {[}see Eqs. \eqref{eq:j_veck_lambda_perp}
and \eqref{eq:j_lab_w_unsolved-1}{]}

\begin{eqnarray*}
\vec{j}^{\mathrm{L}}\left(k\right) & = & 3\int\mathrm{d}\Omega_{k}^{\mathrm{M}}\int\mathrm{d}\lambda\cos^{2}\beta\,\vec{j}^{\mathrm{L}}\left(\vec{k}^{\mathrm{M}},\lambda\right)\\
 & = & 3\int\mathrm{d}\Omega_{k}^{\mathrm{M}}\int\mathrm{d}\lambda\,\left|\hat{d}_{\mathrm{eff}}^{\mathrm{L}}\t\hat{\tilde{\mathcal{E}}}_{\mathrm{eff}}^{\mathrm{L}}\right|^{2}\left|\vec{D}^{\mathrm{L}}\t\vec{\tilde{\mathcal{E}}}^{\mathrm{L}}\right|^{2}\vec{k}^{\mathrm{L}},
\end{eqnarray*}

which simply shows that the anisotropic orientation average of $\vec{j}^{\mathrm{L}}(\vec{k}^{\mathrm{M}},\lambda)$
is equivalent to the isotropic averaging of $\vert\hat{d}_{\mathrm{p}}^{\mathrm{L}}\t\hat{\tilde{\mathcal{E}}}_{\mathrm{p}}^{\mathrm{L}}\vert^{2}\vert\vec{D}^{\mathrm{L}}\t\vec{\tilde{\mathcal{E}}}^{\mathrm{L}}\vert^{2}$,
where we introduced an effective bound-bound transition dipole $\hat{d}_{\mathrm{eff}}^{\mathrm{M}}=\hat{z}^{\mathrm{M}}$
and the effective field which interacts with it $\hat{\tilde{\mathcal{E}}}_{\mathrm{eff}}^{\mathrm{L}}=\hat{z}^{\mathrm{L}}$,
in order to make evident that, mathematically, we are dealing with
a particular case of the generalized PXECD effect considered in \cite{ordonez_generalized_2018},
where first a pump pulse of arbitrary polarization excites the system
into a superposition of two excited states and then a probe pulse
of arbitrary polarization photoionizes the system from intermediate
state. In the present case the effective pump pulse excites the system
from an effective ground state into a single excited state (the actual
ground state) through the interaction $\hat{d}_{\mathrm{eff}}^{\mathrm{L}}\t\hat{\tilde{\mathcal{E}}}_{\mathrm{eff}}^{\mathrm{L}}$
and then the probe pulse (the actual pulse) photoionizes the system
from the excited state. That is, we only have to deal with Eq. (31)
in \cite{ordonez_generalized_2018}, which in our case reads as 

\begin{align}
\vec{j}^{\mathrm{L}}\left(k\right) & =3\int\mathrm{d}\Omega_{k}^{\mathrm{M}}\int\mathrm{d}\lambda\,\left|\hat{d}_{\mathrm{eff}}^{\mathrm{L}}\t\hat{\tilde{\mathcal{E}}}_{\mathrm{eff}}^{\mathrm{L}}\right|^{2}\left|\vec{D}^{\mathrm{L}}\t\vec{\tilde{\mathcal{E}}}^{\mathrm{L}}\right|^{2}\vec{k}^{\mathrm{L}}\nonumber \\
 & =\frac{1}{5}\Re\bigg\{\int\mathrm{d}\Omega_{k}^{\mathrm{M}}\left[\left(\hat{d}_{\mathrm{eff}}^{\mathrm{M}}\x\vec{D}^{\mathrm{M}*}\right)\t\vec{D}^{\mathrm{M}}\right]\left(\hat{d}_{\mathrm{eff}}^{\mathrm{M}}\t\vec{k}^{\mathrm{M}}\right)\left[\left(\hat{\tilde{\mathcal{E}}}_{\mathrm{eff}}^{\mathrm{L}}\x\tilde{\vec{\mathcal{E}}}^{\mathrm{L}*}\right)\t\tilde{\vec{\mathcal{E}}}^{\mathrm{L}}\right]\hat{\tilde{\mathcal{E}}}_{\mathrm{eff}}^{\mathrm{L}}\nonumber \\
 & +\int\mathrm{d}\Omega_{k}^{\mathrm{M}}\left[\left(\hat{d}_{\mathrm{eff}}^{\mathrm{M}}\x\vec{D}^{\mathrm{M}*}\right)\t\vec{k}^{\mathrm{M}}\right]\left(\hat{d}_{\mathrm{eff}}^{\mathrm{M}}\t\vec{D}^{\mathrm{M}}\right)\left(\hat{\tilde{\mathcal{E}}}_{\mathrm{eff}}^{\mathrm{L}}\t\tilde{\vec{\mathcal{E}}}^{\mathrm{L}}\right)\left(\hat{\tilde{\mathcal{E}}}_{\mathrm{eff}}^{\mathrm{L}}\x\tilde{\vec{\mathcal{E}}}^{\mathrm{L}*}\right)\nonumber \\
 & +\int\mathrm{d}\Omega_{k}^{\mathrm{M}}\left[\left(\hat{d}_{\mathrm{eff}}^{\mathrm{M}}\x\vec{D}^{\mathrm{M}}\right)\t\vec{k}^{\mathrm{M}}\right]\left(\hat{d}_{\mathrm{eff}}^{\mathrm{M}}\t\vec{D}^{\mathrm{M}*}\right)\left(\hat{\tilde{\mathcal{E}}}_{\mathrm{eff}}^{\mathrm{L}}\t\tilde{\vec{\mathcal{E}}}^{\mathrm{L}*}\right)\left(\hat{\tilde{\mathcal{E}}}_{\mathrm{eff}}^{\mathrm{L}}\x\tilde{\vec{\mathcal{E}}}^{\mathrm{L}}\right)\bigg\}\nonumber \\
 & +\frac{1}{10}\int\mathrm{d}\Omega_{k}^{\mathrm{M}}\left[\left(\vec{D}^{\mathrm{M}*}\x\vec{D}^{\mathrm{M}}\right)\t\vec{k}^{\mathrm{M}}\right]\left(\tilde{\vec{\mathcal{E}}}^{\mathrm{L}*}\x\tilde{\vec{\mathcal{E}}}^{\mathrm{L}}\right),\label{eq:PXECD_diag}
\end{align}

If the molecular alignment (which we have already set along $\hat{z}^{\mathrm{L}}$)
is perpendicular to the polarization plane we set $\tilde{\vec{\mathcal{E}}}^{\mathrm{L}}=\left(\hat{x}^{\mathrm{L}}+\sigma\i\hat{y}^{\mathrm{L}}\right)/\sqrt{2}$.
The second and third terms vanish because $(\hat{\tilde{\mathcal{E}}}_{\mathrm{eff}}^{\mathrm{L}}\t\tilde{\vec{\mathcal{E}}}^{\mathrm{L}})=0$
and Eq. \eqref{eq:PXECD_diag} yields

\begin{multline}
\vec{j}^{\mathrm{L}}\left(k\right)=\frac{\sigma\left|\tilde{\mathcal{E}}\right|^{2}}{2}\bigg\{ \frac{1}{5}\int\mathrm{d}\Omega_{k}^{\mathrm{M}}\left[\left(\mathrm{i}\vec{D}^{\mathrm{M}*}\x\vec{D}^{\mathrm{M}}\right)\t\vec{k}^{\mathrm{M}}\right]\\
+\frac{2}{5}\int\mathrm{d}\Omega_{k}^{\mathrm{M}}\left[\left(\i\vec{D}^{\mathrm{M}*}\x\vec{D}^{\mathrm{M}}\right)_{z}k_{z}^{\mathrm{M}}\right]\bigg\} \hat{z}^{\mathrm{L}},\label{eq:j_perpendicular_alignment_PXECD}
\end{multline}

On the other hand, for the case in which molecular alignment is in
the plane of the light polarization we set $\tilde{\vec{\mathcal{E}}}^{\mathrm{L}}=\left(\hat{y}^\mathrm{L}+\sigma\i\hat{z}^\mathrm{L}\right)/\sqrt{2}$.
The first term vanishes because $[(\hat{\tilde{\mathcal{E}}}_{\mathrm{eff}}^{\mathrm{L}}\x\tilde{\vec{\mathcal{E}}}^{\mathrm{L}*})\t\tilde{\vec{\mathcal{E}}}^{\mathrm{L}}]=0$,
and with the help of the vector identities $(\vec{a}\x\vec{b})\t(\vec{c}\x\vec{d})$$=(\vec{a}\t\vec{c})(\vec{b}\t\vec{d})$$-(\vec{a}\t\vec{d})(\vec{b}\t\vec{c})$
and $(\vec{a}\x\vec{b})\x\vec{c}=$$(\vec{a}\t\vec{c})\vec{b}-(\vec{b}\t\vec{c})\vec{a}$
we obtain 

\begin{multline}
\vec{j}^{\mathrm{L}}\left(k\right)=\frac{\sigma\left|\tilde{\mathcal{E}}\right|^{2}}{2}\bigg\{ \frac{2}{5}\int\mathrm{d}\Omega_{k}^{\mathrm{M}}\left[\left(\i\vec{D}^{\mathrm{M}*}\x\vec{D}^{\mathrm{M}}\right)\t\vec{k}^{\mathrm{M}}\right]\\
-\frac{1}{5}\int\mathrm{d}\Omega_{k}^{\mathrm{M}}\left(\i\vec{D}^{\mathrm{M}*}\x\vec{D}^{\mathrm{M}}\right)_{z}k_{z}^{\mathrm{M}}\bigg\} \hat{x}^{\mathrm{L}}.\label{eq:j_parallel_alignment_PXECD}
\end{multline}

In both cases, Eq. \eqref{eq:j_perpendicular_alignment_PXECD} and
\eqref{eq:j_parallel_alignment_PXECD} show that $\vec{j}^{\mathrm{L}}\left(k\right)$
is along the direction perpendicular to the light polarization plane
and that there is an imbalance in the scalar product $(\i\vec{D}^{\mathrm{M}*}\x\vec{D}^{\mathrm{M}})\t\vec{k}^{\mathrm{M}}$
that singles out the molecular axis being aligned. Equations \eqref{eq:j_perpendicular_alignment_PXECD}
and \eqref{eq:j_parallel_alignment_PXECD} coincide with Eqs. \eqref{eq:j_lab_w_solved_perpendicular}
and \eqref{eq:j_lab_w_solved_parallel}, respectively, when we set
$w\left(\beta\right)=3\cos^{2}\beta$ and consequently $w_{i}=3/5$
in Eqs. \eqref{eq:j_lab_w_solved_perpendicular} and \eqref{eq:j_lab_w_solved_parallel}. 

Now we proceed to the general derivation where the only assumption
on $w\left(\beta\right)$ is that it is symmetric with respect to
$\beta=\pi/2$, which simply imposes the condition of alignment. Since
symmetry implies that the incoherent terms corresponding to linear
polarization along $\hat{x}^{\mathrm{L}}$ and $\hat{y}^{\mathrm{L}}$
in Eq. \eqref{eq:j_veck_lambda_perp} vanish\footnote{Consider the analog of Fig. \ref{fig:symmetry_aligned} for linearly
polarized light along $x$ ($y$). The total system becomes symmetric
with respect to rotations of $\pi$ around $x$ ($y$) and therefore
there can be no asymmetry along $z$.}, we will focus exclusively on the coherent term. For the case in
which the molecular alignment is perpendicular to the light polarization
plane, the relevant integral over orientations is of the form {[}see
Eq. \eqref{eq:j_veck_lambda_perp}{]}

\begin{equation}
\int\mathrm{d}\lambda\,w\left(\beta\right)\left(\vec{a}^{\mathrm{L}}\t\hat{z}^{\mathrm{L}}\right)\vec{b}^{\mathrm{L}},
\end{equation}

where $\int\mathrm{d}\lambda\equiv\int_{0}^{2\pi}\mathrm{d}\alpha\int_{0}^{\pi}\mathrm{d}\beta\int_{0}^{2\pi}\mathrm{d}\gamma\,\sin\beta/8\pi^{2}$,
$\vec{a}$ and $\vec{b}$ are vectors fixed in the molecular frame.
To transform a vector from the molecular frame to the lab frame we
use $\vec{v}^{\mathrm{L}}=R\left(\lambda\right)\vec{v}^{\mathrm{M}}$,
where 

\begin{equation}
R\left(\lambda\right)=\left(\begin{array}{ccc}
-\mathrm{s}\alpha\,\mathrm{s}\gamma+\mathrm{c}\alpha\,\mathrm{c}\beta\,\mathrm{c}\gamma & -\mathrm{s}\alpha\,\mathrm{c}\gamma-\mathrm{s}\gamma\,\mathrm{c}\alpha\,\mathrm{c}\beta & \mathrm{s}\beta\,\mathrm{c}\alpha\\
\mathrm{s}\alpha\,\mathrm{c}\beta\,\mathrm{c}\gamma+\mathrm{s}\gamma\,\mathrm{c}\alpha & -\mathrm{s}\alpha\,\mathrm{s}\gamma\,\mathrm{c}\beta+\mathrm{c}\alpha\,\mathrm{c}\gamma & \mathrm{s}\alpha\,\mathrm{s}\beta\\
-\mathrm{s}\beta\,\mathrm{c}\gamma & \mathrm{s}\beta\,\mathrm{s}\gamma & \mathrm{c}\beta
\end{array}\right),
\end{equation}

and s and c stand for $\sin$ and $\cos$, respectively.  With the help of $R\left(\lambda\right)$ we calculate the expression
$\left(\vec{a}^{\mathrm{L}}\t\hat{z}^{\mathrm{L}}\right)\vec{b}^{\mathrm{L}}$
in terms of the molecular frame components of $\vec{a}$ and $\vec{b}$
and then note that most of the terms vanish after integration over
$\alpha$ and $\gamma$. The non-vanishing terms read as

\begin{eqnarray}
 &  & \int\mathrm{d}\lambda\,w\left(\beta\right)\left(\vec{a}^{\mathrm{L}}\t\hat{z}^{\mathrm{L}}\right)\vec{b}^{\mathrm{L}}\nonumber \\
 & = & \bigg\{\left[\int\mathrm{d}\lambda w\left(\beta\right)\sin^{2}\beta\cos^{2}\gamma\right]a_{x}^{\mathrm{M}}b_{x}^{\mathrm{M}}+\left[\int\mathrm{d}\lambda w\left(\beta\right)\sin^{2}\beta\sin^{2}\gamma\right]a_{y}^{\mathrm{M}}b_{y}^{\mathrm{M}}\nonumber \\
 &  & +\left[\int\mathrm{d}\lambda w\left(\beta\right)\cos^{2}\beta\right]a_{z}^{\mathrm{M}}b_{z}^{\mathrm{M}}\bigg\}\hat{z}^{\mathrm{L}}\nonumber \\
 & = & \bigg\{ \left[\frac{1}{2}\int_{0}^{\pi}\mathrm{d}\beta w\left(\beta\right)\sin^{3}\beta\right]\frac{1}{2}\left(a_{x}^{\mathrm{M}}b_{x}^{\mathrm{M}}+a_{y}^{\mathrm{M}}b_{y}^{\mathrm{M}}\right) \nonumber \\
 & & +\left[\frac{1}{2}\int_{0}^{\pi}\mathrm{d}\beta w\left(\beta\right)\sin\beta\cos^{2}\beta\right]a_{z}^{\mathrm{M}}b_{z}^{\mathrm{M}}\bigg\} \hat{z}^{\mathrm{L}}\nonumber \\
 & = & \left[\frac{1}{3}w_{i}\left(\vec{a}^{\mathrm{M}}\t\vec{b}^{\mathrm{M}}\right)+\left(1-w_{i}\right)a_{z}^{\mathrm{M}}b_{z}^{\mathrm{M}}\right]\hat{z}^{\mathrm{L}},\label{eq:(a.z)b}
\end{eqnarray}

where we defined 

\begin{equation}
w_{i}\equiv\frac{3}{4}\int_{0}^{\pi}\mathrm{d}\beta w\left(\beta\right)\sin^{3}\beta,\label{eq:w_i}
\end{equation}

and we assumed that $w\left(\beta\right)$ is normalized so that $\int\mathrm{d}\lambda w\left(\beta\right)=1$,
which implies that

\begin{equation}
\frac{1}{2}\int_{0}^{\pi}\mathrm{d}\beta\sin(\beta)w(\beta)=1.\label{eq:w(beta)_norm}
\end{equation}

In the case in which molecular alignment is in the plane of the light
polarization the relevant integral is of the form {[}see Eq. \eqref{eq:j_veck_lambda_parallel}{]}

\begin{equation}
\int\mathrm{d}\lambda\,w\left(\beta\right)\left(\vec{a}^{\mathrm{L}}\t\hat{x}^{\mathrm{L}}\right)\vec{b}^{\mathrm{L}},
\end{equation}

and we proceed analogously as before to find that the only terms that
do not vanish after integration yield

\begin{eqnarray}
 &  & \int\mathrm{d}\lambda\,w\left(\beta\right)\left(\vec{a}^{\mathrm{L}}\t\hat{x}^{\mathrm{L}}\right)\vec{b}^{\mathrm{L}}\nonumber \\
 & = & \frac{1}{2}\int_{0}^{\pi}\mathrm{d}\beta\sin\beta\,w\left(\beta\right)\left[\frac{1}{4}\left(1+\cos^{2}\beta\right)\left(a_{x}^{\mathrm{M}}b_{x}^{\mathrm{M}}+a_{y}^{\mathrm{M}}b_{y}^{\mathrm{M}}\right)+\frac{1}{2}\sin^{2}\beta a_{z}^{\mathrm{M}}b_{z}^{\mathrm{M}}\right]\hat{x}^{\mathrm{L}}\nonumber \\
 & = & \frac{1}{2}\int_{0}^{\pi}\mathrm{d}\beta\sin\beta\,w\left(\beta\right)\left[\frac{1}{4}\left(2-\sin^{2}\beta\right)\vec{a}^{\mathrm{M}}\t\vec{b}^{\mathrm{M}}-\frac{1}{4}\left(2-3\sin^{2}\beta\right)a_{z}^{\mathrm{M}}b_{z}^{\mathrm{M}}\right]\hat{x}^{\mathrm{L}}\nonumber \\
 & = & \left[\frac{1}{2}\left(1-\frac{w_{i}}{3}\right)\vec{a}^{\mathrm{M}}\t\vec{b}^{\mathrm{M}}-\frac{1}{2}\left(1-w_{i}\right)a_{z}^{\mathrm{M}}b_{z}^{\mathrm{M}}\right]\hat{x}^{\mathrm{L}}.\label{eq:(a.x)b}
\end{eqnarray}

\bibliographystyle{apsrev4-1}
\bibliography{MyLibrary}

\begin{thebibliography}{69}%
\makeatletter
\providecommand \@ifxundefined [1]{%
 \@ifx{#1\undefined}
}%
\providecommand \@ifnum [1]{%
 \ifnum #1\expandafter \@firstoftwo
 \else \expandafter \@secondoftwo
 \fi
}%
\providecommand \@ifx [1]{%
 \ifx #1\expandafter \@firstoftwo
 \else \expandafter \@secondoftwo
 \fi
}%
\providecommand \natexlab [1]{#1}%
\providecommand \enquote  [1]{``#1''}%
\providecommand \bibnamefont  [1]{#1}%
\providecommand \bibfnamefont [1]{#1}%
\providecommand \citenamefont [1]{#1}%
\providecommand \href@noop [0]{\@secondoftwo}%
\providecommand \href [0]{\begingroup \@sanitize@url \@href}%
\providecommand \@href[1]{\@@startlink{#1}\@@href}%
\providecommand \@@href[1]{\endgroup#1\@@endlink}%
\providecommand \@sanitize@url [0]{\catcode `\\12\catcode `\$12\catcode
  `\&12\catcode `\#12\catcode `\^12\catcode `\_12\catcode `\%12\relax}%
\providecommand \@@startlink[1]{}%
\providecommand \@@endlink[0]{}%
\providecommand \url  [0]{\begingroup\@sanitize@url \@url }%
\providecommand \@url [1]{\endgroup\@href {#1}{\urlprefix }}%
\providecommand \urlprefix  [0]{URL }%
\providecommand \Eprint [0]{\href }%
\providecommand \doibase [0]{http://dx.doi.org/}%
\providecommand \selectlanguage [0]{\@gobble}%
\providecommand \bibinfo  [0]{\@secondoftwo}%
\providecommand \bibfield  [0]{\@secondoftwo}%
\providecommand \translation [1]{[#1]}%
\providecommand \BibitemOpen [0]{}%
\providecommand \bibitemStop [0]{}%
\providecommand \bibitemNoStop [0]{.\EOS\space}%
\providecommand \EOS [0]{\spacefactor3000\relax}%
\providecommand \BibitemShut  [1]{\csname bibitem#1\endcsname}%
\let\auto@bib@innerbib\@empty
%</preamble>
\bibitem [{\citenamefont {Ordonez}\ and\ \citenamefont
  {Smirnova}(2018{\natexlab{a}})}]{ordonez_2018_hydrogen}%
  \BibitemOpen
  \bibfield  {author} {\bibinfo {author} {\bibfnamefont {A.~F.}\ \bibnamefont
  {Ordonez}}\ and\ \bibinfo {author} {\bibfnamefont {O.}~\bibnamefont
  {Smirnova}},\ }\href {http://arxiv.org/abs/1806.09049} {\bibfield  {journal}
  {\bibinfo  {journal} {arXiv:1806.09049 [physics, physics:quant-ph]}\ }
  (\bibinfo {year} {2018}{\natexlab{a}})},\ \bibinfo {note} {arXiv:
  1806.09049}\BibitemShut {NoStop}%
\bibitem [{\citenamefont {Ritchie}(1976)}]{ritchie_theory_1976}%
  \BibitemOpen
  \bibfield  {author} {\bibinfo {author} {\bibfnamefont {B.}~\bibnamefont
  {Ritchie}},\ }\href
  {http://journals.aps.org/pra/abstract/10.1103/PhysRevA.13.1411} {\bibfield
  {journal} {\bibinfo  {journal} {Physical Review A}\ }\textbf {\bibinfo
  {volume} {13}},\ \bibinfo {pages} {1411} (\bibinfo {year}
  {1976})}\BibitemShut {NoStop}%
\bibitem [{\citenamefont {Powis}(2000{\natexlab{a}})}]{powis00}%
  \BibitemOpen
  \bibfield  {author} {\bibinfo {author} {\bibfnamefont {I.}~\bibnamefont
  {Powis}},\ }\href {http://dx.doi.org/10.1063/1.480581} {\bibfield  {journal}
  {\bibinfo  {journal} {The Journal of Chemical Physics}\ }\textbf {\bibinfo
  {volume} {112}},\ \bibinfo {pages} {301} (\bibinfo {year}
  {2000}{\natexlab{a}})}\BibitemShut {NoStop}%
\bibitem [{\citenamefont {B\"owering}\ \emph {et~al.}(2001)\citenamefont
  {B\"owering}, \citenamefont {Lischke}, \citenamefont {Schmidtke},
  \citenamefont {M\"uller}, \citenamefont {Khalil},\ and\ \citenamefont
  {Heinzmann}}]{bowering_asymmetry_2001}%
  \BibitemOpen
  \bibfield  {author} {\bibinfo {author} {\bibfnamefont {N.}~\bibnamefont
  {B\"owering}}, \bibinfo {author} {\bibfnamefont {T.}~\bibnamefont {Lischke}},
  \bibinfo {author} {\bibfnamefont {B.}~\bibnamefont {Schmidtke}}, \bibinfo
  {author} {\bibfnamefont {N.}~\bibnamefont {M\"uller}}, \bibinfo {author}
  {\bibfnamefont {T.}~\bibnamefont {Khalil}}, \ and\ \bibinfo {author}
  {\bibfnamefont {U.}~\bibnamefont {Heinzmann}},\ }\href {\doibase
  10.1103/PhysRevLett.86.1187} {\bibfield  {journal} {\bibinfo  {journal}
  {Physical Review Letters}\ }\textbf {\bibinfo {volume} {86}},\ \bibinfo
  {pages} {1187} (\bibinfo {year} {2001})}\BibitemShut {NoStop}%
\bibitem [{\citenamefont {Condon}(1937)}]{condon_theories_1937}%
  \BibitemOpen
  \bibfield  {author} {\bibinfo {author} {\bibfnamefont {E.~U.}\ \bibnamefont
  {Condon}},\ }\href {\doibase 10.1103/RevModPhys.9.432} {\bibfield  {journal}
  {\bibinfo  {journal} {Reviews of Modern Physics}\ }\textbf {\bibinfo {volume}
  {9}},\ \bibinfo {pages} {432} (\bibinfo {year} {1937})}\BibitemShut {NoStop}%
\bibitem [{\citenamefont {Ordonez}\ and\ \citenamefont
  {Smirnova}(2018{\natexlab{b}})}]{ordonez_generalized_2018}%
  \BibitemOpen
  \bibfield  {author} {\bibinfo {author} {\bibfnamefont {A.~F.}\ \bibnamefont
  {Ordonez}}\ and\ \bibinfo {author} {\bibfnamefont {O.}~\bibnamefont
  {Smirnova}},\ }\href {\doibase 10.1103/PhysRevA.98.063428} {\bibfield
  {journal} {\bibinfo  {journal} {Physical Review A}\ }\textbf {\bibinfo
  {volume} {98}},\ \bibinfo {pages} {063428} (\bibinfo {year}
  {2018}{\natexlab{b}})}\BibitemShut {NoStop}%
\bibitem [{\citenamefont {Patterson}\ \emph {et~al.}(2013)\citenamefont
  {Patterson}, \citenamefont {Schnell},\ and\ \citenamefont
  {Doyle}}]{patterson_enantiomer-specific_2013}%
  \BibitemOpen
  \bibfield  {author} {\bibinfo {author} {\bibfnamefont {D.}~\bibnamefont
  {Patterson}}, \bibinfo {author} {\bibfnamefont {M.}~\bibnamefont {Schnell}},
  \ and\ \bibinfo {author} {\bibfnamefont {J.~M.}\ \bibnamefont {Doyle}},\
  }\href {\doibase 10.1038/nature12150} {\bibfield  {journal} {\bibinfo
  {journal} {Nature}\ }\textbf {\bibinfo {volume} {497}},\ \bibinfo {pages}
  {475} (\bibinfo {year} {2013})}\BibitemShut {NoStop}%
\bibitem [{\citenamefont {Patterson}\ and\ \citenamefont
  {Doyle}(2013)}]{patterson_sensitive_2013}%
  \BibitemOpen
  \bibfield  {author} {\bibinfo {author} {\bibfnamefont {D.}~\bibnamefont
  {Patterson}}\ and\ \bibinfo {author} {\bibfnamefont {J.~M.}\ \bibnamefont
  {Doyle}},\ }\href {\doibase 10.1103/PhysRevLett.111.023008} {\bibfield
  {journal} {\bibinfo  {journal} {Physical Review Letters}\ }\textbf {\bibinfo
  {volume} {111}},\ \bibinfo {pages} {023008} (\bibinfo {year}
  {2013})}\BibitemShut {NoStop}%
\bibitem [{\citenamefont {Yachmenev}\ and\ \citenamefont
  {Yurchenko}(2016)}]{yurchenko_2016}%
  \BibitemOpen
  \bibfield  {author} {\bibinfo {author} {\bibfnamefont {A.}~\bibnamefont
  {Yachmenev}}\ and\ \bibinfo {author} {\bibfnamefont {S.~N.}\ \bibnamefont
  {Yurchenko}},\ }\href {\doibase 10.1103/PhysRevLett.117.033001} {\bibfield
  {journal} {\bibinfo  {journal} {Phys. Rev. Lett.}\ }\textbf {\bibinfo
  {volume} {117}},\ \bibinfo {pages} {033001} (\bibinfo {year}
  {2016})}\BibitemShut {NoStop}%
\bibitem [{\citenamefont {Eibenberger}\ \emph {et~al.}(2017)\citenamefont
  {Eibenberger}, \citenamefont {Doyle},\ and\ \citenamefont
  {Patterson}}]{patterson_2017}%
  \BibitemOpen
  \bibfield  {author} {\bibinfo {author} {\bibfnamefont {S.}~\bibnamefont
  {Eibenberger}}, \bibinfo {author} {\bibfnamefont {J.}~\bibnamefont {Doyle}},
  \ and\ \bibinfo {author} {\bibfnamefont {D.}~\bibnamefont {Patterson}},\
  }\href {\doibase 10.1103/PhysRevLett.118.123002} {\bibfield  {journal}
  {\bibinfo  {journal} {Phys. Rev. Lett.}\ }\textbf {\bibinfo {volume} {118}},\
  \bibinfo {pages} {123002} (\bibinfo {year} {2017})}\BibitemShut {NoStop}%
\bibitem [{\citenamefont {Fischer}\ \emph {et~al.}(2001)\citenamefont
  {Fischer}, \citenamefont {Buckingham},\ and\ \citenamefont
  {Albrecht}}]{fischer2001isotropic}%
  \BibitemOpen
  \bibfield  {author} {\bibinfo {author} {\bibfnamefont {P.}~\bibnamefont
  {Fischer}}, \bibinfo {author} {\bibfnamefont {A.~D.}\ \bibnamefont
  {Buckingham}}, \ and\ \bibinfo {author} {\bibfnamefont {A.~C.}\ \bibnamefont
  {Albrecht}},\ }\href {\doibase 10.1103/PhysRevA.64.053816} {\bibfield
  {journal} {\bibinfo  {journal} {Phys. Rev. A}\ }\textbf {\bibinfo {volume}
  {64}},\ \bibinfo {pages} {053816} (\bibinfo {year} {2001})}\BibitemShut
  {NoStop}%
\bibitem [{\citenamefont {Beaulieu}\ \emph {et~al.}(2018)\citenamefont
  {Beaulieu}, \citenamefont {Comby}, \citenamefont {Descamps}, \citenamefont
  {Fabre}, \citenamefont {Garcia}, \citenamefont {G{\'e}neaux}, \citenamefont
  {Harvey}, \citenamefont {L{\'e}gar{\'e}}, \citenamefont {Mas{\'i}n},
  \citenamefont {Nahon}, \citenamefont {Ordonez}, \citenamefont {Petit},
  \citenamefont {Pons}, \citenamefont {Mairesse}, \citenamefont {Smirnova},\
  and\ \citenamefont {Blanchet}}]{beaulieu_PXCD}%
  \BibitemOpen
  \bibfield  {author} {\bibinfo {author} {\bibfnamefont {S.}~\bibnamefont
  {Beaulieu}}, \bibinfo {author} {\bibfnamefont {A.}~\bibnamefont {Comby}},
  \bibinfo {author} {\bibfnamefont {D.}~\bibnamefont {Descamps}}, \bibinfo
  {author} {\bibfnamefont {B.}~\bibnamefont {Fabre}}, \bibinfo {author}
  {\bibfnamefont {G.~A.}\ \bibnamefont {Garcia}}, \bibinfo {author}
  {\bibfnamefont {R.}~\bibnamefont {G{\'e}neaux}}, \bibinfo {author}
  {\bibfnamefont {A.~G.}\ \bibnamefont {Harvey}}, \bibinfo {author}
  {\bibfnamefont {F.}~\bibnamefont {L{\'e}gar{\'e}}}, \bibinfo {author}
  {\bibfnamefont {Z.}~\bibnamefont {Mas{\'i}n}}, \bibinfo {author}
  {\bibfnamefont {L.}~\bibnamefont {Nahon}}, \bibinfo {author} {\bibfnamefont
  {A.~F.}\ \bibnamefont {Ordonez}}, \bibinfo {author} {\bibfnamefont
  {S.}~\bibnamefont {Petit}}, \bibinfo {author} {\bibfnamefont
  {B.}~\bibnamefont {Pons}}, \bibinfo {author} {\bibfnamefont {Y.}~\bibnamefont
  {Mairesse}}, \bibinfo {author} {\bibfnamefont {O.}~\bibnamefont {Smirnova}},
  \ and\ \bibinfo {author} {\bibfnamefont {V.}~\bibnamefont {Blanchet}},\
  }\href {\doibase 10.1038/s41567-017-0038-z} {\bibfield  {journal} {\bibinfo
  {journal} {Nature Physics}\ }\textbf {\bibinfo {volume} {14}},\ \bibinfo
  {pages} {484} (\bibinfo {year} {2018})}\BibitemShut {NoStop}%
\bibitem [{\citenamefont {Garcia}\ \emph {et~al.}(2003)\citenamefont {Garcia},
  \citenamefont {Nahon}, \citenamefont {Lebech}, \citenamefont {Houver},
  \citenamefont {Dowek},\ and\ \citenamefont {Powis}}]{garcia_2003}%
  \BibitemOpen
  \bibfield  {author} {\bibinfo {author} {\bibfnamefont {G.~A.}\ \bibnamefont
  {Garcia}}, \bibinfo {author} {\bibfnamefont {L.}~\bibnamefont {Nahon}},
  \bibinfo {author} {\bibfnamefont {M.}~\bibnamefont {Lebech}}, \bibinfo
  {author} {\bibfnamefont {J.-C.}\ \bibnamefont {Houver}}, \bibinfo {author}
  {\bibfnamefont {D.}~\bibnamefont {Dowek}}, \ and\ \bibinfo {author}
  {\bibfnamefont {I.}~\bibnamefont {Powis}},\ }\href {\doibase
  10.1063/1.1621379} {\bibfield  {journal} {\bibinfo  {journal} {The Journal of
  Chemical Physics}\ }\textbf {\bibinfo {volume} {119}},\ \bibinfo {pages}
  {8781} (\bibinfo {year} {2003})},\ \Eprint
  {http://arxiv.org/abs/https://doi.org/10.1063/1.1621379}
  {https://doi.org/10.1063/1.1621379} \BibitemShut {NoStop}%
\bibitem [{\citenamefont {Turchini}\ \emph {et~al.}(2004)\citenamefont
  {Turchini}, \citenamefont {Zema}, \citenamefont {Contini}, \citenamefont
  {Alberti}, \citenamefont {Alagia}, \citenamefont {Stranges}, \citenamefont
  {Fronzoni}, \citenamefont {Stener}, \citenamefont {Decleva},\ and\
  \citenamefont {Prosperi}}]{turchini_2004}%
  \BibitemOpen
  \bibfield  {author} {\bibinfo {author} {\bibfnamefont {S.}~\bibnamefont
  {Turchini}}, \bibinfo {author} {\bibfnamefont {N.}~\bibnamefont {Zema}},
  \bibinfo {author} {\bibfnamefont {G.}~\bibnamefont {Contini}}, \bibinfo
  {author} {\bibfnamefont {G.}~\bibnamefont {Alberti}}, \bibinfo {author}
  {\bibfnamefont {M.}~\bibnamefont {Alagia}}, \bibinfo {author} {\bibfnamefont
  {S.}~\bibnamefont {Stranges}}, \bibinfo {author} {\bibfnamefont
  {G.}~\bibnamefont {Fronzoni}}, \bibinfo {author} {\bibfnamefont
  {M.}~\bibnamefont {Stener}}, \bibinfo {author} {\bibfnamefont
  {P.}~\bibnamefont {Decleva}}, \ and\ \bibinfo {author} {\bibfnamefont
  {T.}~\bibnamefont {Prosperi}},\ }\href {\doibase 10.1103/PhysRevA.70.014502}
  {\bibfield  {journal} {\bibinfo  {journal} {Phys. Rev. A}\ }\textbf {\bibinfo
  {volume} {70}},\ \bibinfo {pages} {014502} (\bibinfo {year}
  {2004})}\BibitemShut {NoStop}%
\bibitem [{\citenamefont {Hergenhahn}\ \emph {et~al.}(2004)\citenamefont
  {Hergenhahn}, \citenamefont {Rennie}, \citenamefont {Kugeler}, \citenamefont
  {Marburger}, \citenamefont {Lischke}, \citenamefont {Powis},\ and\
  \citenamefont {Garcia}}]{Hergenhahn_2004}%
  \BibitemOpen
  \bibfield  {author} {\bibinfo {author} {\bibfnamefont {U.}~\bibnamefont
  {Hergenhahn}}, \bibinfo {author} {\bibfnamefont {E.~E.}\ \bibnamefont
  {Rennie}}, \bibinfo {author} {\bibfnamefont {O.}~\bibnamefont {Kugeler}},
  \bibinfo {author} {\bibfnamefont {S.}~\bibnamefont {Marburger}}, \bibinfo
  {author} {\bibfnamefont {T.}~\bibnamefont {Lischke}}, \bibinfo {author}
  {\bibfnamefont {I.}~\bibnamefont {Powis}}, \ and\ \bibinfo {author}
  {\bibfnamefont {G.}~\bibnamefont {Garcia}},\ }\href {\doibase
  10.1063/1.1651474} {\bibfield  {journal} {\bibinfo  {journal} {The Journal of
  Chemical Physics}\ }\textbf {\bibinfo {volume} {120}},\ \bibinfo {pages}
  {4553} (\bibinfo {year} {2004})},\ \Eprint
  {http://arxiv.org/abs/https://doi.org/10.1063/1.1651474}
  {https://doi.org/10.1063/1.1651474} \BibitemShut {NoStop}%
\bibitem [{\citenamefont {Lischke}\ \emph {et~al.}(2004)\citenamefont
  {Lischke}, \citenamefont {B\"owering}, \citenamefont {Schmidtke},
  \citenamefont {M\"uller}, \citenamefont {Khalil},\ and\ \citenamefont
  {Heinzmann}}]{Lischke_2004}%
  \BibitemOpen
  \bibfield  {author} {\bibinfo {author} {\bibfnamefont {T.}~\bibnamefont
  {Lischke}}, \bibinfo {author} {\bibfnamefont {N.}~\bibnamefont {B\"owering}},
  \bibinfo {author} {\bibfnamefont {B.}~\bibnamefont {Schmidtke}}, \bibinfo
  {author} {\bibfnamefont {N.}~\bibnamefont {M\"uller}}, \bibinfo {author}
  {\bibfnamefont {T.}~\bibnamefont {Khalil}}, \ and\ \bibinfo {author}
  {\bibfnamefont {U.}~\bibnamefont {Heinzmann}},\ }\href {\doibase
  10.1103/PhysRevA.70.022507} {\bibfield  {journal} {\bibinfo  {journal} {Phys.
  Rev. A}\ }\textbf {\bibinfo {volume} {70}},\ \bibinfo {pages} {022507}
  (\bibinfo {year} {2004})}\BibitemShut {NoStop}%
\bibitem [{\citenamefont {Stranges}\ \emph {et~al.}(2005)\citenamefont
  {Stranges}, \citenamefont {Turchini}, \citenamefont {Alagia}, \citenamefont
  {Alberti}, \citenamefont {Contini}, \citenamefont {Decleva}, \citenamefont
  {Fronzoni}, \citenamefont {Stener}, \citenamefont {Zema},\ and\ \citenamefont
  {Prosperi}}]{Stranges_2005}%
  \BibitemOpen
  \bibfield  {author} {\bibinfo {author} {\bibfnamefont {S.}~\bibnamefont
  {Stranges}}, \bibinfo {author} {\bibfnamefont {S.}~\bibnamefont {Turchini}},
  \bibinfo {author} {\bibfnamefont {M.}~\bibnamefont {Alagia}}, \bibinfo
  {author} {\bibfnamefont {G.}~\bibnamefont {Alberti}}, \bibinfo {author}
  {\bibfnamefont {G.}~\bibnamefont {Contini}}, \bibinfo {author} {\bibfnamefont
  {P.}~\bibnamefont {Decleva}}, \bibinfo {author} {\bibfnamefont
  {G.}~\bibnamefont {Fronzoni}}, \bibinfo {author} {\bibfnamefont
  {M.}~\bibnamefont {Stener}}, \bibinfo {author} {\bibfnamefont
  {N.}~\bibnamefont {Zema}}, \ and\ \bibinfo {author} {\bibfnamefont
  {T.}~\bibnamefont {Prosperi}},\ }\href {\doibase 10.1063/1.1940632}
  {\bibfield  {journal} {\bibinfo  {journal} {The Journal of Chemical Physics}\
  }\textbf {\bibinfo {volume} {122}},\ \bibinfo {pages} {244303} (\bibinfo
  {year} {2005})},\ \Eprint
  {http://arxiv.org/abs/https://doi.org/10.1063/1.1940632}
  {https://doi.org/10.1063/1.1940632} \BibitemShut {NoStop}%
\bibitem [{\citenamefont {Giardini}\ \emph {et~al.}(2005)\citenamefont
  {Giardini}, \citenamefont {Catone}, \citenamefont {Stranges}, \citenamefont
  {Satta}, \citenamefont {Tacconi}, \citenamefont {Piccirillo}, \citenamefont
  {Turchini}, \citenamefont {Zema}, \citenamefont {Contini}, \citenamefont
  {Prosperi}, \citenamefont {Decleva}, \citenamefont {Tommaso}, \citenamefont
  {Fronzoni}, \citenamefont {Stener}, \citenamefont {Filippi},\ and\
  \citenamefont {Speranza}}]{Giardini_2005}%
  \BibitemOpen
  \bibfield  {author} {\bibinfo {author} {\bibfnamefont {A.}~\bibnamefont
  {Giardini}}, \bibinfo {author} {\bibfnamefont {D.}~\bibnamefont {Catone}},
  \bibinfo {author} {\bibfnamefont {S.}~\bibnamefont {Stranges}}, \bibinfo
  {author} {\bibfnamefont {M.}~\bibnamefont {Satta}}, \bibinfo {author}
  {\bibfnamefont {M.}~\bibnamefont {Tacconi}}, \bibinfo {author} {\bibfnamefont
  {S.}~\bibnamefont {Piccirillo}}, \bibinfo {author} {\bibfnamefont
  {S.}~\bibnamefont {Turchini}}, \bibinfo {author} {\bibfnamefont
  {N.}~\bibnamefont {Zema}}, \bibinfo {author} {\bibfnamefont {G.}~\bibnamefont
  {Contini}}, \bibinfo {author} {\bibfnamefont {T.}~\bibnamefont {Prosperi}},
  \bibinfo {author} {\bibfnamefont {P.}~\bibnamefont {Decleva}}, \bibinfo
  {author} {\bibfnamefont {D.~D.}\ \bibnamefont {Tommaso}}, \bibinfo {author}
  {\bibfnamefont {G.}~\bibnamefont {Fronzoni}}, \bibinfo {author}
  {\bibfnamefont {M.}~\bibnamefont {Stener}}, \bibinfo {author} {\bibfnamefont
  {A.}~\bibnamefont {Filippi}}, \ and\ \bibinfo {author} {\bibfnamefont
  {M.}~\bibnamefont {Speranza}},\ }\href {\doibase 10.1002/cphc.200400483}
  {\bibfield  {journal} {\bibinfo  {journal} {ChemPhysChem}\ }\textbf {\bibinfo
  {volume} {6}},\ \bibinfo {pages} {1164} (\bibinfo {year} {2005})},\ \Eprint
  {http://arxiv.org/abs/https://onlinelibrary.wiley.com/doi/pdf/10.1002/cphc.200400483}
  {https://onlinelibrary.wiley.com/doi/pdf/10.1002/cphc.200400483} \BibitemShut
  {NoStop}%
\bibitem [{\citenamefont {Harding}\ \emph {et~al.}(2005)\citenamefont
  {Harding}, \citenamefont {Mikajlo}, \citenamefont {Powis}, \citenamefont
  {Barth}, \citenamefont {Joshi}, \citenamefont {Ulrich},\ and\ \citenamefont
  {Hergenhahn}}]{Harding_2005}%
  \BibitemOpen
  \bibfield  {author} {\bibinfo {author} {\bibfnamefont {C.~J.}\ \bibnamefont
  {Harding}}, \bibinfo {author} {\bibfnamefont {E.}~\bibnamefont {Mikajlo}},
  \bibinfo {author} {\bibfnamefont {I.}~\bibnamefont {Powis}}, \bibinfo
  {author} {\bibfnamefont {S.}~\bibnamefont {Barth}}, \bibinfo {author}
  {\bibfnamefont {S.}~\bibnamefont {Joshi}}, \bibinfo {author} {\bibfnamefont
  {V.}~\bibnamefont {Ulrich}}, \ and\ \bibinfo {author} {\bibfnamefont
  {U.}~\bibnamefont {Hergenhahn}},\ }\href {\doibase 10.1063/1.2136150}
  {\bibfield  {journal} {\bibinfo  {journal} {The Journal of Chemical Physics}\
  }\textbf {\bibinfo {volume} {123}},\ \bibinfo {pages} {234310} (\bibinfo
  {year} {2005})},\ \Eprint
  {http://arxiv.org/abs/https://doi.org/10.1063/1.2136150}
  {https://doi.org/10.1063/1.2136150} \BibitemShut {NoStop}%
\bibitem [{\citenamefont {Nahon}\ \emph {et~al.}(2006)\citenamefont {Nahon},
  \citenamefont {Garcia}, \citenamefont {Harding}, \citenamefont {Mikajlo},\
  and\ \citenamefont {Powis}}]{nahon_determination_2006}%
  \BibitemOpen
  \bibfield  {author} {\bibinfo {author} {\bibfnamefont {L.}~\bibnamefont
  {Nahon}}, \bibinfo {author} {\bibfnamefont {G.~A.}\ \bibnamefont {Garcia}},
  \bibinfo {author} {\bibfnamefont {C.~J.}\ \bibnamefont {Harding}}, \bibinfo
  {author} {\bibfnamefont {E.}~\bibnamefont {Mikajlo}}, \ and\ \bibinfo
  {author} {\bibfnamefont {I.}~\bibnamefont {Powis}},\ }\href {\doibase
  10.1063/1.2336432} {\bibfield  {journal} {\bibinfo  {journal} {The Journal of
  Chemical Physics}\ }\textbf {\bibinfo {volume} {125}},\ \bibinfo {pages}
  {114309} (\bibinfo {year} {2006})}\BibitemShut {NoStop}%
\bibitem [{\citenamefont {Contini}\ \emph {et~al.}(2007)\citenamefont
  {Contini}, \citenamefont {Zema}, \citenamefont {Turchini}, \citenamefont
  {Catone}, \citenamefont {Prosperi}, \citenamefont {Carravetta}, \citenamefont
  {Bolognesi}, \citenamefont {Avaldi},\ and\ \citenamefont
  {Feyer}}]{Contini_2007}%
  \BibitemOpen
  \bibfield  {author} {\bibinfo {author} {\bibfnamefont {G.}~\bibnamefont
  {Contini}}, \bibinfo {author} {\bibfnamefont {N.}~\bibnamefont {Zema}},
  \bibinfo {author} {\bibfnamefont {S.}~\bibnamefont {Turchini}}, \bibinfo
  {author} {\bibfnamefont {D.}~\bibnamefont {Catone}}, \bibinfo {author}
  {\bibfnamefont {T.}~\bibnamefont {Prosperi}}, \bibinfo {author}
  {\bibfnamefont {V.}~\bibnamefont {Carravetta}}, \bibinfo {author}
  {\bibfnamefont {P.}~\bibnamefont {Bolognesi}}, \bibinfo {author}
  {\bibfnamefont {L.}~\bibnamefont {Avaldi}}, \ and\ \bibinfo {author}
  {\bibfnamefont {V.}~\bibnamefont {Feyer}},\ }\href {\doibase
  10.1063/1.2779324} {\bibfield  {journal} {\bibinfo  {journal} {The Journal of
  Chemical Physics}\ }\textbf {\bibinfo {volume} {127}},\ \bibinfo {pages}
  {124310} (\bibinfo {year} {2007})},\ \Eprint
  {http://arxiv.org/abs/https://doi.org/10.1063/1.2779324}
  {https://doi.org/10.1063/1.2779324} \BibitemShut {NoStop}%
\bibitem [{\citenamefont {Garcia}\ \emph {et~al.}(2008)\citenamefont {Garcia},
  \citenamefont {Nahon}, \citenamefont {Harding},\ and\ \citenamefont
  {Powis}}]{Garcia_2008}%
  \BibitemOpen
  \bibfield  {author} {\bibinfo {author} {\bibfnamefont {G.~A.}\ \bibnamefont
  {Garcia}}, \bibinfo {author} {\bibfnamefont {L.}~\bibnamefont {Nahon}},
  \bibinfo {author} {\bibfnamefont {C.~J.}\ \bibnamefont {Harding}}, \ and\
  \bibinfo {author} {\bibfnamefont {I.}~\bibnamefont {Powis}},\ }\href
  {\doibase 10.1039/B714095A} {\bibfield  {journal} {\bibinfo  {journal} {Phys.
  Chem. Chem. Phys.}\ }\textbf {\bibinfo {volume} {10}},\ \bibinfo {pages}
  {1628} (\bibinfo {year} {2008})}\BibitemShut {NoStop}%
\bibitem [{\citenamefont {Ulrich}\ \emph {et~al.}(2008)\citenamefont {Ulrich},
  \citenamefont {Barth}, \citenamefont {Joshi}, \citenamefont {Hergenhahn},
  \citenamefont {Mikajlo}, \citenamefont {Harding},\ and\ \citenamefont
  {Powis}}]{ulrich_giant_2008}%
  \BibitemOpen
  \bibfield  {author} {\bibinfo {author} {\bibfnamefont {V.}~\bibnamefont
  {Ulrich}}, \bibinfo {author} {\bibfnamefont {S.}~\bibnamefont {Barth}},
  \bibinfo {author} {\bibfnamefont {S.}~\bibnamefont {Joshi}}, \bibinfo
  {author} {\bibfnamefont {U.}~\bibnamefont {Hergenhahn}}, \bibinfo {author}
  {\bibfnamefont {E.}~\bibnamefont {Mikajlo}}, \bibinfo {author} {\bibfnamefont
  {C.~J.}\ \bibnamefont {Harding}}, \ and\ \bibinfo {author} {\bibfnamefont
  {I.}~\bibnamefont {Powis}},\ }\href {\doibase 10.1021/jp709761u} {\bibfield
  {journal} {\bibinfo  {journal} {The Journal of Physical Chemistry A}\
  }\textbf {\bibinfo {volume} {112}},\ \bibinfo {pages} {3544} (\bibinfo {year}
  {2008})}\BibitemShut {NoStop}%
\bibitem [{\citenamefont {Powis}\ \emph
  {et~al.}(2008{\natexlab{a}})\citenamefont {Powis}, \citenamefont {Harding},
  \citenamefont {Garcia},\ and\ \citenamefont {Nahon}}]{Powis_2008_CPC_9}%
  \BibitemOpen
  \bibfield  {author} {\bibinfo {author} {\bibfnamefont {I.}~\bibnamefont
  {Powis}}, \bibinfo {author} {\bibfnamefont {C.~J.}\ \bibnamefont {Harding}},
  \bibinfo {author} {\bibfnamefont {G.~A.}\ \bibnamefont {Garcia}}, \ and\
  \bibinfo {author} {\bibfnamefont {L.}~\bibnamefont {Nahon}},\ }\href
  {\doibase 10.1002/cphc.200700748} {\bibfield  {journal} {\bibinfo  {journal}
  {ChemPhysChem}\ }\textbf {\bibinfo {volume} {9}},\ \bibinfo {pages} {475}
  (\bibinfo {year} {2008}{\natexlab{a}})},\ \Eprint
  {http://arxiv.org/abs/https://onlinelibrary.wiley.com/doi/pdf/10.1002/cphc.200700748}
  {https://onlinelibrary.wiley.com/doi/pdf/10.1002/cphc.200700748} \BibitemShut
  {NoStop}%
\bibitem [{\citenamefont {Powis}\ \emph
  {et~al.}(2008{\natexlab{b}})\citenamefont {Powis}, \citenamefont {Harding},
  \citenamefont {Barth}, \citenamefont {Joshi}, \citenamefont {Ulrich},\ and\
  \citenamefont {Hergenhahn}}]{Powis_2008_PRA_78}%
  \BibitemOpen
  \bibfield  {author} {\bibinfo {author} {\bibfnamefont {I.}~\bibnamefont
  {Powis}}, \bibinfo {author} {\bibfnamefont {C.~J.}\ \bibnamefont {Harding}},
  \bibinfo {author} {\bibfnamefont {S.}~\bibnamefont {Barth}}, \bibinfo
  {author} {\bibfnamefont {S.}~\bibnamefont {Joshi}}, \bibinfo {author}
  {\bibfnamefont {V.}~\bibnamefont {Ulrich}}, \ and\ \bibinfo {author}
  {\bibfnamefont {U.}~\bibnamefont {Hergenhahn}},\ }\href {\doibase
  10.1103/PhysRevA.78.052501} {\bibfield  {journal} {\bibinfo  {journal} {Phys.
  Rev. A}\ }\textbf {\bibinfo {volume} {78}},\ \bibinfo {pages} {052501}
  (\bibinfo {year} {2008}{\natexlab{b}})}\BibitemShut {NoStop}%
\bibitem [{\citenamefont {Turchini}\ \emph {et~al.}(2009)\citenamefont
  {Turchini}, \citenamefont {Catone}, \citenamefont {Contini}, \citenamefont
  {Zema}, \citenamefont {Irrera}, \citenamefont {Stener}, \citenamefont
  {Tommaso}, \citenamefont {Decleva},\ and\ \citenamefont
  {Prosperi}}]{Turchini_2009}%
  \BibitemOpen
  \bibfield  {author} {\bibinfo {author} {\bibfnamefont {S.}~\bibnamefont
  {Turchini}}, \bibinfo {author} {\bibfnamefont {D.}~\bibnamefont {Catone}},
  \bibinfo {author} {\bibfnamefont {G.}~\bibnamefont {Contini}}, \bibinfo
  {author} {\bibfnamefont {N.}~\bibnamefont {Zema}}, \bibinfo {author}
  {\bibfnamefont {S.}~\bibnamefont {Irrera}}, \bibinfo {author} {\bibfnamefont
  {M.}~\bibnamefont {Stener}}, \bibinfo {author} {\bibfnamefont {D.~D.}\
  \bibnamefont {Tommaso}}, \bibinfo {author} {\bibfnamefont {P.}~\bibnamefont
  {Decleva}}, \ and\ \bibinfo {author} {\bibfnamefont {T.}~\bibnamefont
  {Prosperi}},\ }\href {\doibase 10.1002/cphc.200800862} {\bibfield  {journal}
  {\bibinfo  {journal} {ChemPhysChem}\ }\textbf {\bibinfo {volume} {10}},\
  \bibinfo {pages} {1839} (\bibinfo {year} {2009})}\BibitemShut {NoStop}%
\bibitem [{\citenamefont {Garcia}\ \emph {et~al.}(2010)\citenamefont {Garcia},
  \citenamefont {Soldi-Lose}, \citenamefont {Nahon},\ and\ \citenamefont
  {Powis}}]{Garcia_2010}%
  \BibitemOpen
  \bibfield  {author} {\bibinfo {author} {\bibfnamefont {G.~A.}\ \bibnamefont
  {Garcia}}, \bibinfo {author} {\bibfnamefont {H.}~\bibnamefont {Soldi-Lose}},
  \bibinfo {author} {\bibfnamefont {L.}~\bibnamefont {Nahon}}, \ and\ \bibinfo
  {author} {\bibfnamefont {I.}~\bibnamefont {Powis}},\ }\href {\doibase
  10.1021/jp909344r} {\bibfield  {journal} {\bibinfo  {journal} {The Journal of
  Physical Chemistry A}\ }\textbf {\bibinfo {volume} {114}},\ \bibinfo {pages}
  {847} (\bibinfo {year} {2010})},\ \bibinfo {note} {pMID: 20038111},\ \Eprint
  {http://arxiv.org/abs/https://doi.org/10.1021/jp909344r}
  {https://doi.org/10.1021/jp909344r} \BibitemShut {NoStop}%
\bibitem [{\citenamefont {Nahon}\ \emph {et~al.}(2010)\citenamefont {Nahon},
  \citenamefont {Garcia}, \citenamefont {Soldi-Lose}, \citenamefont {Daly},\
  and\ \citenamefont {Powis}}]{Nahon_2010}%
  \BibitemOpen
  \bibfield  {author} {\bibinfo {author} {\bibfnamefont {L.}~\bibnamefont
  {Nahon}}, \bibinfo {author} {\bibfnamefont {G.~A.}\ \bibnamefont {Garcia}},
  \bibinfo {author} {\bibfnamefont {H.}~\bibnamefont {Soldi-Lose}}, \bibinfo
  {author} {\bibfnamefont {S.}~\bibnamefont {Daly}}, \ and\ \bibinfo {author}
  {\bibfnamefont {I.}~\bibnamefont {Powis}},\ }\href {\doibase
  10.1103/PhysRevA.82.032514} {\bibfield  {journal} {\bibinfo  {journal} {Phys.
  Rev. A}\ }\textbf {\bibinfo {volume} {82}},\ \bibinfo {pages} {032514}
  (\bibinfo {year} {2010})}\BibitemShut {NoStop}%
\bibitem [{\citenamefont {Daly}\ \emph {et~al.}(2011)\citenamefont {Daly},
  \citenamefont {Powis}, \citenamefont {Garcia}, \citenamefont {Soldi-Lose},\
  and\ \citenamefont {Nahon}}]{Daly_2011}%
  \BibitemOpen
  \bibfield  {author} {\bibinfo {author} {\bibfnamefont {S.}~\bibnamefont
  {Daly}}, \bibinfo {author} {\bibfnamefont {I.}~\bibnamefont {Powis}},
  \bibinfo {author} {\bibfnamefont {G.~A.}\ \bibnamefont {Garcia}}, \bibinfo
  {author} {\bibfnamefont {H.}~\bibnamefont {Soldi-Lose}}, \ and\ \bibinfo
  {author} {\bibfnamefont {L.}~\bibnamefont {Nahon}},\ }\href {\doibase
  10.1063/1.3536500} {\bibfield  {journal} {\bibinfo  {journal} {The Journal of
  Chemical Physics}\ }\textbf {\bibinfo {volume} {134}},\ \bibinfo {pages}
  {064306} (\bibinfo {year} {2011})},\ \Eprint
  {http://arxiv.org/abs/https://doi.org/10.1063/1.3536500}
  {https://doi.org/10.1063/1.3536500} \BibitemShut {NoStop}%
\bibitem [{\citenamefont {Catone}\ \emph {et~al.}(2012)\citenamefont {Catone},
  \citenamefont {Stener}, \citenamefont {Decleva}, \citenamefont {Contini},
  \citenamefont {Zema}, \citenamefont {Prosperi}, \citenamefont {Feyer},
  \citenamefont {Prince},\ and\ \citenamefont {Turchini}}]{Catone_2012}%
  \BibitemOpen
  \bibfield  {author} {\bibinfo {author} {\bibfnamefont {D.}~\bibnamefont
  {Catone}}, \bibinfo {author} {\bibfnamefont {M.}~\bibnamefont {Stener}},
  \bibinfo {author} {\bibfnamefont {P.}~\bibnamefont {Decleva}}, \bibinfo
  {author} {\bibfnamefont {G.}~\bibnamefont {Contini}}, \bibinfo {author}
  {\bibfnamefont {N.}~\bibnamefont {Zema}}, \bibinfo {author} {\bibfnamefont
  {T.}~\bibnamefont {Prosperi}}, \bibinfo {author} {\bibfnamefont
  {V.}~\bibnamefont {Feyer}}, \bibinfo {author} {\bibfnamefont {K.~C.}\
  \bibnamefont {Prince}}, \ and\ \bibinfo {author} {\bibfnamefont
  {S.}~\bibnamefont {Turchini}},\ }\href {\doibase
  10.1103/PhysRevLett.108.083001} {\bibfield  {journal} {\bibinfo  {journal}
  {Phys. Rev. Lett.}\ }\textbf {\bibinfo {volume} {108}},\ \bibinfo {pages}
  {083001} (\bibinfo {year} {2012})}\BibitemShut {NoStop}%
\bibitem [{\citenamefont {Catone}\ \emph {et~al.}(2013)\citenamefont {Catone},
  \citenamefont {Turchini}, \citenamefont {Stener}, \citenamefont {Decleva},
  \citenamefont {Contini}, \citenamefont {Prosperi}, \citenamefont {Feyer},
  \citenamefont {Prince},\ and\ \citenamefont {Zema}}]{Catone_2013}%
  \BibitemOpen
  \bibfield  {author} {\bibinfo {author} {\bibfnamefont {D.}~\bibnamefont
  {Catone}}, \bibinfo {author} {\bibfnamefont {S.}~\bibnamefont {Turchini}},
  \bibinfo {author} {\bibfnamefont {M.}~\bibnamefont {Stener}}, \bibinfo
  {author} {\bibfnamefont {P.}~\bibnamefont {Decleva}}, \bibinfo {author}
  {\bibfnamefont {G.}~\bibnamefont {Contini}}, \bibinfo {author} {\bibfnamefont
  {T.}~\bibnamefont {Prosperi}}, \bibinfo {author} {\bibfnamefont
  {V.}~\bibnamefont {Feyer}}, \bibinfo {author} {\bibfnamefont {K.~C.}\
  \bibnamefont {Prince}}, \ and\ \bibinfo {author} {\bibfnamefont
  {N.}~\bibnamefont {Zema}},\ }\href {\doibase 10.1007/s12210-013-0245-1}
  {\bibfield  {journal} {\bibinfo  {journal} {Rendiconti Lincei}\ }\textbf
  {\bibinfo {volume} {24}},\ \bibinfo {pages} {269} (\bibinfo {year}
  {2013})}\BibitemShut {NoStop}%
\bibitem [{\citenamefont {Garcia}\ \emph {et~al.}(2013)\citenamefont {Garcia},
  \citenamefont {Nahon}, \citenamefont {Daly},\ and\ \citenamefont
  {Powis}}]{garcia13}%
  \BibitemOpen
  \bibfield  {author} {\bibinfo {author} {\bibfnamefont {G.~A.}\ \bibnamefont
  {Garcia}}, \bibinfo {author} {\bibfnamefont {L.}~\bibnamefont {Nahon}},
  \bibinfo {author} {\bibfnamefont {S.}~\bibnamefont {Daly}}, \ and\ \bibinfo
  {author} {\bibfnamefont {I.}~\bibnamefont {Powis}},\ }\href {\doibase
  10.1038/ncomms3132} {\bibfield  {journal} {\bibinfo  {journal} {Nature
  Communications}\ }\textbf {\bibinfo {volume} {4}} (\bibinfo {year} {2013}),\
  10.1038/ncomms3132}\BibitemShut {NoStop}%
\bibitem [{\citenamefont {Turchini}\ \emph {et~al.}(2013)\citenamefont
  {Turchini}, \citenamefont {Catone}, \citenamefont {Zema}, \citenamefont
  {Contini}, \citenamefont {Prosperi}, \citenamefont {Decleva}, \citenamefont
  {Stener}, \citenamefont {Rondino}, \citenamefont {Piccirillo}, \citenamefont
  {Prince},\ and\ \citenamefont {Speranza}}]{Turchini_2013}%
  \BibitemOpen
  \bibfield  {author} {\bibinfo {author} {\bibfnamefont {S.}~\bibnamefont
  {Turchini}}, \bibinfo {author} {\bibfnamefont {D.}~\bibnamefont {Catone}},
  \bibinfo {author} {\bibfnamefont {N.}~\bibnamefont {Zema}}, \bibinfo {author}
  {\bibfnamefont {G.}~\bibnamefont {Contini}}, \bibinfo {author} {\bibfnamefont
  {T.}~\bibnamefont {Prosperi}}, \bibinfo {author} {\bibfnamefont
  {P.}~\bibnamefont {Decleva}}, \bibinfo {author} {\bibfnamefont
  {M.}~\bibnamefont {Stener}}, \bibinfo {author} {\bibfnamefont
  {F.}~\bibnamefont {Rondino}}, \bibinfo {author} {\bibfnamefont
  {S.}~\bibnamefont {Piccirillo}}, \bibinfo {author} {\bibfnamefont {K.~C.}\
  \bibnamefont {Prince}}, \ and\ \bibinfo {author} {\bibfnamefont
  {M.}~\bibnamefont {Speranza}},\ }\href {\doibase 10.1002/cphc.201200975}
  {\bibfield  {journal} {\bibinfo  {journal} {ChemPhysChem}\ }\textbf {\bibinfo
  {volume} {14}},\ \bibinfo {pages} {1723} (\bibinfo {year}
  {2013})}\BibitemShut {NoStop}%
\bibitem [{\citenamefont {Tia}\ \emph {et~al.}(2013)\citenamefont {Tia},
  \citenamefont {Cunha~de Miranda}, \citenamefont {Daly}, \citenamefont
  {Gaie-Levrel}, \citenamefont {Garcia}, \citenamefont {Powis},\ and\
  \citenamefont {Nahon}}]{Tia_2013}%
  \BibitemOpen
  \bibfield  {author} {\bibinfo {author} {\bibfnamefont {M.}~\bibnamefont
  {Tia}}, \bibinfo {author} {\bibfnamefont {B.}~\bibnamefont {Cunha~de
  Miranda}}, \bibinfo {author} {\bibfnamefont {S.}~\bibnamefont {Daly}},
  \bibinfo {author} {\bibfnamefont {F.}~\bibnamefont {Gaie-Levrel}}, \bibinfo
  {author} {\bibfnamefont {G.~A.}\ \bibnamefont {Garcia}}, \bibinfo {author}
  {\bibfnamefont {I.}~\bibnamefont {Powis}}, \ and\ \bibinfo {author}
  {\bibfnamefont {L.}~\bibnamefont {Nahon}},\ }\href {\doibase
  10.1021/jz4014129} {\bibfield  {journal} {\bibinfo  {journal} {The Journal of
  Physical Chemistry Letters}\ }\textbf {\bibinfo {volume} {4}},\ \bibinfo
  {pages} {2698} (\bibinfo {year} {2013})},\ \Eprint
  {http://arxiv.org/abs/https://doi.org/10.1021/jz4014129}
  {https://doi.org/10.1021/jz4014129} \BibitemShut {NoStop}%
\bibitem [{\citenamefont {Powis}\ \emph {et~al.}(2014)\citenamefont {Powis},
  \citenamefont {Daly}, \citenamefont {Tia}, \citenamefont {Cunha~de Miranda},
  \citenamefont {Garcia},\ and\ \citenamefont {Nahon}}]{Powis_2014}%
  \BibitemOpen
  \bibfield  {author} {\bibinfo {author} {\bibfnamefont {I.}~\bibnamefont
  {Powis}}, \bibinfo {author} {\bibfnamefont {S.}~\bibnamefont {Daly}},
  \bibinfo {author} {\bibfnamefont {M.}~\bibnamefont {Tia}}, \bibinfo {author}
  {\bibfnamefont {B.}~\bibnamefont {Cunha~de Miranda}}, \bibinfo {author}
  {\bibfnamefont {G.~A.}\ \bibnamefont {Garcia}}, \ and\ \bibinfo {author}
  {\bibfnamefont {L.}~\bibnamefont {Nahon}},\ }\href {\doibase
  10.1039/C3CP53248H} {\bibfield  {journal} {\bibinfo  {journal} {Phys. Chem.
  Chem. Phys.}\ }\textbf {\bibinfo {volume} {16}},\ \bibinfo {pages} {467}
  (\bibinfo {year} {2014})}\BibitemShut {NoStop}%
\bibitem [{\citenamefont {Tia}\ \emph {et~al.}(2014)\citenamefont {Tia},
  \citenamefont {Cunha~de Miranda}, \citenamefont {Daly}, \citenamefont
  {Gaie-Levrel}, \citenamefont {Garcia}, \citenamefont {Nahon},\ and\
  \citenamefont {Powis}}]{Tia_2014}%
  \BibitemOpen
  \bibfield  {author} {\bibinfo {author} {\bibfnamefont {M.}~\bibnamefont
  {Tia}}, \bibinfo {author} {\bibfnamefont {B.}~\bibnamefont {Cunha~de
  Miranda}}, \bibinfo {author} {\bibfnamefont {S.}~\bibnamefont {Daly}},
  \bibinfo {author} {\bibfnamefont {F.}~\bibnamefont {Gaie-Levrel}}, \bibinfo
  {author} {\bibfnamefont {G.~A.}\ \bibnamefont {Garcia}}, \bibinfo {author}
  {\bibfnamefont {L.}~\bibnamefont {Nahon}}, \ and\ \bibinfo {author}
  {\bibfnamefont {I.}~\bibnamefont {Powis}},\ }\href {\doibase
  10.1021/jp5016142} {\bibfield  {journal} {\bibinfo  {journal} {The Journal of
  Physical Chemistry A}\ }\textbf {\bibinfo {volume} {118}},\ \bibinfo {pages}
  {2765} (\bibinfo {year} {2014})},\ \bibinfo {note} {pMID: 24654892},\ \Eprint
  {http://arxiv.org/abs/https://doi.org/10.1021/jp5016142}
  {https://doi.org/10.1021/jp5016142} \BibitemShut {NoStop}%
\bibitem [{\citenamefont {Nahon}\ \emph {et~al.}(2016)\citenamefont {Nahon},
  \citenamefont {Nag}, \citenamefont {Garcia}, \citenamefont {Myrgorodska},
  \citenamefont {Meierhenrich}, \citenamefont {Beaulieu}, \citenamefont
  {Wanie}, \citenamefont {Blanchet}, \citenamefont {Geneaux},\ and\
  \citenamefont {Powis}}]{Nahon2016_det}%
  \BibitemOpen
  \bibfield  {author} {\bibinfo {author} {\bibfnamefont {L.}~\bibnamefont
  {Nahon}}, \bibinfo {author} {\bibfnamefont {L.}~\bibnamefont {Nag}}, \bibinfo
  {author} {\bibfnamefont {G.~A.}\ \bibnamefont {Garcia}}, \bibinfo {author}
  {\bibfnamefont {I.}~\bibnamefont {Myrgorodska}}, \bibinfo {author}
  {\bibfnamefont {U.}~\bibnamefont {Meierhenrich}}, \bibinfo {author}
  {\bibfnamefont {S.}~\bibnamefont {Beaulieu}}, \bibinfo {author}
  {\bibfnamefont {V.}~\bibnamefont {Wanie}}, \bibinfo {author} {\bibfnamefont
  {V.}~\bibnamefont {Blanchet}}, \bibinfo {author} {\bibfnamefont
  {R.}~\bibnamefont {Geneaux}}, \ and\ \bibinfo {author} {\bibfnamefont
  {I.}~\bibnamefont {Powis}},\ }\href {\doibase 10.1039/C6CP01293K} {\bibfield
  {journal} {\bibinfo  {journal} {Phys. Chem. Chem. Phys.}\ }\textbf {\bibinfo
  {volume} {18}},\ \bibinfo {pages} {12696} (\bibinfo {year}
  {2016})}\BibitemShut {NoStop}%
\bibitem [{\citenamefont {Garcia}\ \emph {et~al.}(2016)\citenamefont {Garcia},
  \citenamefont {Dossmann}, \citenamefont {Nahon}, \citenamefont {Daly},\ and\
  \citenamefont {Powis}}]{Garcia_2016}%
  \BibitemOpen
  \bibfield  {author} {\bibinfo {author} {\bibfnamefont {G.~A.}\ \bibnamefont
  {Garcia}}, \bibinfo {author} {\bibfnamefont {H.}~\bibnamefont {Dossmann}},
  \bibinfo {author} {\bibfnamefont {L.}~\bibnamefont {Nahon}}, \bibinfo
  {author} {\bibfnamefont {S.}~\bibnamefont {Daly}}, \ and\ \bibinfo {author}
  {\bibfnamefont {I.}~\bibnamefont {Powis}},\ }\href {\doibase
  10.1002/cphc.201601250} {\bibfield  {journal} {\bibinfo  {journal}
  {ChemPhysChem}\ }\textbf {\bibinfo {volume} {18}},\ \bibinfo {pages} {500}
  (\bibinfo {year} {2016})},\ \Eprint
  {http://arxiv.org/abs/https://onlinelibrary.wiley.com/doi/pdf/10.1002/cphc.201601250}
  {https://onlinelibrary.wiley.com/doi/pdf/10.1002/cphc.201601250} \BibitemShut
  {NoStop}%
\bibitem [{\citenamefont {Catone}\ \emph {et~al.}(2017)\citenamefont {Catone},
  \citenamefont {Turchini}, \citenamefont {Contini}, \citenamefont {Prosperi},
  \citenamefont {Stener}, \citenamefont {Decleva},\ and\ \citenamefont
  {Zema}}]{Catone_2017}%
  \BibitemOpen
  \bibfield  {author} {\bibinfo {author} {\bibfnamefont {D.}~\bibnamefont
  {Catone}}, \bibinfo {author} {\bibfnamefont {S.}~\bibnamefont {Turchini}},
  \bibinfo {author} {\bibfnamefont {G.}~\bibnamefont {Contini}}, \bibinfo
  {author} {\bibfnamefont {T.}~\bibnamefont {Prosperi}}, \bibinfo {author}
  {\bibfnamefont {M.}~\bibnamefont {Stener}}, \bibinfo {author} {\bibfnamefont
  {P.}~\bibnamefont {Decleva}}, \ and\ \bibinfo {author} {\bibfnamefont
  {N.}~\bibnamefont {Zema}},\ }\href {\doibase
  https://doi.org/10.1016/j.chemphys.2016.09.004} {\bibfield  {journal}
  {\bibinfo  {journal} {Chemical Physics}\ }\textbf {\bibinfo {volume} {482}},\
  \bibinfo {pages} {294 } (\bibinfo {year} {2017})},\ \bibinfo {note}
  {electrons and nuclei in motion - correlation and dynamics in molecules (on
  the occasion of the 70th birthday of Lorenz S. Cederbaum)}\BibitemShut
  {NoStop}%
\bibitem [{\citenamefont {Cherepkov}(1982)}]{cherepkov_circular_1982}%
  \BibitemOpen
  \bibfield  {author} {\bibinfo {author} {\bibfnamefont {N.~A.}\ \bibnamefont
  {Cherepkov}},\ }\href
  {http://www.sciencedirect.com/science/article/pii/0009261482836002}
  {\bibfield  {journal} {\bibinfo  {journal} {Chemical Physics Letters}\
  }\textbf {\bibinfo {volume} {87}},\ \bibinfo {pages} {344} (\bibinfo {year}
  {1982})}\BibitemShut {NoStop}%
\bibitem [{\citenamefont {Powis}(2000{\natexlab{b}})}]{Powis_lAlanine_2000}%
  \BibitemOpen
  \bibfield  {author} {\bibinfo {author} {\bibfnamefont {I.}~\bibnamefont
  {Powis}},\ }\href {\doibase 10.1021/jp9933119} {\bibfield  {journal}
  {\bibinfo  {journal} {The Journal of Physical Chemistry A}\ }\textbf
  {\bibinfo {volume} {104}},\ \bibinfo {pages} {878} (\bibinfo {year}
  {2000}{\natexlab{b}})}\BibitemShut {NoStop}%
\bibitem [{\citenamefont {Stener}\ \emph {et~al.}(2004)\citenamefont {Stener},
  \citenamefont {Fronzoni}, \citenamefont {Tomasso},\ and\ \citenamefont
  {Decleva}}]{stener_density_2004}%
  \BibitemOpen
  \bibfield  {author} {\bibinfo {author} {\bibfnamefont {M.}~\bibnamefont
  {Stener}}, \bibinfo {author} {\bibfnamefont {G.}~\bibnamefont {Fronzoni}},
  \bibinfo {author} {\bibfnamefont {D.~D.}\ \bibnamefont {Tomasso}}, \ and\
  \bibinfo {author} {\bibfnamefont {P.}~\bibnamefont {Decleva}},\ }\href
  {\doibase 10.1063/1.1640617} {\bibfield  {journal} {\bibinfo  {journal} {The
  Journal of Chemical Physics}\ }\textbf {\bibinfo {volume} {120}},\ \bibinfo
  {pages} {3284} (\bibinfo {year} {2004})}\BibitemShut {NoStop}%
\bibitem [{\citenamefont {Harding}\ and\ \citenamefont
  {Powis}(2006)}]{Harding_2006}%
  \BibitemOpen
  \bibfield  {author} {\bibinfo {author} {\bibfnamefont {C.~J.}\ \bibnamefont
  {Harding}}\ and\ \bibinfo {author} {\bibfnamefont {I.}~\bibnamefont
  {Powis}},\ }\href {\doibase 10.1063/1.2402175} {\bibfield  {journal}
  {\bibinfo  {journal} {The Journal of Chemical Physics}\ }\textbf {\bibinfo
  {volume} {125}},\ \bibinfo {pages} {234306} (\bibinfo {year} {2006})},\
  \Eprint {http://arxiv.org/abs/https://doi.org/10.1063/1.2402175}
  {https://doi.org/10.1063/1.2402175} \BibitemShut {NoStop}%
\bibitem [{\citenamefont {Tommaso}\ \emph {et~al.}(2006)\citenamefont
  {Tommaso}, \citenamefont {Stener}, \citenamefont {Fronzoni},\ and\
  \citenamefont {Decleva}}]{Tommaso_2006}%
  \BibitemOpen
  \bibfield  {author} {\bibinfo {author} {\bibfnamefont {D.~D.}\ \bibnamefont
  {Tommaso}}, \bibinfo {author} {\bibfnamefont {M.}~\bibnamefont {Stener}},
  \bibinfo {author} {\bibfnamefont {G.}~\bibnamefont {Fronzoni}}, \ and\
  \bibinfo {author} {\bibfnamefont {P.}~\bibnamefont {Decleva}},\ }\href
  {\doibase 10.1002/cphc.200500602} {\bibfield  {journal} {\bibinfo  {journal}
  {ChemPhysChem}\ }\textbf {\bibinfo {volume} {7}},\ \bibinfo {pages} {924}
  (\bibinfo {year} {2006})},\ \Eprint
  {http://arxiv.org/abs/https://onlinelibrary.wiley.com/doi/pdf/10.1002/cphc.200500602}
  {https://onlinelibrary.wiley.com/doi/pdf/10.1002/cphc.200500602} \BibitemShut
  {NoStop}%
\bibitem [{\citenamefont {Stener}\ \emph {et~al.}(2006)\citenamefont {Stener},
  \citenamefont {Di~Tommaso}, \citenamefont {Fronzoni}, \citenamefont
  {Decleva},\ and\ \citenamefont {Powis}}]{stener2006theoretical}%
  \BibitemOpen
  \bibfield  {author} {\bibinfo {author} {\bibfnamefont {M.}~\bibnamefont
  {Stener}}, \bibinfo {author} {\bibfnamefont {D.}~\bibnamefont {Di~Tommaso}},
  \bibinfo {author} {\bibfnamefont {G.}~\bibnamefont {Fronzoni}}, \bibinfo
  {author} {\bibfnamefont {P.}~\bibnamefont {Decleva}}, \ and\ \bibinfo
  {author} {\bibfnamefont {I.}~\bibnamefont {Powis}},\ }\href {\doibase
  10.1063/1.2150438} {\bibfield  {journal} {\bibinfo  {journal} {The Journal of
  Chemical Physics}\ }\textbf {\bibinfo {volume} {124}},\ \bibinfo {pages}
  {024326} (\bibinfo {year} {2006})}\BibitemShut {NoStop}%
\bibitem [{\citenamefont {Dreissigacker}\ and\ \citenamefont
  {Lein}(2014)}]{dreissigacker_photoelectron_2014}%
  \BibitemOpen
  \bibfield  {author} {\bibinfo {author} {\bibfnamefont {I.}~\bibnamefont
  {Dreissigacker}}\ and\ \bibinfo {author} {\bibfnamefont {M.}~\bibnamefont
  {Lein}},\ }\href {\doibase 10.1103/PhysRevA.89.053406} {\bibfield  {journal}
  {\bibinfo  {journal} {Physical Review A}\ }\textbf {\bibinfo {volume} {89}},\
  \bibinfo {pages} {053406} (\bibinfo {year} {2014})}\BibitemShut {NoStop}%
\bibitem [{\citenamefont {Artemyev}\ \emph {et~al.}(2015)\citenamefont
  {Artemyev}, \citenamefont {M\"uller}, \citenamefont {Hochstuhl},\ and\
  \citenamefont {Demekhin}}]{Artemyev_2015}%
  \BibitemOpen
  \bibfield  {author} {\bibinfo {author} {\bibfnamefont {A.~N.}\ \bibnamefont
  {Artemyev}}, \bibinfo {author} {\bibfnamefont {A.~D.}\ \bibnamefont
  {M\"uller}}, \bibinfo {author} {\bibfnamefont {D.}~\bibnamefont {Hochstuhl}},
  \ and\ \bibinfo {author} {\bibfnamefont {P.~V.}\ \bibnamefont {Demekhin}},\
  }\href {\doibase 10.1063/1.4922690} {\bibfield  {journal} {\bibinfo
  {journal} {The Journal of Chemical Physics}\ }\textbf {\bibinfo {volume}
  {142}},\ \bibinfo {pages} {244105} (\bibinfo {year} {2015})},\ \Eprint
  {http://arxiv.org/abs/https://doi.org/10.1063/1.4922690}
  {https://doi.org/10.1063/1.4922690} \BibitemShut {NoStop}%
\bibitem [{\citenamefont {Goetz}\ \emph {et~al.}(2017)\citenamefont {Goetz},
  \citenamefont {Isaev}, \citenamefont {Nikoobakht}, \citenamefont {Berger},\
  and\ \citenamefont {Koch}}]{Koch2017}%
  \BibitemOpen
  \bibfield  {author} {\bibinfo {author} {\bibfnamefont {R.~E.}\ \bibnamefont
  {Goetz}}, \bibinfo {author} {\bibfnamefont {T.~A.}\ \bibnamefont {Isaev}},
  \bibinfo {author} {\bibfnamefont {B.}~\bibnamefont {Nikoobakht}}, \bibinfo
  {author} {\bibfnamefont {R.}~\bibnamefont {Berger}}, \ and\ \bibinfo {author}
  {\bibfnamefont {C.~P.}\ \bibnamefont {Koch}},\ }\href {\doibase
  10.1063/1.4973456} {\bibfield  {journal} {\bibinfo  {journal} {The Journal of
  Chemical Physics}\ }\textbf {\bibinfo {volume} {146}},\ \bibinfo {pages}
  {024306} (\bibinfo {year} {2017})}\BibitemShut {NoStop}%
\bibitem [{\citenamefont {Ilchen}\ \emph {et~al.}(2017)\citenamefont {Ilchen},
  \citenamefont {Hartmann}, \citenamefont {Rupprecht}, \citenamefont
  {Artemyev}, \citenamefont {Coffee}, \citenamefont {Li}, \citenamefont
  {Ohldag}, \citenamefont {Ogasawara}, \citenamefont {Osipov}, \citenamefont
  {Ray}, \citenamefont {Schmidt}, \citenamefont {Wolf}, \citenamefont
  {Ehresmann}, \citenamefont {Moeller}, \citenamefont {Knie},\ and\
  \citenamefont {Demekhin}}]{Ilchen_2017}%
  \BibitemOpen
  \bibfield  {author} {\bibinfo {author} {\bibfnamefont {M.}~\bibnamefont
  {Ilchen}}, \bibinfo {author} {\bibfnamefont {G.}~\bibnamefont {Hartmann}},
  \bibinfo {author} {\bibfnamefont {P.}~\bibnamefont {Rupprecht}}, \bibinfo
  {author} {\bibfnamefont {A.~N.}\ \bibnamefont {Artemyev}}, \bibinfo {author}
  {\bibfnamefont {R.~N.}\ \bibnamefont {Coffee}}, \bibinfo {author}
  {\bibfnamefont {Z.}~\bibnamefont {Li}}, \bibinfo {author} {\bibfnamefont
  {H.}~\bibnamefont {Ohldag}}, \bibinfo {author} {\bibfnamefont
  {H.}~\bibnamefont {Ogasawara}}, \bibinfo {author} {\bibfnamefont
  {T.}~\bibnamefont {Osipov}}, \bibinfo {author} {\bibfnamefont
  {D.}~\bibnamefont {Ray}}, \bibinfo {author} {\bibfnamefont {P.}~\bibnamefont
  {Schmidt}}, \bibinfo {author} {\bibfnamefont {T.~J.~A.}\ \bibnamefont
  {Wolf}}, \bibinfo {author} {\bibfnamefont {A.}~\bibnamefont {Ehresmann}},
  \bibinfo {author} {\bibfnamefont {S.}~\bibnamefont {Moeller}}, \bibinfo
  {author} {\bibfnamefont {A.}~\bibnamefont {Knie}}, \ and\ \bibinfo {author}
  {\bibfnamefont {P.~V.}\ \bibnamefont {Demekhin}},\ }\href {\doibase
  10.1103/PhysRevA.95.053423} {\bibfield  {journal} {\bibinfo  {journal} {Phys.
  Rev. A}\ }\textbf {\bibinfo {volume} {95}},\ \bibinfo {pages} {053423}
  (\bibinfo {year} {2017})}\BibitemShut {NoStop}%
\bibitem [{\citenamefont {Tia}\ \emph {et~al.}(2017)\citenamefont {Tia},
  \citenamefont {Pitzer}, \citenamefont {Kastirke}, \citenamefont {Gatzke},
  \citenamefont {Kim}, \citenamefont {Trinter}, \citenamefont {Rist},
  \citenamefont {Hartung}, \citenamefont {Trabert}, \citenamefont {Siebert},
  \citenamefont {Henrichs}, \citenamefont {Becht}, \citenamefont {Zeller},
  \citenamefont {Gassert}, \citenamefont {Wiegandt}, \citenamefont {Wallauer},
  \citenamefont {Kuhlins}, \citenamefont {Schober}, \citenamefont {Bauer},
  \citenamefont {Wechselberger}, \citenamefont {Burzynski}, \citenamefont
  {Neff}, \citenamefont {Weller}, \citenamefont {Metz}, \citenamefont
  {Kircher}, \citenamefont {Waitz}, \citenamefont {Williams}, \citenamefont
  {Schmidt}, \citenamefont {M\"uller}, \citenamefont {Knie}, \citenamefont
  {Hans}, \citenamefont {Ben~Ltaief}, \citenamefont {Ehresmann}, \citenamefont
  {Berger}, \citenamefont {Fukuzawa}, \citenamefont {Ueda}, \citenamefont
  {Schmidt-B\"ocking}, \citenamefont {D\"orner}, \citenamefont {Jahnke},
  \citenamefont {Demekhin},\ and\ \citenamefont {Sch\"offler}}]{Tia_2017}%
  \BibitemOpen
  \bibfield  {author} {\bibinfo {author} {\bibfnamefont {M.}~\bibnamefont
  {Tia}}, \bibinfo {author} {\bibfnamefont {M.}~\bibnamefont {Pitzer}},
  \bibinfo {author} {\bibfnamefont {G.}~\bibnamefont {Kastirke}}, \bibinfo
  {author} {\bibfnamefont {J.}~\bibnamefont {Gatzke}}, \bibinfo {author}
  {\bibfnamefont {H.-K.}\ \bibnamefont {Kim}}, \bibinfo {author} {\bibfnamefont
  {F.}~\bibnamefont {Trinter}}, \bibinfo {author} {\bibfnamefont
  {J.}~\bibnamefont {Rist}}, \bibinfo {author} {\bibfnamefont {A.}~\bibnamefont
  {Hartung}}, \bibinfo {author} {\bibfnamefont {D.}~\bibnamefont {Trabert}},
  \bibinfo {author} {\bibfnamefont {J.}~\bibnamefont {Siebert}}, \bibinfo
  {author} {\bibfnamefont {K.}~\bibnamefont {Henrichs}}, \bibinfo {author}
  {\bibfnamefont {J.}~\bibnamefont {Becht}}, \bibinfo {author} {\bibfnamefont
  {S.}~\bibnamefont {Zeller}}, \bibinfo {author} {\bibfnamefont
  {H.}~\bibnamefont {Gassert}}, \bibinfo {author} {\bibfnamefont
  {F.}~\bibnamefont {Wiegandt}}, \bibinfo {author} {\bibfnamefont
  {R.}~\bibnamefont {Wallauer}}, \bibinfo {author} {\bibfnamefont
  {A.}~\bibnamefont {Kuhlins}}, \bibinfo {author} {\bibfnamefont
  {C.}~\bibnamefont {Schober}}, \bibinfo {author} {\bibfnamefont
  {T.}~\bibnamefont {Bauer}}, \bibinfo {author} {\bibfnamefont
  {N.}~\bibnamefont {Wechselberger}}, \bibinfo {author} {\bibfnamefont
  {P.}~\bibnamefont {Burzynski}}, \bibinfo {author} {\bibfnamefont
  {J.}~\bibnamefont {Neff}}, \bibinfo {author} {\bibfnamefont {M.}~\bibnamefont
  {Weller}}, \bibinfo {author} {\bibfnamefont {D.}~\bibnamefont {Metz}},
  \bibinfo {author} {\bibfnamefont {M.}~\bibnamefont {Kircher}}, \bibinfo
  {author} {\bibfnamefont {M.}~\bibnamefont {Waitz}}, \bibinfo {author}
  {\bibfnamefont {J.~B.}\ \bibnamefont {Williams}}, \bibinfo {author}
  {\bibfnamefont {L.~P.~H.}\ \bibnamefont {Schmidt}}, \bibinfo {author}
  {\bibfnamefont {A.~D.}\ \bibnamefont {M\"uller}}, \bibinfo {author}
  {\bibfnamefont {A.}~\bibnamefont {Knie}}, \bibinfo {author} {\bibfnamefont
  {A.}~\bibnamefont {Hans}}, \bibinfo {author} {\bibfnamefont {L.}~\bibnamefont
  {Ben~Ltaief}}, \bibinfo {author} {\bibfnamefont {A.}~\bibnamefont
  {Ehresmann}}, \bibinfo {author} {\bibfnamefont {R.}~\bibnamefont {Berger}},
  \bibinfo {author} {\bibfnamefont {H.}~\bibnamefont {Fukuzawa}}, \bibinfo
  {author} {\bibfnamefont {K.}~\bibnamefont {Ueda}}, \bibinfo {author}
  {\bibfnamefont {H.}~\bibnamefont {Schmidt-B\"ocking}}, \bibinfo {author}
  {\bibfnamefont {R.}~\bibnamefont {D\"orner}}, \bibinfo {author}
  {\bibfnamefont {T.}~\bibnamefont {Jahnke}}, \bibinfo {author} {\bibfnamefont
  {P.~V.}\ \bibnamefont {Demekhin}}, \ and\ \bibinfo {author} {\bibfnamefont
  {M.}~\bibnamefont {Sch\"offler}},\ }\href {\doibase
  10.1021/acs.jpclett.7b01000} {\bibfield  {journal} {\bibinfo  {journal} {The
  Journal of Physical Chemistry Letters}\ }\textbf {\bibinfo {volume} {8}},\
  \bibinfo {pages} {2780} (\bibinfo {year} {2017})},\ \bibinfo {note} {pMID:
  28582620},\ \Eprint
  {http://arxiv.org/abs/https://doi.org/10.1021/acs.jpclett.7b01000}
  {https://doi.org/10.1021/acs.jpclett.7b01000} \BibitemShut {NoStop}%
\bibitem [{\citenamefont {Daly}\ \emph {et~al.}(2017)\citenamefont {Daly},
  \citenamefont {Powis}, \citenamefont {Garcia}, \citenamefont {Tia},\ and\
  \citenamefont {Nahon}}]{Daly_2017}%
  \BibitemOpen
  \bibfield  {author} {\bibinfo {author} {\bibfnamefont {S.}~\bibnamefont
  {Daly}}, \bibinfo {author} {\bibfnamefont {I.}~\bibnamefont {Powis}},
  \bibinfo {author} {\bibfnamefont {G.~A.}\ \bibnamefont {Garcia}}, \bibinfo
  {author} {\bibfnamefont {M.}~\bibnamefont {Tia}}, \ and\ \bibinfo {author}
  {\bibfnamefont {L.}~\bibnamefont {Nahon}},\ }\href {\doibase
  10.1063/1.4983139} {\bibfield  {journal} {\bibinfo  {journal} {The Journal of
  Chemical Physics}\ }\textbf {\bibinfo {volume} {147}},\ \bibinfo {pages}
  {013937} (\bibinfo {year} {2017})},\ \Eprint
  {http://arxiv.org/abs/https://doi.org/10.1063/1.4983139}
  {https://doi.org/10.1063/1.4983139} \BibitemShut {NoStop}%
\bibitem [{\citenamefont {Miles}\ \emph {et~al.}(2017)\citenamefont {Miles},
  \citenamefont {Fernandes}, \citenamefont {Young}, \citenamefont {Bond},
  \citenamefont {Crane}, \citenamefont {Ghafur}, \citenamefont {Townsend},
  \citenamefont {S\'a},\ and\ \citenamefont {Greenwood}}]{Miles_2017}%
  \BibitemOpen
  \bibfield  {author} {\bibinfo {author} {\bibfnamefont {J.}~\bibnamefont
  {Miles}}, \bibinfo {author} {\bibfnamefont {D.}~\bibnamefont {Fernandes}},
  \bibinfo {author} {\bibfnamefont {A.}~\bibnamefont {Young}}, \bibinfo
  {author} {\bibfnamefont {C.}~\bibnamefont {Bond}}, \bibinfo {author}
  {\bibfnamefont {S.}~\bibnamefont {Crane}}, \bibinfo {author} {\bibfnamefont
  {O.}~\bibnamefont {Ghafur}}, \bibinfo {author} {\bibfnamefont
  {D.}~\bibnamefont {Townsend}}, \bibinfo {author} {\bibfnamefont
  {J.}~\bibnamefont {S\'a}}, \ and\ \bibinfo {author} {\bibfnamefont
  {J.}~\bibnamefont {Greenwood}},\ }\href {\doibase
  https://doi.org/10.1016/j.aca.2017.06.051} {\bibfield  {journal} {\bibinfo
  {journal} {Analytica Chimica Acta}\ }\textbf {\bibinfo {volume} {984}},\
  \bibinfo {pages} {134 } (\bibinfo {year} {2017})}\BibitemShut {NoStop}%
\bibitem [{\citenamefont {Lux}\ \emph {et~al.}(2012)\citenamefont {Lux},
  \citenamefont {Wollenhaupt}, \citenamefont {Bolze}, \citenamefont {Liang},
  \citenamefont {Koehler}, \citenamefont {Sarpe},\ and\ \citenamefont
  {Baumert}}]{lux_circular_2012}%
  \BibitemOpen
  \bibfield  {author} {\bibinfo {author} {\bibfnamefont {C.}~\bibnamefont
  {Lux}}, \bibinfo {author} {\bibfnamefont {M.}~\bibnamefont {Wollenhaupt}},
  \bibinfo {author} {\bibfnamefont {T.}~\bibnamefont {Bolze}}, \bibinfo
  {author} {\bibfnamefont {Q.}~\bibnamefont {Liang}}, \bibinfo {author}
  {\bibfnamefont {J.}~\bibnamefont {Koehler}}, \bibinfo {author} {\bibfnamefont
  {C.}~\bibnamefont {Sarpe}}, \ and\ \bibinfo {author} {\bibfnamefont
  {T.}~\bibnamefont {Baumert}},\ }\href {\doibase 10.1002/anie.201109035}
  {\bibfield  {journal} {\bibinfo  {journal} {Angewandte Chemie International
  Edition}\ }\textbf {\bibinfo {volume} {51}},\ \bibinfo {pages} {5001}
  (\bibinfo {year} {2012})}\BibitemShut {NoStop}%
\bibitem [{\citenamefont {Lehmann}\ \emph {et~al.}(2013)\citenamefont
  {Lehmann}, \citenamefont {Ram}, \citenamefont {Powis},\ and\ \citenamefont
  {Jansenn}}]{lehmann_imaging_2013}%
  \BibitemOpen
  \bibfield  {author} {\bibinfo {author} {\bibfnamefont {C.~S.}\ \bibnamefont
  {Lehmann}}, \bibinfo {author} {\bibfnamefont {R.~B.}\ \bibnamefont {Ram}},
  \bibinfo {author} {\bibfnamefont {I.}~\bibnamefont {Powis}}, \ and\ \bibinfo
  {author} {\bibfnamefont {M.~H.~M.}\ \bibnamefont {Jansenn}},\ }\href
  {\doibase 10.1063/1.4844295} {\bibfield  {journal} {\bibinfo  {journal} {The
  Journal of Chemical Physics}\ }\textbf {\bibinfo {volume} {139}},\ \bibinfo
  {pages} {234307} (\bibinfo {year} {2013})}\BibitemShut {NoStop}%
\bibitem [{\citenamefont {Rafiee~Fanood}\ \emph {et~al.}(2014)\citenamefont
  {Rafiee~Fanood}, \citenamefont {Powis},\ and\ \citenamefont
  {Janssen}}]{Rafiee_2014}%
  \BibitemOpen
  \bibfield  {author} {\bibinfo {author} {\bibfnamefont {M.~M.}\ \bibnamefont
  {Rafiee~Fanood}}, \bibinfo {author} {\bibfnamefont {I.}~\bibnamefont
  {Powis}}, \ and\ \bibinfo {author} {\bibfnamefont {M.~H.~M.}\ \bibnamefont
  {Janssen}},\ }\href {\doibase 10.1021/jp5113125} {\bibfield  {journal}
  {\bibinfo  {journal} {The Journal of Physical Chemistry A}\ }\textbf
  {\bibinfo {volume} {118}},\ \bibinfo {pages} {11541} (\bibinfo {year}
  {2014})},\ \bibinfo {note} {pMID: 25402546},\ \Eprint
  {http://arxiv.org/abs/https://doi.org/10.1021/jp5113125}
  {https://doi.org/10.1021/jp5113125} \BibitemShut {NoStop}%
\bibitem [{\citenamefont {Rafiee~Fanood}\ \emph {et~al.}(2015)\citenamefont
  {Rafiee~Fanood}, \citenamefont {Janssen},\ and\ \citenamefont
  {Powis}}]{Rafiee_2015}%
  \BibitemOpen
  \bibfield  {author} {\bibinfo {author} {\bibfnamefont {M.~M.}\ \bibnamefont
  {Rafiee~Fanood}}, \bibinfo {author} {\bibfnamefont {M.~H.~M.}\ \bibnamefont
  {Janssen}}, \ and\ \bibinfo {author} {\bibfnamefont {I.}~\bibnamefont
  {Powis}},\ }\href {\doibase 10.1039/C5CP00583C} {\bibfield  {journal}
  {\bibinfo  {journal} {Phys. Chem. Chem. Phys.}\ }\textbf {\bibinfo {volume}
  {17}},\ \bibinfo {pages} {8614} (\bibinfo {year} {2015})}\BibitemShut
  {NoStop}%
\bibitem [{\citenamefont {Lux}\ \emph {et~al.}(2015)\citenamefont {Lux},
  \citenamefont {Wollenhaupt}, \citenamefont {Sarpe},\ and\ \citenamefont
  {Baumert}}]{lux_photoelectron_2015}%
  \BibitemOpen
  \bibfield  {author} {\bibinfo {author} {\bibfnamefont {C.}~\bibnamefont
  {Lux}}, \bibinfo {author} {\bibfnamefont {M.}~\bibnamefont {Wollenhaupt}},
  \bibinfo {author} {\bibfnamefont {C.}~\bibnamefont {Sarpe}}, \ and\ \bibinfo
  {author} {\bibfnamefont {T.}~\bibnamefont {Baumert}},\ }\href {\doibase
  10.1002/cphc.201402643} {\bibfield  {journal} {\bibinfo  {journal}
  {ChemPhysChem}\ }\textbf {\bibinfo {volume} {16}},\ \bibinfo {pages} {115}
  (\bibinfo {year} {2015})}\BibitemShut {NoStop}%
\bibitem [{\citenamefont {Lux}\ \emph {et~al.}(2016)\citenamefont {Lux},
  \citenamefont {Senftleben}, \citenamefont {Sarpe}, \citenamefont
  {Wollenhaupt},\ and\ \citenamefont {Baumert}}]{Lux_2016}%
  \BibitemOpen
  \bibfield  {author} {\bibinfo {author} {\bibfnamefont {C.}~\bibnamefont
  {Lux}}, \bibinfo {author} {\bibfnamefont {A.}~\bibnamefont {Senftleben}},
  \bibinfo {author} {\bibfnamefont {C.}~\bibnamefont {Sarpe}}, \bibinfo
  {author} {\bibfnamefont {M.}~\bibnamefont {Wollenhaupt}}, \ and\ \bibinfo
  {author} {\bibfnamefont {T.}~\bibnamefont {Baumert}},\ }\href
  {http://stacks.iop.org/0953-4075/49/i=2/a=02LT01} {\bibfield  {journal}
  {\bibinfo  {journal} {Journal of Physics B: Atomic, Molecular and Optical
  Physics}\ }\textbf {\bibinfo {volume} {49}},\ \bibinfo {pages} {02LT01}
  (\bibinfo {year} {2016})}\BibitemShut {NoStop}%
\bibitem [{\citenamefont {Fanood}\ \emph {et~al.}(2016)\citenamefont {Fanood},
  \citenamefont {Janssen},\ and\ \citenamefont {Powis}}]{rafiee2016wavelength}%
  \BibitemOpen
  \bibfield  {author} {\bibinfo {author} {\bibfnamefont {M.~M.~R.}\
  \bibnamefont {Fanood}}, \bibinfo {author} {\bibfnamefont {M.~H.~M.}\
  \bibnamefont {Janssen}}, \ and\ \bibinfo {author} {\bibfnamefont
  {I.}~\bibnamefont {Powis}},\ }\href {\doibase 10.1063/1.4963229} {\bibfield
  {journal} {\bibinfo  {journal} {The Journal of Chemical Physics}\ }\textbf
  {\bibinfo {volume} {145}},\ \bibinfo {pages} {124320} (\bibinfo {year}
  {2016})}\BibitemShut {NoStop}%
\bibitem [{\citenamefont {Kastner}\ \emph {et~al.}(2016)\citenamefont
  {Kastner}, \citenamefont {Lux}, \citenamefont {Ring}, \citenamefont
  {Z\"ullighoven}, \citenamefont {Sarpe}, \citenamefont {Senftleben},\ and\
  \citenamefont {Baumert}}]{Kastner_2016}%
  \BibitemOpen
  \bibfield  {author} {\bibinfo {author} {\bibfnamefont {A.}~\bibnamefont
  {Kastner}}, \bibinfo {author} {\bibfnamefont {C.}~\bibnamefont {Lux}},
  \bibinfo {author} {\bibfnamefont {T.}~\bibnamefont {Ring}}, \bibinfo {author}
  {\bibfnamefont {S.}~\bibnamefont {Z\"ullighoven}}, \bibinfo {author}
  {\bibfnamefont {C.}~\bibnamefont {Sarpe}}, \bibinfo {author} {\bibfnamefont
  {A.}~\bibnamefont {Senftleben}}, \ and\ \bibinfo {author} {\bibfnamefont
  {T.}~\bibnamefont {Baumert}},\ }\href {\doibase 10.1002/cphc.201501067}
  {\bibfield  {journal} {\bibinfo  {journal} {ChemPhysChem}\ }\textbf {\bibinfo
  {volume} {17}},\ \bibinfo {pages} {1119} (\bibinfo {year} {2016})},\ \Eprint
  {http://arxiv.org/abs/https://onlinelibrary.wiley.com/doi/pdf/10.1002/cphc.201501067}
  {https://onlinelibrary.wiley.com/doi/pdf/10.1002/cphc.201501067} \BibitemShut
  {NoStop}%
\bibitem [{\citenamefont {Beaulieu}\ \emph {et~al.}(2017)\citenamefont
  {Beaulieu}, \citenamefont {Comby}, \citenamefont {Clergerie}, \citenamefont
  {Caillat}, \citenamefont {Descamps}, \citenamefont {Dudovich}, \citenamefont
  {Fabre}, \citenamefont {G{\'e}neaux}, \citenamefont {L{\'e}gar{\'e}},
  \citenamefont {Petit}, \citenamefont {Pons}, \citenamefont {Porat},
  \citenamefont {Ruchon}, \citenamefont {Ta{\"\i}eb}, \citenamefont
  {Blanchet},\ and\ \citenamefont {Mairesse}}]{beaulieu_science_2017}%
  \BibitemOpen
  \bibfield  {author} {\bibinfo {author} {\bibfnamefont {S.}~\bibnamefont
  {Beaulieu}}, \bibinfo {author} {\bibfnamefont {A.}~\bibnamefont {Comby}},
  \bibinfo {author} {\bibfnamefont {A.}~\bibnamefont {Clergerie}}, \bibinfo
  {author} {\bibfnamefont {J.}~\bibnamefont {Caillat}}, \bibinfo {author}
  {\bibfnamefont {D.}~\bibnamefont {Descamps}}, \bibinfo {author}
  {\bibfnamefont {N.}~\bibnamefont {Dudovich}}, \bibinfo {author}
  {\bibfnamefont {B.}~\bibnamefont {Fabre}}, \bibinfo {author} {\bibfnamefont
  {R.}~\bibnamefont {G{\'e}neaux}}, \bibinfo {author} {\bibfnamefont
  {F.}~\bibnamefont {L{\'e}gar{\'e}}}, \bibinfo {author} {\bibfnamefont
  {S.}~\bibnamefont {Petit}}, \bibinfo {author} {\bibfnamefont
  {B.}~\bibnamefont {Pons}}, \bibinfo {author} {\bibfnamefont {G.}~\bibnamefont
  {Porat}}, \bibinfo {author} {\bibfnamefont {T.}~\bibnamefont {Ruchon}},
  \bibinfo {author} {\bibfnamefont {R.}~\bibnamefont {Ta{\"\i}eb}}, \bibinfo
  {author} {\bibfnamefont {V.}~\bibnamefont {Blanchet}}, \ and\ \bibinfo
  {author} {\bibfnamefont {Y.}~\bibnamefont {Mairesse}},\ }\href {\doibase
  10.1126/science.aao5624} {\bibfield  {journal} {\bibinfo  {journal}
  {Science}\ }\textbf {\bibinfo {volume} {358}},\ \bibinfo {pages} {1288}
  (\bibinfo {year} {2017})}\BibitemShut {NoStop}%
\bibitem [{\citenamefont {Kastner}\ \emph {et~al.}(2017)\citenamefont
  {Kastner}, \citenamefont {Ring}, \citenamefont {Kr\"uger}, \citenamefont
  {Park}, \citenamefont {Sch\"afer}, \citenamefont {Senftleben},\ and\
  \citenamefont {Baumert}}]{Kastner_2017}%
  \BibitemOpen
  \bibfield  {author} {\bibinfo {author} {\bibfnamefont {A.}~\bibnamefont
  {Kastner}}, \bibinfo {author} {\bibfnamefont {T.}~\bibnamefont {Ring}},
  \bibinfo {author} {\bibfnamefont {B.~C.}\ \bibnamefont {Kr\"uger}}, \bibinfo
  {author} {\bibfnamefont {G.~B.}\ \bibnamefont {Park}}, \bibinfo {author}
  {\bibfnamefont {T.}~\bibnamefont {Sch\"afer}}, \bibinfo {author}
  {\bibfnamefont {A.}~\bibnamefont {Senftleben}}, \ and\ \bibinfo {author}
  {\bibfnamefont {T.}~\bibnamefont {Baumert}},\ }\href {\doibase
  10.1063/1.4982614} {\bibfield  {journal} {\bibinfo  {journal} {The Journal of
  Chemical Physics}\ }\textbf {\bibinfo {volume} {147}},\ \bibinfo {pages}
  {013926} (\bibinfo {year} {2017})},\ \Eprint
  {http://arxiv.org/abs/https://doi.org/10.1063/1.4982614}
  {https://doi.org/10.1063/1.4982614} \BibitemShut {NoStop}%
\bibitem [{\citenamefont {Comby}\ \emph {et~al.}(2016)\citenamefont {Comby},
  \citenamefont {Beaulieu}, \citenamefont {Boggio-Pasqua}, \citenamefont
  {Descamps}, \citenamefont {Legare}, \citenamefont {Nahon}, \citenamefont
  {Petit}, \citenamefont {Pons}, \citenamefont {Fabre}, \citenamefont
  {Mairesse},\ and\ \citenamefont {Blanchet}}]{comby_relaxation_2016}%
  \BibitemOpen
  \bibfield  {author} {\bibinfo {author} {\bibfnamefont {A.}~\bibnamefont
  {Comby}}, \bibinfo {author} {\bibfnamefont {S.}~\bibnamefont {Beaulieu}},
  \bibinfo {author} {\bibfnamefont {M.}~\bibnamefont {Boggio-Pasqua}}, \bibinfo
  {author} {\bibfnamefont {D.}~\bibnamefont {Descamps}}, \bibinfo {author}
  {\bibfnamefont {F.}~\bibnamefont {Legare}}, \bibinfo {author} {\bibfnamefont
  {L.}~\bibnamefont {Nahon}}, \bibinfo {author} {\bibfnamefont
  {S.}~\bibnamefont {Petit}}, \bibinfo {author} {\bibfnamefont
  {B.}~\bibnamefont {Pons}}, \bibinfo {author} {\bibfnamefont {B.}~\bibnamefont
  {Fabre}}, \bibinfo {author} {\bibfnamefont {Y.}~\bibnamefont {Mairesse}}, \
  and\ \bibinfo {author} {\bibfnamefont {V.}~\bibnamefont {Blanchet}},\ }\href
  {\doibase 10.1021/acs.jpclett.6b02065} {\bibfield  {journal} {\bibinfo
  {journal} {The Journal of Physical Chemistry Letters}\ }\textbf {\bibinfo
  {volume} {7}},\ \bibinfo {pages} {4514} (\bibinfo {year} {2016})}\BibitemShut
  {NoStop}%
\bibitem [{\citenamefont {Beaulieu}\ \emph
  {et~al.}(2016{\natexlab{a}})\citenamefont {Beaulieu}, \citenamefont {Ferre},
  \citenamefont {Geneaux}, \citenamefont {Canonge}, \citenamefont {Descamps},
  \citenamefont {Fabre}, \citenamefont {Fedorov}, \citenamefont {Legare},
  \citenamefont {Petit}, \citenamefont {Ruchon}, \citenamefont {{V Blanchet}},
  \citenamefont {Mairesse},\ and\ \citenamefont
  {Pons}}]{beaulieu_universality_2016}%
  \BibitemOpen
  \bibfield  {author} {\bibinfo {author} {\bibfnamefont {S.}~\bibnamefont
  {Beaulieu}}, \bibinfo {author} {\bibfnamefont {A.}~\bibnamefont {Ferre}},
  \bibinfo {author} {\bibfnamefont {R.}~\bibnamefont {Geneaux}}, \bibinfo
  {author} {\bibfnamefont {R.}~\bibnamefont {Canonge}}, \bibinfo {author}
  {\bibfnamefont {D.}~\bibnamefont {Descamps}}, \bibinfo {author}
  {\bibfnamefont {B.}~\bibnamefont {Fabre}}, \bibinfo {author} {\bibfnamefont
  {N.}~\bibnamefont {Fedorov}}, \bibinfo {author} {\bibfnamefont
  {F.}~\bibnamefont {Legare}}, \bibinfo {author} {\bibfnamefont
  {S.}~\bibnamefont {Petit}}, \bibinfo {author} {\bibfnamefont
  {T.}~\bibnamefont {Ruchon}}, \bibinfo {author} {\bibnamefont {{V Blanchet}}},
  \bibinfo {author} {\bibfnamefont {Y.}~\bibnamefont {Mairesse}}, \ and\
  \bibinfo {author} {\bibfnamefont {B.}~\bibnamefont {Pons}},\ }\href {\doibase
  10.1088/1367-2630/18/10/102002} {\bibfield  {journal} {\bibinfo  {journal}
  {New Journal of Physics}\ }\textbf {\bibinfo {volume} {18}},\ \bibinfo
  {pages} {102002} (\bibinfo {year} {2016}{\natexlab{a}})}\BibitemShut
  {NoStop}%
\bibitem [{\citenamefont {Beaulieu}\ \emph
  {et~al.}(2016{\natexlab{b}})\citenamefont {Beaulieu}, \citenamefont {Comby},
  \citenamefont {Fabre}, \citenamefont {Descamps}, \citenamefont {Ferre},
  \citenamefont {Garcia}, \citenamefont {Geneaux}, \citenamefont {Legare},
  \citenamefont {Nahon}, \citenamefont {Petit}, \citenamefont {Ruchon},
  \citenamefont {Pons}, \citenamefont {Blanchet},\ and\ \citenamefont
  {Mairesse}}]{Beaulieu_2016_Faraday}%
  \BibitemOpen
  \bibfield  {author} {\bibinfo {author} {\bibfnamefont {S.}~\bibnamefont
  {Beaulieu}}, \bibinfo {author} {\bibfnamefont {A.}~\bibnamefont {Comby}},
  \bibinfo {author} {\bibfnamefont {B.}~\bibnamefont {Fabre}}, \bibinfo
  {author} {\bibfnamefont {D.}~\bibnamefont {Descamps}}, \bibinfo {author}
  {\bibfnamefont {A.}~\bibnamefont {Ferre}}, \bibinfo {author} {\bibfnamefont
  {G.}~\bibnamefont {Garcia}}, \bibinfo {author} {\bibfnamefont
  {R.}~\bibnamefont {Geneaux}}, \bibinfo {author} {\bibfnamefont
  {F.}~\bibnamefont {Legare}}, \bibinfo {author} {\bibfnamefont
  {L.}~\bibnamefont {Nahon}}, \bibinfo {author} {\bibfnamefont
  {S.}~\bibnamefont {Petit}}, \bibinfo {author} {\bibfnamefont
  {T.}~\bibnamefont {Ruchon}}, \bibinfo {author} {\bibfnamefont
  {B.}~\bibnamefont {Pons}}, \bibinfo {author} {\bibfnamefont {V.}~\bibnamefont
  {Blanchet}}, \ and\ \bibinfo {author} {\bibfnamefont {Y.}~\bibnamefont
  {Mairesse}},\ }\href {\doibase 10.1039/C6FD00113K} {\bibfield  {journal}
  {\bibinfo  {journal} {Faraday Discuss.}\ }\textbf {\bibinfo {volume} {194}},\
  \bibinfo {pages} {325} (\bibinfo {year} {2016}{\natexlab{b}})}\BibitemShut
  {NoStop}%
\bibitem [{\citenamefont {Yao}\ \emph {et~al.}(2008)\citenamefont {Yao},
  \citenamefont {Xiao},\ and\ \citenamefont {Niu}}]{yao2008}%
  \BibitemOpen
  \bibfield  {author} {\bibinfo {author} {\bibfnamefont {W.}~\bibnamefont
  {Yao}}, \bibinfo {author} {\bibfnamefont {D.}~\bibnamefont {Xiao}}, \ and\
  \bibinfo {author} {\bibfnamefont {Q.}~\bibnamefont {Niu}},\ }\href {\doibase
  10.1103/PhysRevB.77.235406} {\bibfield  {journal} {\bibinfo  {journal} {Phys.
  Rev. B}\ }\textbf {\bibinfo {volume} {77}},\ \bibinfo {pages} {235406}
  (\bibinfo {year} {2008})}\BibitemShut {NoStop}%
\bibitem [{\citenamefont {Dubs}\ \emph
  {et~al.}(1985{\natexlab{a}})\citenamefont {Dubs}, \citenamefont {Dixit},\
  and\ \citenamefont {McKoy}}]{dubs_circular_1985}%
  \BibitemOpen
  \bibfield  {author} {\bibinfo {author} {\bibfnamefont {R.~L.}\ \bibnamefont
  {Dubs}}, \bibinfo {author} {\bibfnamefont {S.~N.}\ \bibnamefont {Dixit}}, \
  and\ \bibinfo {author} {\bibfnamefont {V.}~\bibnamefont {McKoy}},\ }\href
  {\doibase 10.1103/PhysRevLett.54.1249} {\bibfield  {journal} {\bibinfo
  {journal} {Physical Review Letters}\ }\textbf {\bibinfo {volume} {54}},\
  \bibinfo {pages} {1249} (\bibinfo {year} {1985}{\natexlab{a}})}\BibitemShut
  {NoStop}%
\bibitem [{\citenamefont {Dubs}\ \emph
  {et~al.}(1985{\natexlab{b}})\citenamefont {Dubs}, \citenamefont {Dixit},\
  and\ \citenamefont {McKoy}}]{dubs_circular_1985-1}%
  \BibitemOpen
  \bibfield  {author} {\bibinfo {author} {\bibfnamefont {R.~L.}\ \bibnamefont
  {Dubs}}, \bibinfo {author} {\bibfnamefont {S.~N.}\ \bibnamefont {Dixit}}, \
  and\ \bibinfo {author} {\bibfnamefont {V.}~\bibnamefont {McKoy}},\ }\href
  {\doibase 10.1103/PhysRevB.32.8389} {\bibfield  {journal} {\bibinfo
  {journal} {Physical Review B}\ }\textbf {\bibinfo {volume} {32}},\ \bibinfo
  {pages} {8389} (\bibinfo {year} {1985}{\natexlab{b}})}\BibitemShut {NoStop}%
\bibitem [{\citenamefont {Holmegaard}\ \emph {et~al.}(2010)\citenamefont
  {Holmegaard}, \citenamefont {Hansen}, \citenamefont {Kalh{\o}j},
  \citenamefont {Louise~Kragh}, \citenamefont {Stapelfeldt}, \citenamefont
  {Filsinger}, \citenamefont {K\"{u}pper}, \citenamefont {Meijer},
  \citenamefont {Dimitrovski}, \citenamefont {Abu-samha}, \citenamefont
  {Martiny},\ and\ \citenamefont
  {Bojer~Madsen}}]{holmegaard_photoelectron_2010}%
  \BibitemOpen
  \bibfield  {author} {\bibinfo {author} {\bibfnamefont {L.}~\bibnamefont
  {Holmegaard}}, \bibinfo {author} {\bibfnamefont {J.~L.}\ \bibnamefont
  {Hansen}}, \bibinfo {author} {\bibfnamefont {L.}~\bibnamefont {Kalh{\o}j}},
  \bibinfo {author} {\bibfnamefont {S.}~\bibnamefont {Louise~Kragh}}, \bibinfo
  {author} {\bibfnamefont {H.}~\bibnamefont {Stapelfeldt}}, \bibinfo {author}
  {\bibfnamefont {F.}~\bibnamefont {Filsinger}}, \bibinfo {author}
  {\bibfnamefont {J.}~\bibnamefont {K\"{u}pper}}, \bibinfo {author}
  {\bibfnamefont {G.}~\bibnamefont {Meijer}}, \bibinfo {author} {\bibfnamefont
  {D.}~\bibnamefont {Dimitrovski}}, \bibinfo {author} {\bibfnamefont
  {M.}~\bibnamefont {Abu-samha}}, \bibinfo {author} {\bibfnamefont {C.~P.~J.}\
  \bibnamefont {Martiny}}, \ and\ \bibinfo {author} {\bibfnamefont
  {L.}~\bibnamefont {Bojer~Madsen}},\ }\href {\doibase 10.1038/nphys1666}
  {\bibfield  {journal} {\bibinfo  {journal} {Nature Physics}\ }\textbf
  {\bibinfo {volume} {6}},\ \bibinfo {pages} {428} (\bibinfo {year}
  {2010})}\BibitemShut {NoStop}%
\end{thebibliography}%

\end{document}